\documentclass[journal]{IEEEtran}

\ifCLASSINFOpdf
\else
\fi
\usepackage{amsmath}
\usepackage{graphicx}
\usepackage{graphicx}
\usepackage{xcolor}
\usepackage[bottom]{footmisc}
\usepackage{epstopdf}
\usepackage{cite}

\usepackage{tikz}

\usetikzlibrary{arrows,shapes,positioning,shadows,trees}

\tikzset{
	basic/.style  = {draw, text width=5cm, drop shadow, font=\sffamily, rectangle},
	root/.style   = {basic, rounded corners=2pt, thin, align=center,
		fill=blue!20},
	level 2/.style = {draw, rounded corners=8pt, thin,align=center, fill=blue!10,
		text width=3cm},
	level 3/.style = {draw, thin, align=center, fill=green!10, text width=8em}
}

\usepackage{long2}
\usepackage{pdflscape}
\usepackage{longtable}
\usepackage{dingbat}
\usepackage{pifont}
\newcommand{\cmark}{\ding{51}}%
\newcommand{\xmark}{\ding{55}}%
\usepackage[right]{lineno}
\usepackage{algorithm}
\usepackage{algorithmic}
\usepackage{microtype}

\makeatletter
\def\BState{\State\hskip-\ALG@thistlm}
\makeatother

\begin{document}
	\raggedbottom
	%
	\title{Security, Privacy and Trust for Smart Mobile-Internet of Things (M-IoT): A Survey}
	%
	%
	%
	
	\author{Vishal Sharma, Ilsun You*, Karl Andersson, Francesco Palmieri, Mubashir Husain Rehmani, Jaedeok Lim
		\thanks{V. Sharma and I. You (Corresponding Author) are with Department of Information Security Engineering, Soonchunhyang University, Asan-si 31538, Republic of Korea, Email: vishal\_sharma2012@hotmail.com, ilsunu@gmail.com. K. Andersson is with the Luleå University of Technology, Skellefteå, Sweden, Email: karl.andersson@ltu.se. F. Palmieri is with the University of Salerno, Fisciano, Italy, Email: fpalmieri@unisa.it. M.H. Rehmani is with the Cork Institute of Technology (CIT), Ireland, Email: mshrehmani@gmail.com. J. Lim is with the Information Security Research Division, Electronics and Telecommunications Research Institute (ETRI), Gwangju, South Korea, Email: jdscol92@etri.re.kr}
		{\thanks{Manuscript received .....; revised ........}}}

	\maketitle
	
	\begin{abstract}
		\textcolor{black}{With an enormous range of applications, Internet of Things (IoT) has magnetized industries and academicians from everywhere. IoT facilitates operations through ubiquitous connectivity by providing Internet access to all the devices with computing capabilities. With the evolution of wireless infrastructure, the focus from simple IoT has been shifted to smart, connected and mobile IoT (M-IoT) devices and platforms, which can enable low-complexity, low-cost and efficient computing through sensors, machines, and even crowdsourcing. All these devices can be grouped under a common term of M-IoT. Even though the positive impact on applications has been tremendous, security, privacy and trust are still the major concerns for such networks and an insufficient enforcement of these requirements introduces non-negligible threats to M-IoT devices and platforms. Thus, it is important to understand the range of solutions which are available for providing a secure, privacy-compliant, and trustworthy mechanism for M-IoT. There is no direct survey available, which focuses on security, privacy, trust, secure protocols, physical layer security and handover protections in M-IoT. This paper covers such requisites and presents comparisons of state-the-art solutions for IoT which are applicable to security, privacy, and trust in smart and connected M-IoT networks. Apart from these, various challenges, applications, advantages, technologies, standards, open issues, and roadmap for security, privacy and trust are also discussed in this paper.}
	\end{abstract}
	

	%
	\IEEEpeerreviewmaketitle
	
	\section{\textcolor{black}{Introduction}}
	\textcolor{black}{Mobile-Internet of Things (M-IoT) offers vendors a utility for providing smart services to their users by forming a highly sustainable, secure and cost-effective network~\cite{farris2018federated,liu2018neighbor,ghasempour2016optimizing}}. The smart M-IoT paves a way for incorporating a large set of services like healthcare, business monitoring, strategic planning, public safety communications, weather forecasting, navigation, reconnaissance, and data acquisition~\cite{misra2019detour,afzal2019enabling,celik2019program}. Security and efficiency of these services are the main objectives of organizations aiming at the spread of smart M-IoT. \textcolor{black}{M-IoT focuses on user-specific commercialization, where users pay as per their active applications while offering them with flexible and dynamic procedures for the selection of a service~\cite{cheng2017traffic,goudos2017survey,ghasempour2016optimum}}.
	In order to enhance the security, utility and lifetime of services, most of the established business enterprises are looking forward to procuring long range and low power solutions for connecting billions of devices to their core networks without much dependence on the existing infrastructure. Such an ideology allows for easier management and configuration of M-IoT networks and associated devices. \textcolor{black}{Solutions like Low Power Wide Area Network (LPWAN), Long Range Wide Area Network (LoRaWAN) and Narrow Band-IoT (NB-IoT) are efficient in deploying M-IoT networks~\cite{3G3,sharma2018lorawan,adelantado2017understanding,neumann2016indoor}}. However, at the moment, both the technologies are rival to each other and their applicability and use cases are subject to the decisions of deploying companies and the regulations of the countries involved in their development. With better reach and ease of deployment over existing cellular setup, NB-IoT and Long Term Evolution for Machines (LTE-M) are under consideration as their unification will enhance the types of applications for M-IoT by adopting service strategies similar to mobile networks~\cite{jermyn2015scalability}~\cite{jover2015connection}.
	
	The major interests of some leading organizations have been towards the establishment of a different spectrum which is also obtained as a dedicated range from their allocated space or frequency band. Technologies like Software-Defined Networking (SDN) and Network-Function Virtualization (NFV) provide an altogether different way for deploying these networks in a secure way~\cite{chakrabarty2015black,ojo2016sdn,sharma2017efficient,bi2019software,muthanna2019secure}. With a centralized controller, a common node helps to monitor the network, whereas network slicing through NFV will help to distribute the implementation and management of SDNs. M-IoT can operate as a separate slice, and a local or global controller can manage the related activities. Procedures like secondary authentication and group authentication can be seen as potential solutions for ensuring security in smart M-IoT. However, the effective implementation of rules and policies at the control layer due to the configuration complexity and artifacts requires intelligent solutions that can be assured by using certain aspects of optimization, machine learning or artificial intelligence.
	
	\begin{figure}
		\centering
		\includegraphics[width=250px]{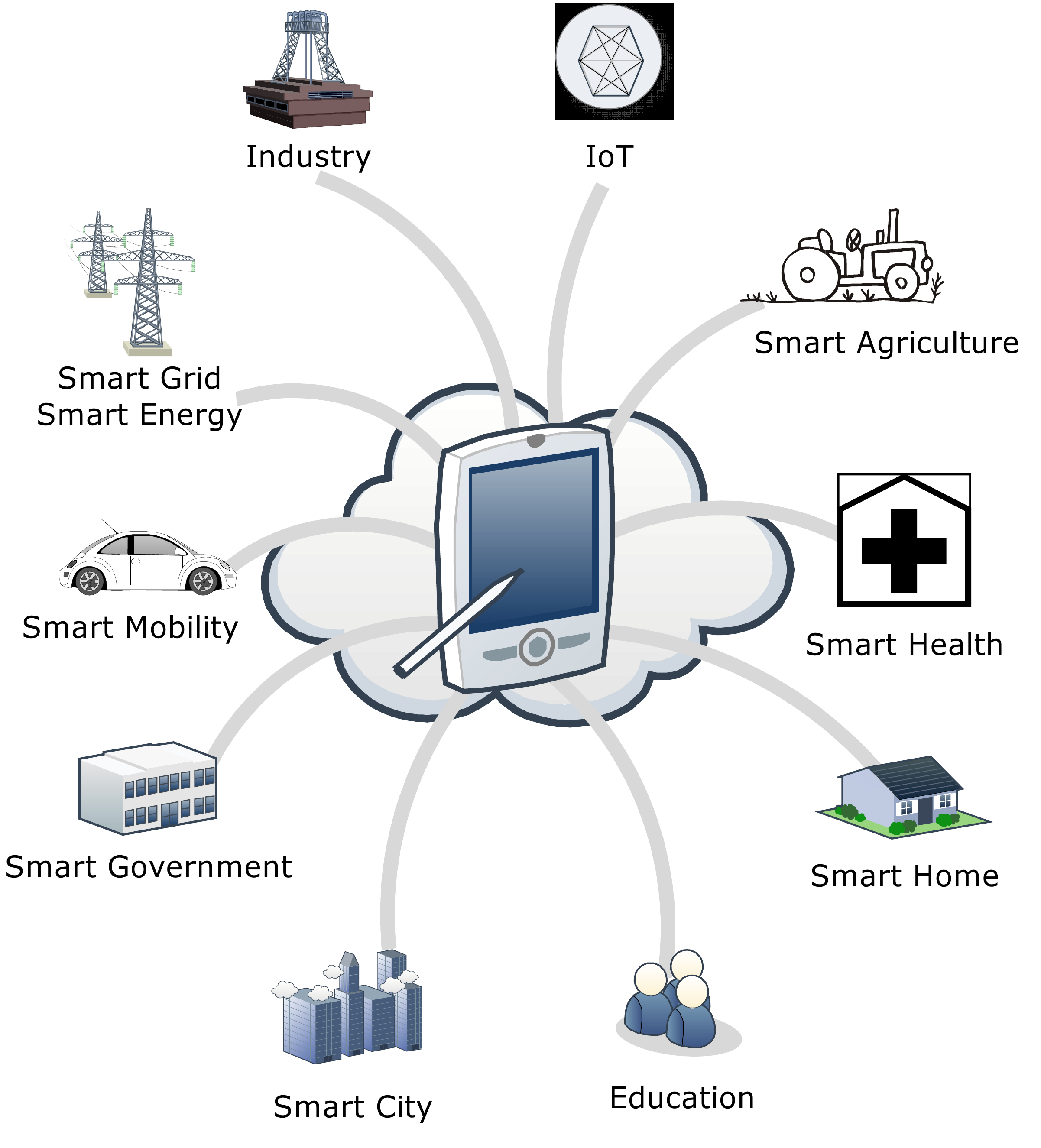}
		\caption{An overview of M-IoT applications.}\label{fig1}
	\end{figure}
	
	\textcolor{black}{In smart M-IoT, security refers to the protection of the infrastructure from potentially hazardous components and users, which may exploit the network with vulnerabilities, based on the known/unknown cyber attacks. For privacy, it deals with the preservation of lawfulness in sharing the information about-and-between the involved devices. Since smart M-IoT will be dealing with a lot of connected components, maintenance of isolation in traffic patterns and establishing anonymity of users becomes an utmost requirement. Trust refers to the faithfulness in the identification of devices for communication. It further involves the reputation-building between the devices and the infrastructure leading a way to make the network secure while preserving its privacy.}
	
	Current market trends have shown that despite several solutions for establishing M-IoT communications, the end to end security will be one of the major concerns for the mobile operators. Identification of new cyber threats, which consist in zero-day attacks is another major requirement of the security industry~\cite{aksu2018advertising}~\cite{ammar2018internet}. It is estimated that M-IoT will hit the market by 2025 with maximum revenue being generated from the security, privacy and trust-based services. Even the major role players will be a low power long range communication models, which can be evaluated around 15+ billion dollars at the same time~\cite{sawng2017technology}. Thus, it is required that the existing state-of-the-art must be followed and evaluated on the basis of performance metrics and parameters that enhance the security, privacy, and trust in M-IoT.
	
	\begin{figure*}[!ht]
		\centering
		\includegraphics[width=400px]{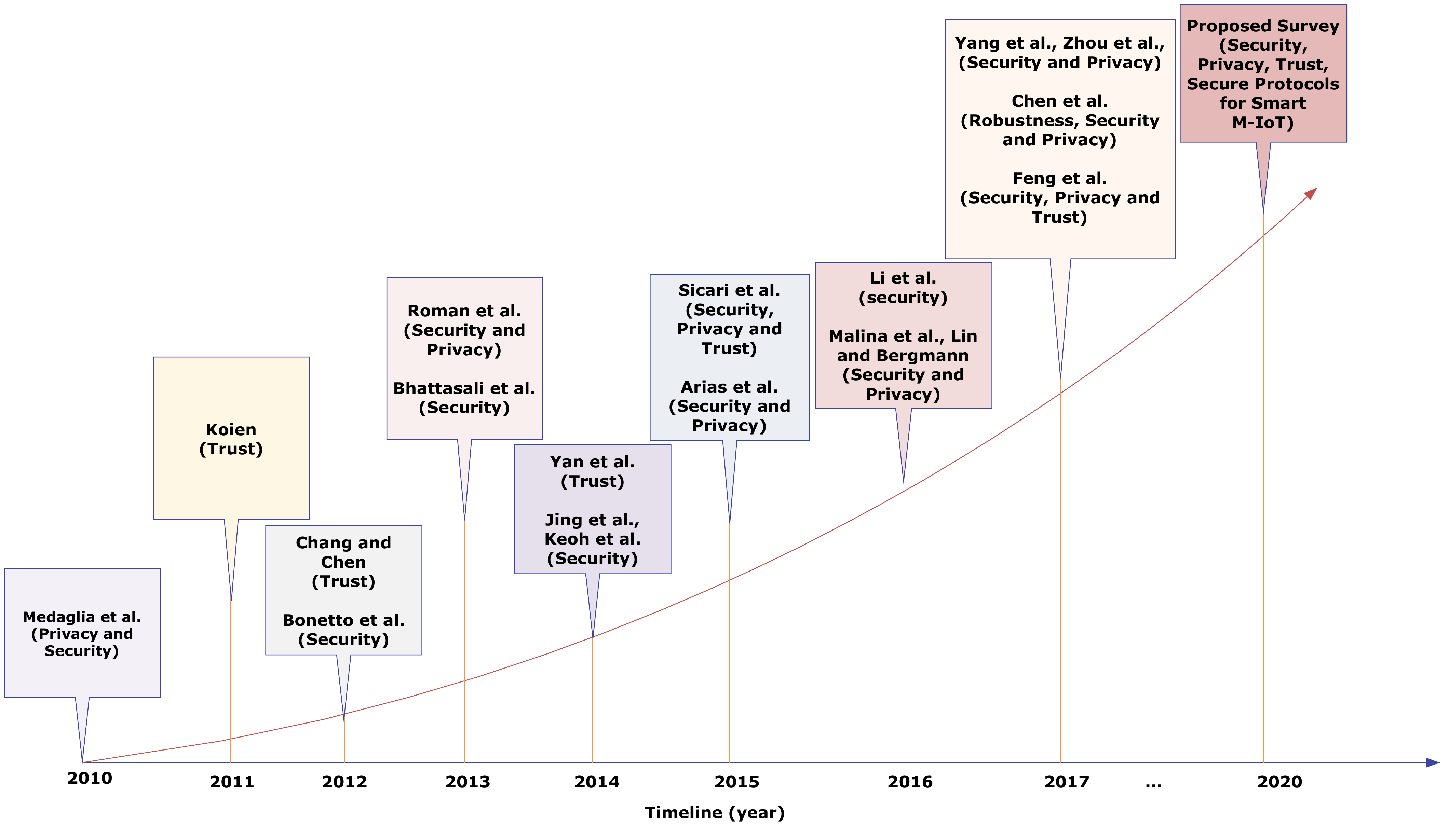}
		\caption{A road map of different studies on security, privacy and trust in IoT and M-IoT between 2010 and 2018.}\label{fig3}
	\end{figure*}
	
	\subsection{Advantages and Applications of Smart M-IoT}
	Smart M-IoT focuses the applications which help in regulating the daily works of their users. \textcolor{black}{Smart M-IoT provides a different set of applications in largely diversified areas such as a smart factory, smart city, smart home, smart grid~\cite{ghasempour2015optimized,ghasempour2016boptimum,ghasempour2016boptimized,chasempour2016boptimizing}, healthcare, personal care, emergencies~\cite{sharma2017energy}, as shown in Fig.~\ref{fig1}}. With smart M-IoT, it becomes easier for both users as well as business organization to accommodate and host services through intelligent architecture with effective security. In terms of market trends, business houses are looking at a huge monetary advantage from smart M-IoT networks and applications. Including these, other advantages and applications of smart M-IoT are as follows:
	\begin{itemize}
		\item Formation of the contextual network through intelligent and rapid data acquisition and processing.
		\item Self-configuring capacity and support for a large set of devices through a common interface.
		\item Support for human to device and device to device communication with lower overheads and low-complexity.
		\item Information management, processing, and validation and data flow management across a wide range of the network.
		\item \textcolor{black}{Support for real-world applications such as driverless cars, urban-surveillance, smart retailing, industrial Internet, and even provisioning of application base for Augmented Reality (AR)/Virtual Reality(VR) services}.
		\item Low-cost deployment and development of personal applications as well as private networks and clouds.
		\item Requires low-frequent maintenance and can be operated through distant mode. On-site evaluations may be subject to special requirements and upgrades.
		\item Supports crowdsourcing as well as edge-computing models by forming an on-demand network in case of public safety communications.
		\item Industrial automation and personalized control formations through light-weight and low-complex Integrated Development Environments (IDEs). Further, M-IoT also helps in tracking the traffic-flows by incorporating transmissions over dynamic nodes, such as drones, smart cars, autonomous bicycles and rail networks.
	\end{itemize}
	
	\begin{table*}[!ht]
		\fontsize{7}{10}\selectfont
		\color{black}
		\centering
		\caption{Abbreviations and Key Terms.}
		\label{st3}
		\begin{tabular}{llll}
			\hline
			\textbf{Abbreviation} & \textbf{Full Form} & \textbf{Abbreviation} & \textbf{Full Form} \\
			\hline
			\hline
			ACL2    &     A Computational Logic for Applicative Common Lisp     &    NFV    &     Network Function Virtualization     \\
			AP    &     Access Point     &    OMA-DM    &     Open Mobile Alliance-Device Management     \\
			AMQP    &     Advanced Message Queuing Protocol     &    PKI & Public Key Infrastructure \\
			AR/VR    &     Augmented Reality/Virtual Reality     &    PWD    &     Password     \\
			AKA    &     Authentication and Key Agreement     &    P2MP    &     Peer to Multi Peers     \\
			AKA'    &     Authentication and Key Agreement Prime     &   P2P    &     Peer to Peer     \\
			AS    &     Authentication Server     &     PSK    &     Pre-Shared Key     \\
			AAA    &     Authentication, Authorization, and Accounting     &  PANA    &     Protocol for Carrying Authentication for Network Access     \\
			AVISPA    &     Automated Validation of Internet Security Protocols and Applications     &   PMIPv6    &     Proxy Mobile IPv6     \\
			BAN    &     Burrows–Abadi–Needham     &     QoE    &     Quality of Experience     \\
			CSI    &     Channel State Information     &    QoS    &     Quality of Service     \\
			CoAP    &     Constrained Application Protocol     &    RPMA    &     Random Phase Multiple Access    \\
			CPS    &     Cyber-Physical Systems    &    RSSI    &     Received Signal Strength Indicator     \\
			DoS    &     Denial of Service     &    RADIUS    &     Remote Access Dial In User Service     \\
			D2D    &     Device to Device     &    RPL    &     Routing Protocol for Low Power and Lossy Networks     \\
			DDoS    &     Distributed Denial of Service     &    RFID    &    Radio-frequency identification     \\
			DNS-SD    &     Domain Name Server-Service Discovery     &    SFTP    &     Secure File Transfer Protocol     \\
			EKE    &     Encrypted Key Exchange     &   SPAM    &     Secure Password Authentication Mechanism     \\
			EAP    &     Extensible Authentication Protocol     &     SSL    &     Secure Sockets Layer     \\
			GTC    &     Generic Token Card     &     SPFP    &     Security Protocol for Fast PMIPv6     \\
			GSM    &     Global System for Mobile Communications     &   SOA    &     Service Oriented Architecture     \\
			HOTA    &     Handover Optimized Ticket-based Authentication     &    SIP    &     Session Initiation Protocol     \\
			HTTPS    &     Hypertext Transfer Protocol Secure     &     SIR    &     Signal-to-Interference Ratio     \\
			IP    &     Internet Protocol     &  SINR    &     Signal-to-Interference-plus-Noise Ratio     \\
			6LoWPAN    &     IPv6 and Low-power Wireless Personal Area Network     &     SDN    &     Software Defined Network     \\
			LoRaWAN & Long Range Wide Area Network &  SQL    &    Structured Query Language     \\
			LEAP    &     Lightweight Extensible Authentication Protocol     &    TR-069    &     Technical Report -069     \\
			LTE-A    &     Long Term Evolution- Advanced     &    TA    &     Ticket-based authentication     \\
			LTE-M    &     Long Term Evolution for Machines     &    TCP    &     Transmission Control Protocol     \\
			LPWAN    &     Low Power Wide Area Network     &    TLS    &     Transport Layer Security     \\
			M2M    &     Machine to Machine     &    TTLS    &     Tunneled Transport Layer Security     \\
			MD    &     Message Digest     &    UNB-IoT    &     Ultra Narrow Band Internet of Things     \\
			MQTT    &     Message Queuing Telemetry Transport     &    UDP    &     User Datagram Protocol     \\
			M-IoT    &     Mobile Internet of Things     &    WEIGHTLESS-N    &     Weightless-Narrow band     \\
			MIMO    &     Multi-Input Multi-Output     &    WEIGHTLESS-P    &     Weightless-Private/Platanus Technology     \\
			NB-IoT    &     Narrow Band Internet of Things     &    WEIGHTLESS-W    &     Weightless-Whitespace     \\
			NFC    &    Near-Field Communication     &    WIMAX    &     Worldwide Interoperability for Microwave Access     \\
			\hline
		\end{tabular}
	\end{table*}

	
	\subsection{Utilities, Contributions and Structure of this Survey}
	\textcolor{black}{This survey covers a majority of the content related to security, privacy, trust-management and protocols for smart M-IoT networks. The content presented in this article is competent compared to the existing surveys and is different in terms of comparative study, which will help its readers follow the parameters and ideology of existing works. Further, this survey can be used by the researchers at any level; especially new researchers can gain a lot from the comparisons and the roadmap sections. Academicians can follow this article to teach new trends related to security of M-IoT and its advancements. This work can help industry researchers to follow what has been done and what can be carried further while deploying applications related to M-IoT. The open challenges presented in the lateral part of this article will help to define problem statements and can be used as a rationale for continuing research on security, privacy and trust aspects of M-IoT.}
	
	\textcolor{black}{This is a comprehensive survey that collectively covers security, privacy, and trust for smart M-IoT, which otherwise are presented as individual topics in the existing surveys. The tabular studies provide a single source to understand the novelty and reach of existing state-of-the-art solutions for smart M-IoT. The roadmap and comparisons with the related survey articles along with key contents to follow for enhancing the knowledge of this subject are given in Section II. Section III presents characteristics, challenges, technologies and standards, an overview of security, privacy and trust along with their methodologies for evaluation. Section IV gives details on secure frameworks for smart M-IoT, Section V discusses the security aware protocols, Section VI presents privacy preservation approaches, Section VII gives details on trust management approaches, Section VIII discusses physical layer security and Section IX gives details on the handover security for smart M-IoT networks. Research Challenges, open issues, and future directions are presented in Section X. Finally, Section XI concludes this article. The details of abbreviations and key terms used throughout the paper are presented in Table~\ref{st3}.}
	
	\begin{table*}[!ht]
		\fontsize{7}{10}\selectfont
		\color{black}
		\centering
		\caption{Comparison with Related Survey Articles.}
		\label{comparison_survey}
		\begin{tabular}{lllllllllll}
			\hline
			\textbf{Article}    &    \textbf{Year}    &    \textbf{Focus}    &    \textbf{Smart M-IoT}    &    \textbf{Security}    &    \textbf{Privacy}    &    \textbf{Trust}    &    \textbf{Classifications}    &    \textbf{Protocol Security}    &    \textbf{Handover Security}    &    \textbf{Framework Security}    \\
			\hline
			\hline
			Medaglia et al.~\cite{medaglia2010overview}    &    2010    &    IoT    &    No    &    Yes    &    Yes    &    No    &    No    &    No    &    No    &    No    \\
			K{\o}ien et al.~\cite{koien2011reflections}    &    2011    &    IoT    &    No    &    Yes    &    No    &    Yes    &    No    &    No    &    No    &    No    \\
			Bonetto et al.~\cite{bonetto2012secure}    &    2012    &    IoT    &    No    &    Yes    &    No    &    No    &    No    &    Yes    &    No    &    No    \\
			Chang and Chen~\cite{chang2012survey}    &    2012    &    IoT    &    No    &    No    &    No    &    Yes    &    No    &    No    &    No    &    No    \\
			Bhattasali et al.~\cite{bhattasali2013study}    &    2013    &    IoT    &    No    &    Yes    &    No    &    No    &    No    &    No    &    No    &    No    \\
			Roman et al.~\cite{roman2013features}    &    2013    &    IoT    &    No    &    Yes    &    Yes    &    No    &    Yes    &    No    &    No    &    No    \\
			Yan et al.~\cite{yan2014survey}    &    2014    &    IoT    &    No    &    No    &    No    &    Yes    &    Yes    &    No    &    No    &    No    \\
			Jing et al.~\cite{jing2014security}    &    2014    &    IoT    &    No    &    Yes    &    Yes    &    Yes    &    Yes    &    Yes    &    No    &    No    \\
			Sicari et al.~\cite{sicari2015security}    &    2015    &    IoT    &    Yes    &    Yes    &    Yes    &    Yes    &    No    &    No    &    No    &    No    \\
			Arias et al.~\cite{arias2015privacy}    &    2015    &    IoT    &    Yes    &    Yes    &    Yes    &    No    &    No    &    No    &    No    &    No    \\
			Malina et al.~\cite{malina2016perspective}    &    2016    &    IoT    &    No    &    Yes    &    Yes    &    No    &    Yes    &    Yes    &    No    &    No    \\
			Li et al.~\cite{li2016internet}    &    2016    &    IoT    &    No    &    Yes    &    Yes    &    Yes    &    Yes    &    No    &    No    &    No    \\
			Zhou et al.~\cite{zhou2017security}    &    2017    &    IoT    &    No    &    Yes    &    Yes    &    No    &    No    &    No    &    No    &    No    \\
			Yang et al.~\cite{7902207}    &    2017    &    IoT    &    No    &    Yes    &    Yes    &    No    &    Yes    &    Yes    &    No    &    No    \\
			Chen et al.~\cite{7903611}    &    2017    &    IoT    &    No    &    Yes    &    Yes    &    No    &    Yes    &    Yes    &    No    &    Yes    \\
			Feng et al.~\cite{feng2017survey}    &    2017    &    MC    &    No    &    Yes    &    Yes    &    Yes    &    Yes    &    No    &    No    &    Yes    \\
			Yang et al.~\cite{8119706}    &    2017    &    IoT    &    Yes    &    Yes    &    No    &    No    &    No    &    Yes    &    No    &    No    \\
			Proposed&2020&M-IoT&Yes&Yes&Yes&Yes&Yes&Yes&Yes&Yes\\
			\hline
		\end{tabular}
	\end{table*}
	\section{Roadmap and Comparison with Related Survey Articles}
	Fig.~\ref{fig3} helps to follow the roadmap of different surveys presented over the period of time that can be used for selecting an appropriate approach for justifying the requirements of M-IoT networks in terms of security, privacy, and trust. In addition to this, Table~\ref{comparison_survey} provides comparative evaluations and reachability of existing studies which are closely related to the survey presented in this article. There are limited works that focus on the details of M-IoT. Only a few of them have written in parts about such requirements and technologies for supporting communications in smart M-IoT. Despite the limited literature in this direction, some of the key and broad surveys have been selected which provides sufficient material to be followed for covering the aspects related to security, privacy, and trust. From the comparisons, it is evident that the closely related survey is the one provided by Feng et al.~\cite{feng2017survey}, but it covers major portions related to Mobile Crowdsourcing (MC), which is not so tightly related to the requirements of smart M-IoT. The other studies in~\cite{medaglia2010overview,koien2011reflections,bonetto2012secure,chang2012survey,bhattasali2013study,roman2013features,yan2014survey,jing2014security,malina2016perspective,li2016internet,zhou2017security,7902207,7903611} do not focus on major considerations which are mandatory to form a highly secure, private and trustworthy M-IoT networks. Sicari et al.~\cite{sicari2015security}, Arias et al.~\cite{arias2015privacy} and Yang et al.~\cite{8119706} have discussed the concepts related to M-IoT, but do not cover enough details on the security, privacy and trust management in smart M-IoT. In addition to these, there are no comparative strategies provided for discussing the protocol and framework security in any of these surveys, which is a major limitation. Further, handoffs are the major part of mobile-oriented networks, which are not evaluated in the existing studies. Thus, the necessity of such a study, in-depth evaluations and conceptual-reachability of the proposed survey will help researchers to gain insight into the requirements of secure communications in smart and connected M-IoT. In addition, Table~\ref{st2} presents some of the other key contributions, which can be followed for understanding the present standings in the security of M-IoT devices and its applications.
	\begin{table}[!ht]
		\fontsize{7}{10}\selectfont
		\color{black}
		\centering
		\caption{Some key contributions to follow for security, privacy and trust in smart M-IoT.}
		\label{st2}
		\begin{tabular}{lll}
			\hline
			\parbox{1.2cm}{\textbf{Approach}} & \textbf{Author} & \parbox{2.2cm}{\textbf{Ideology}} \\
			\hline
			\hline\\
			\parbox{2.1cm}{TRIFECTA} & Sen et al.~\cite{sen2018trifecta} & \begin{tabular}[c]{@{}l@{}}Security, Energy efficiency,\\ and Communication capacity\end{tabular}\\\\
			\parbox{2.1cm}{Jamming mitigation} & Tang et al.~\cite{tang2018jamming} & \begin{tabular}[c]{@{}l@{}}Hierarchical security\\     game\end{tabular} \\\\
			\parbox{2.1cm}{SEGB-AKA} & Parne et al.~\cite{parne2018segb} & \begin{tabular}[c]{@{}l@{}}AKA protocol-based \\ solution\end{tabular} \\\\
			\parbox{2.1cm}{SNAuth protocol} & Dao et al.~\cite{dao2017achievable} & Peer-aware communications \\\\
			\parbox{2.1cm}{Low-cost security for IoT} & Mangia et al.~\cite{mangia2018low} & \begin{tabular}[c]{@{}l@{}}Rakeness-based compressed \\ sensing\end{tabular} \\\\
			\begin{tabular}[c]{@{}l@{}}Enhanced\\ attestation and\\ security\end{tabular} & Wang et al.~\cite{wang2018enabling} & \begin{tabular}[c]{@{}l@{}}Security-enhanced attestation\\   and policy-based\\ measurement mechanism\end{tabular} \\\\
			\parbox{2.1cm}{Traffic-aware patching} & Cheng et al.~\cite{cheng2017traffic} & \begin{tabular}[c]{@{}l@{}}Patching with limited resources \\ and time constraints\end{tabular} \\\\
			\parbox{2.1cm}{Secure game theoretic approach} & Sedjelmaci et al.~\cite{sedjelmaci2017accurate} & Anomaly detection technique \\\\
			\parbox{2.1cm}{Security access protocol} & Giuliano et al.~\cite{giuliano2017security} & Secure key renewal \\\\
			\parbox{2.1cm}{Lightweight masked AES} & Yu et al.~\cite{yu2017lightweight} & Dynamic differential logic \\\\
			\parbox{2.1cm}{Secure NFC-based approach} & Ulz et al.~\cite{ulz2017secured} & RSSI based trilateration algorithm \\\\
			\parbox{2.1cm}{Security situation awareness} & Xu et al.~\cite{xu2017network} & \begin{tabular}[c]{@{}l@{}}Semantic ontology and \\ user-defined rules\end{tabular} \\\\
			\parbox{2.1cm}{Privacy protector} & Luo et al.~\cite{luo2018privacyprotector} & \begin{tabular}[c]{@{}l@{}}Slepian-wolf-coding-based secret\\   sharing\end{tabular}      \\
			\hline
		\end{tabular}
	\end{table}
	
	\begin{figure*}[!ht]
		\centering
		\includegraphics[width=380px]{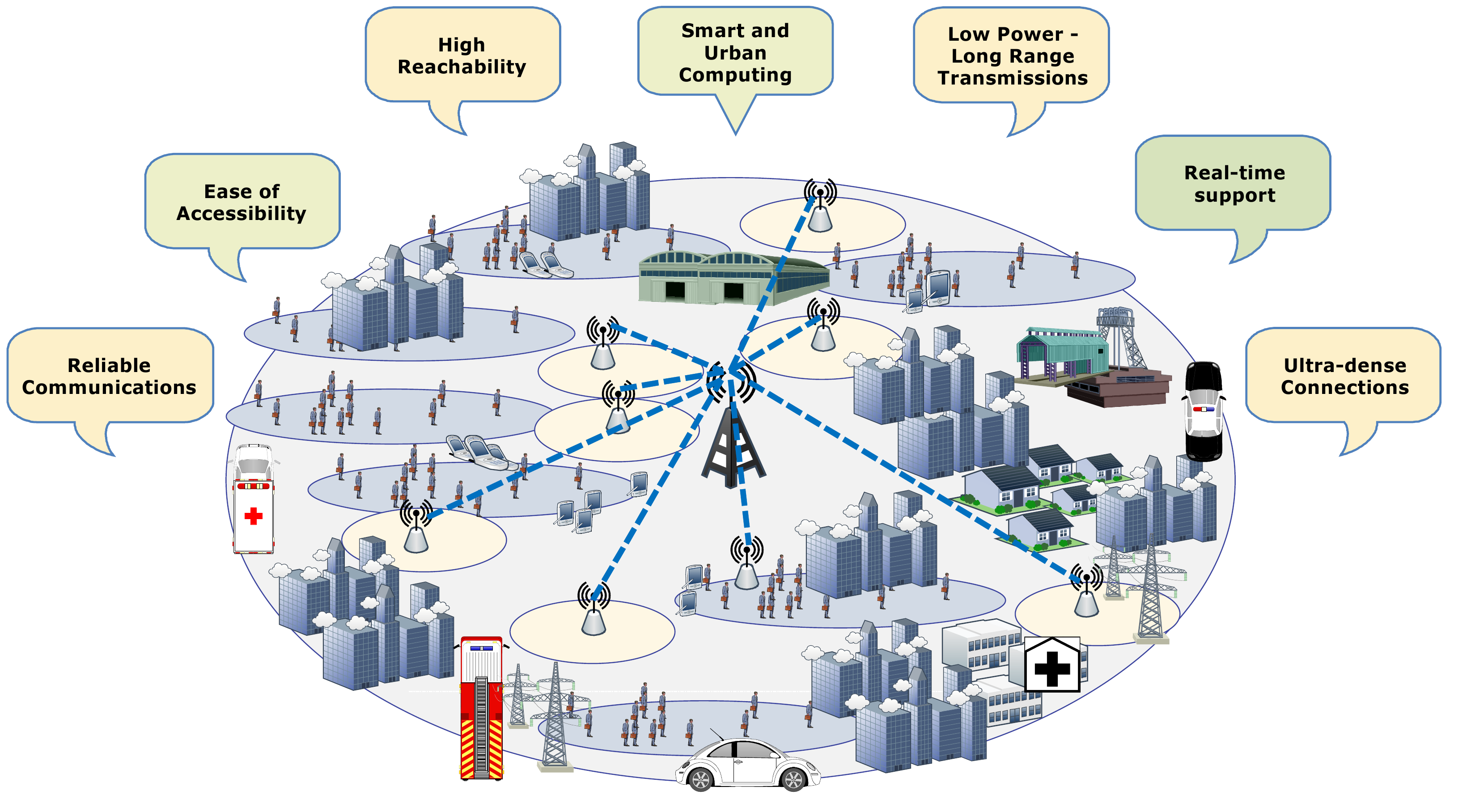}
		\caption{An exemplary illustration of M-IoT scenario and trends in modern day networks. The figure shows crucial aspects and properties to be satisfied for the efficient implementation of M-IoT on the backbone of cellular infrastructure.}\label{fig5}
	\end{figure*}
	
	\section{Smart Mobile IoT Networks and Its Security: An Overview}
	This section presents characteristics and challenges of smart M-IoT. The details are presented on the different types of technology enablers, standards, and general stacks for implementing such a network. 
	\subsection{Characteristics of Smart Mobile IoT Networks}
	Smart M-IoT focuses on reliable and sustainable connectivity between the devices on the move, as shown in Fig.~\ref{fig5}. Smart M-IoT focuses on the establishment of a trust relationship between the devices through an enhanced reputation-cycling. Dependence on \textcolor{black}{Machine to Machine (M2M)} communication~\cite{ghasempour2016finding}, \textcolor{black}{Device to Device} (D2D) marking, in-built-service sharing, and energy conservation are the key characteristics of M-IoT. With the devices operating in a battery constrained environment, M-IoT characterizes on the utilization of technologies that offer a wide range but at low battery consumption. The characteristics of smart M-IoT can be summarized as follows:
	\begin{itemize}
		\item M-IoT includes devices with low power, but operable up to a wide range with lower complexity and lesser resource consumptions.
		\item Supports ultra-dense communication with a unique feature of reliability despite such a huge number of devices operating simultaneously.
		\item M-IoT may be subjected to frequent handovers and may be involved in inter- or intra-handovers depending on their network design and deployment.
		\item Licensed and shared spectrum usability with a primary focus on services similar to short messages. Most of the applications do not require any retraining, and configurations are automatically loaded as a part of application program.
		\item Smart M-IoT applications and services are vendor specific. However, the licensing of narrow bands can be governed by small-scale network organizations with core setups at the big business houses.
		\item M-IoT operations are dependent on the synergy among the mobile operators and rely heavily on the trust-relationship for their security and distributions.
		\item One-tap facilities for all the services, where a user just has to install and load a required feature for experiencing the applications that focus on consumer-electronics, healthcare of smart home automation.
		\item M-IoT needs media independent support for most of the applications as some of the entities may be operating on 3G, while other may have 4G/LTE or even the upcoming 5G accessibility through mmWave functionalities.
		\item Virtualization and privatization of services are the other main characteristics of M-IoT. Virtualization has further been leveraged through the properties of network slicing, which is one of the solutions for distributed security.
		\item Support for immediate acquisition, decision and action are the major features of smart M-IoT. Management of information and building contextual relationships are the other unique characteristics of smart M-IoT.
	\end{itemize}
	\begin{figure*}[!ht]
		\centering
		\includegraphics[width=320px]{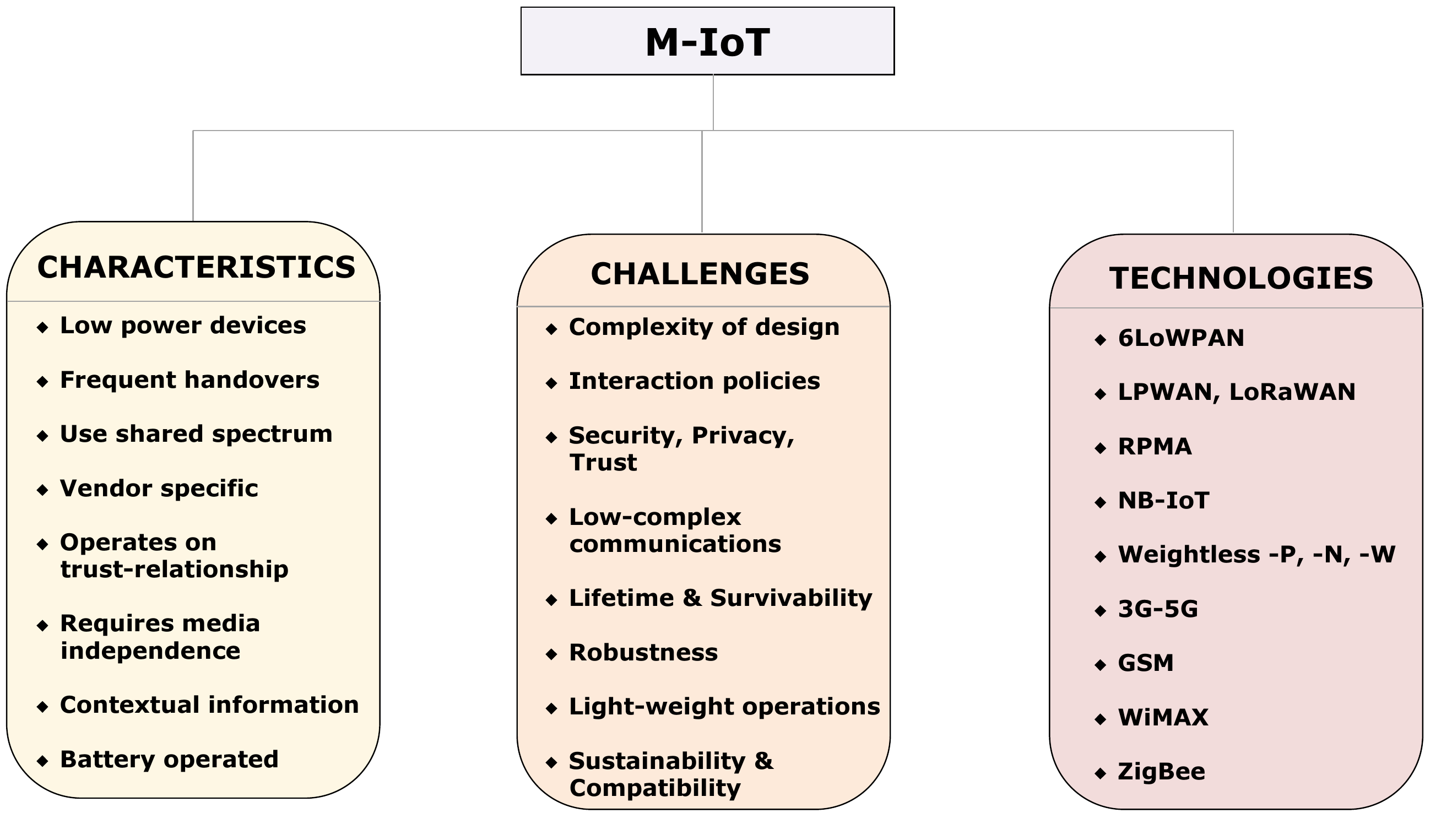}
		\caption{M-IoT Overview: General characteristics of M-IoT networks, challenges in implementation and technologies available for successful deployment of M-IoT.}\label{fig4}
	\end{figure*}
	\begin{table*}[!ht]
		\fontsize{7}{9}\selectfont
		\color{black}
		\centering
		\caption{Types of attacks in M-IoT.}
		\label{st1}
		\begin{tabular}{llll}
			\hline
			\textbf{Type} & \textbf{Attack} & \textbf{Motive} & \textbf{Vulnerability} \\
			\hline
			\hline\\
			\textbf{Passive} & Interception & Information disclose & Insufficient authentication and validation \\\\
			& Release of message & Information disclose & Insufficient authentication and validation \\\\
			& Traffic analysis & Information disclose & Lack of encryption \\\\
			& Sniffing & Information disclose & Insufficient security validation \\\\
			& Keyloggers & Information disclose & Misconfiguration and design flaws \\\\
			&  &  &  \\
			\textbf{Active} & DoS & Information distort and destruct& Buffer overflow, Race condition \\\\
			& DDoS & Information distort and destruct & Buffer overflow, Race condition \\\\
			& Distributed DoS with Reflectors & Information distort and destruct &Buffer overflow, Race condition \\\\
			& Replay attack & Information discovery &Incorrect permissions, User and sever compromise \\\\
			& Masquerading & Information discovery & Insufficient security validation \\\\
			& SQL injection & Information discovery & Incorrect permissions \\\\
			& Man in the middle & Information disclose and discovery &Misconfiguration and design flaws, Insufficient security validation\\\\
			& Modification & Information disclose and disrupt & Misconfiguration and design flaws, Insufficient security  validation\\
			\hline
		\end{tabular}
	\end{table*}
	
	\begin{figure*}[!ht]
		\centering
		\includegraphics[width=290px]{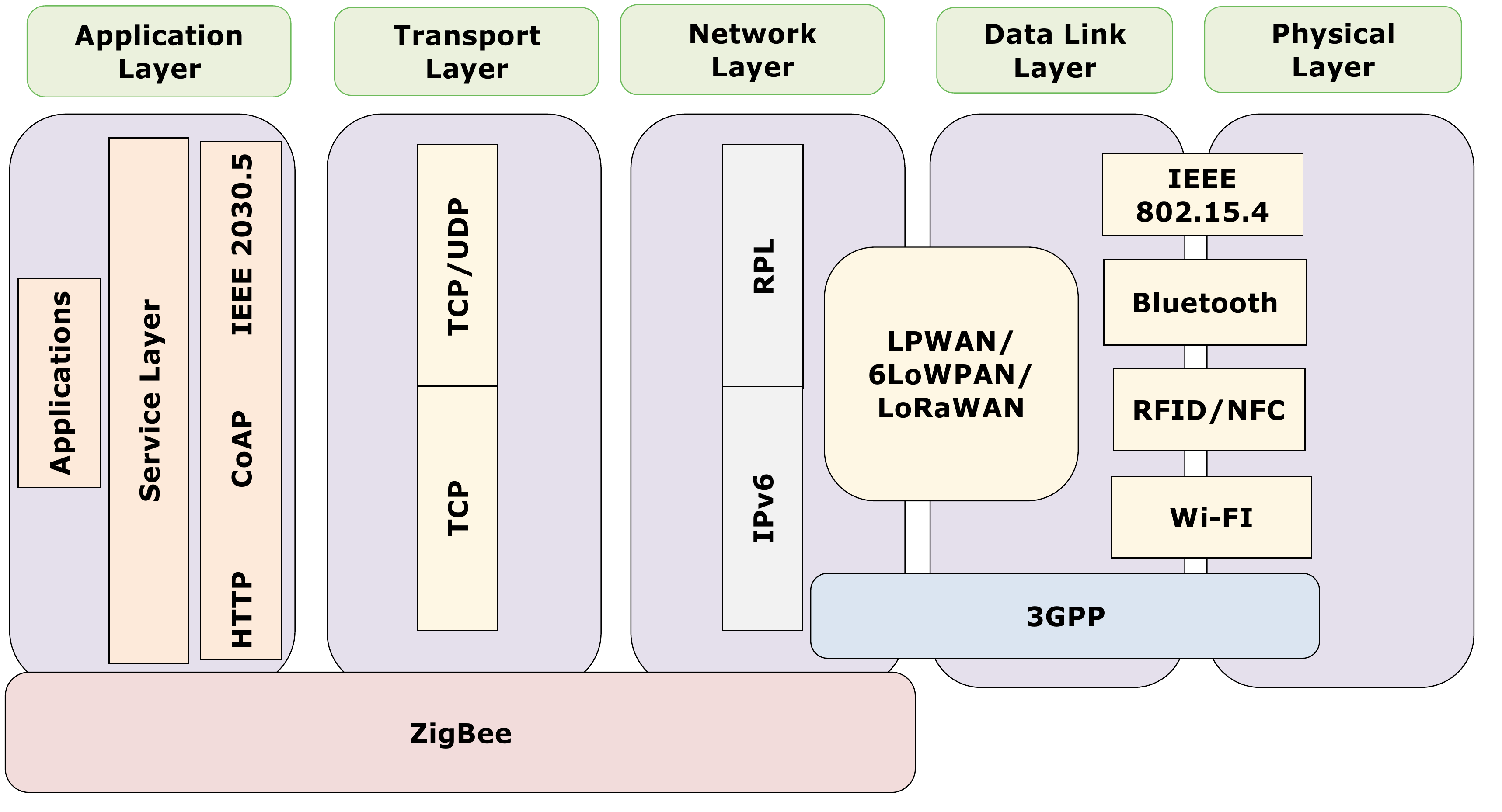}
		\caption{An exemplary overview of a general M-IoT stack.}\label{fig2}
	\end{figure*}
	\begin{figure}[!ht]
		\centering
		\includegraphics[width=240px]{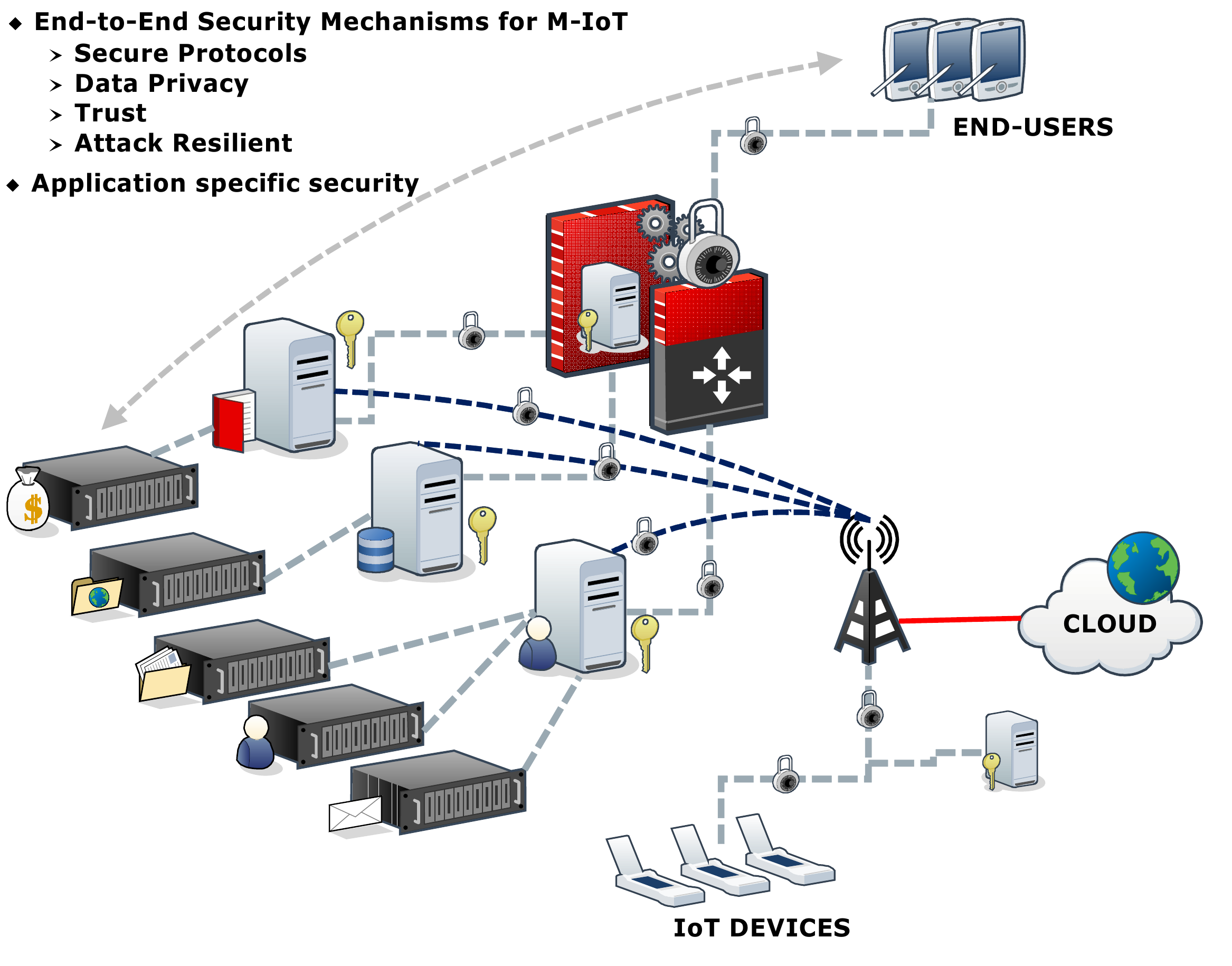}
		\caption{An exemplary illustration of security, trust and privacy aspects in M-IoT. }\label{fig6}
	\end{figure}
	\begin{figure*}[!ht]
		\centering
		\includegraphics[width=330px]{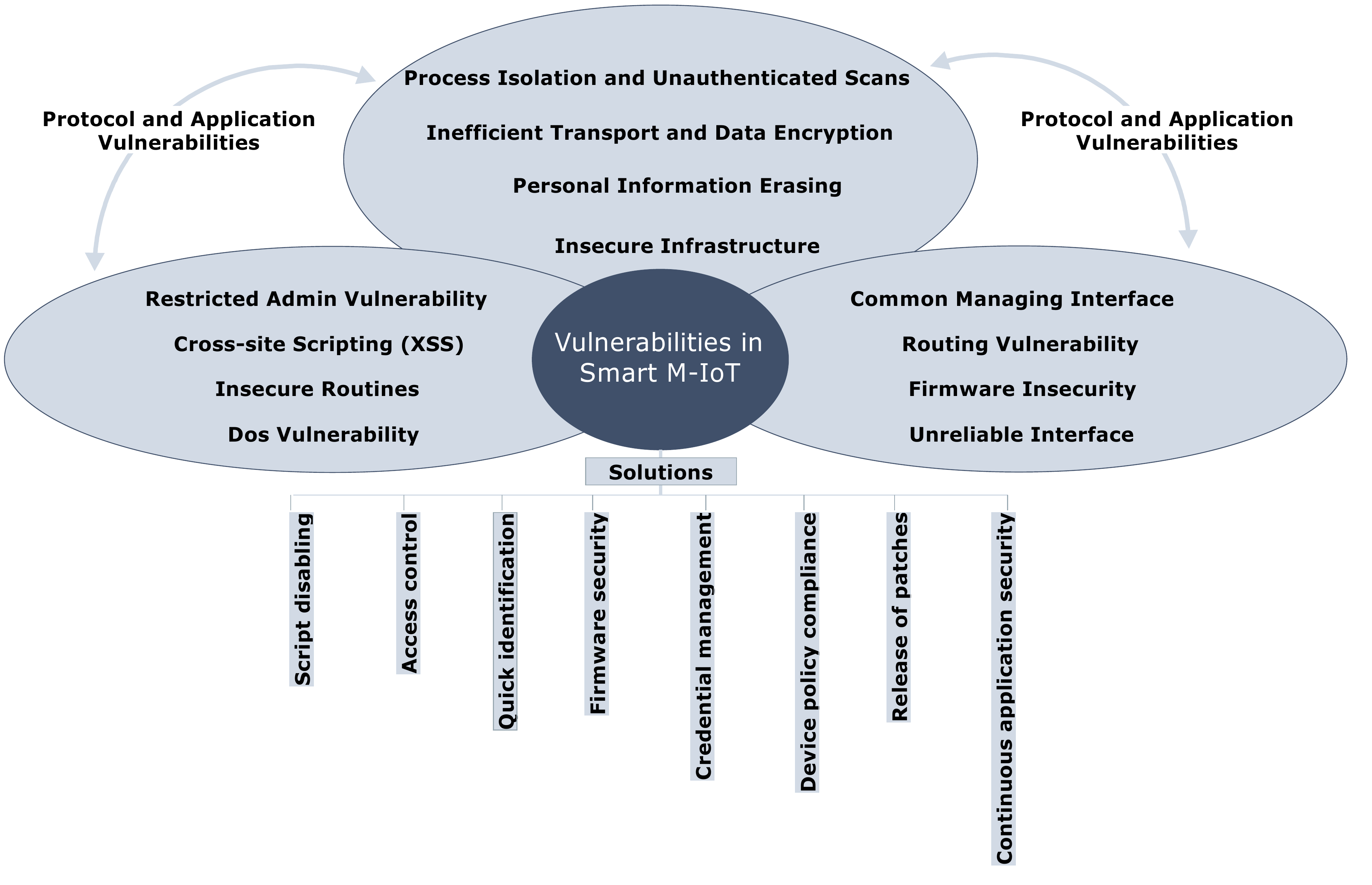}
		\caption{\textcolor{black}{An illustration of vulnerabilities in smart M-IoT with possible remedies.}}\label{fig11}
	\end{figure*}

	\subsection{Challenges of Smart M-IoT}
	Despite a huge set of advantages, there are some crucial challenges associated with the fully-functional usability of smart M-IoT applications. These include,
	\begin{itemize}
		\item Complexity of design: M-IoT faces a major challenge because of design complexity for both its applications as well as network. The applications must be low-complex and must not require extra knowledge for operations by its users. Further, with the requirements of ease of use, M-IoT may cause excessive burden on the developers for designing an easy to follow and deploy environment.
		\item Interaction policies: Smart M-IoT is governed by the rules through which applications interact with each other for facilitating the services to its users. However, the difference in the configuration and operable technology makes it difficult for using common interaction policies for all M-IoT devices. Thus, the formation of rules and generation of interaction policies through consensuses are extremely tedious in M-IoT.
		\item Security: Independently on the technology, security has always been a concern for all types of IoT applications. Prevention against known and unknown attacks and mitigation of zero-day possibilities are the key requirements for security solutions which aim at regulating M-IoT applications~\cite{sharma2017consensus}. Security solutions must be light-weight and should be able to handle the tradeoffs with the performance of a device or the network. Apart from general security, these networks are also subject to crucial requirements of handover security, which can be obtained through existing authentication mechanisms while focusing either on pre-authentication or post-authentication mechanisms depending on the needs and requirements of an application. Management of insider threats and policing are other requirements of security solutions~\cite{kammuller2017insider,sharma2017socializing,li2017application,sharma2017isma}.
		\item Privacy: With most of the applications personalized in M-IoT, leakage of a users' information may pose a huge threat to the entire network and can destroy an individual's belongings. With billions of devices in place, data privacy may be a reason for huge performance overheads in these networks. Thus, it is inevitably important to support data privacy which is otherwise a key challenge for smart M-IoT.
		\item Trust: Security and privacy are established through trust-relationships between the service providers and the users. Trust validations and support for common-reputation systems that can guarantee a low-overhead based mechanism for trust-maintenance are a huge challenge for smart M-IoT networks.
		\item Low-complexity protocols: Different applications need different protocols to communicate, which raises concerns about compatibility issues in terms of protocol selection and arriving at a general agreement during sharing of context between the cross-platform applications. Thus, designing of low-complex protocols with high compatibility and ease of upgrading are the key challenges to handle in smart M-IoT applications.
		\item Lifetime: Since the devices in M-IoT are operable through batteries, it is required that the applications, as well as network architectural support functions, should not cause an excessive computational burden on the devices which may deplete their resources leading to a network shutdown. Thus, enhancement of life, capacity and coverage should be managed in smart M-IoT networks.
	\end{itemize}
	
	Apart from these issues, some of the key attacks in M-IoT, against which effective countermeasures are required, are listed in Table~\ref{st1} and the summaries of characteristics, challenges and technologies are shown in Fig.~\ref{fig4}.
	
	\subsection{M-IoT Technologies, Standards, and Stacks}
	There are a plethora of articles that have discussed various technologies, standards, and stacks which are applicable to M-IoT. However, to make this article self-contained, general information on some of these are presented in this section. \textcolor{black}{For further clarification, an illustration of a general overview of M-IoT stack is shown in Fig.~\ref{fig2}, which can be further studied from~\cite{3G1}~\cite{3G2}; and an exemplary illustration of security, trust and privacy formations in M-IoT is presented in Fig.~\ref{fig6}}. \textcolor{black}{At present, M-IoT is based on low power and wide range technologies, which include 6LoWPAN, LPWAN-based LoRaWAN, Random Phase Multiple Access (RPMA), NB-IoT, Ultra Narrow Band-IoT (UNB-IoT) Weightless-W, Weightless-P, and Weightless-N~\cite{mekki2018comparative,petajajarvi2016evaluation,shelby20116lowpan,trasvina2016network,adelantado2017understanding,ratasuk2016nb}. } Besides these, existing network architectures such as 3G, 4G/LTE, Worldwide Interoperability for Microwave Access (WiMAX), ZigBee, Global System for Mobile Communications (GSM), can be used for supporting applications in M-IoT. The standards for IoT vary depending on the application scenario and the configurations of the devices used in M-IoT. In general, various open projects, organizations, alliances, and IEEE provide a series of standards that primarily focus on supporting smart applications in IoT networks. Some of these are TR-069, OMA-DM, DNS-SD, IEEE series 2413, 21451, 11073, 2200, 2030, 1905, 1900-03, 1701-03, etc. Further details on each of them can be obtained from~\cite{IEEEturl} and~\cite{ioturl}. Apart from these technologies and standards, there are different types of stacks used for supporting smart mobile communications in IoT. However, the general use of stack can be application or network specific and varies as per the configurations of each device. Usually, the stack selections will be affected by the technologies adopted for communications in M-IoT. It is recommended to form compatible and ready-to-integrate models which can be easily deployed in any sort of scenarios irrespective of the device configurations, type and make. Stacks applicable for general IoT can be used for extending services in M-IoT but with modifications to their operating policies as the majority of the traffic flow is maintained on the devices that are non-static in nature~\cite{palattella2013standardized}. \textcolor{black}{Some of the key solutions for IoT stacks include IBM-Watson IoT~\cite{SG1}, Microsoft Azure IoT suite~\cite{SG2}, OpenIoT~\cite{SG3}, OCF~\cite{SG4}, etc.}
	
	\subsection{\textcolor{black}{Vulnerabilities in smart M-IoT}}
	\textcolor{black}{Information security is the major factor driving security in smart M-IoT. These are lead by the studies on vulnerabilities and loopholes at the hardware level, protocol-level, and application-level of M-IoT. Vulnerabilities are studied based on the mode of attack and assessment into different types of classes, related to hardware, protocol, application, software or organizational~\cite{amit2013pinpointing}~\cite{alhazmi2007measuring}. The exploitation of the known vulnerabilities can be prevented by taking several countermeasures against each of the exploits, however, for unknown vulnerabilities, it is tedious to distinguish and resolve until the severity of exploits are unknown~\cite{benton2013openflow}.}
	
	\textcolor{black}{For major of the smart M-IoT, date of release or disclosure plays a crucial role in prevention and it helps to decide the window of prevention. The release of security patches and security updates are further accounted based on the disclosure dates. Usually, increasing the speed of deliverables causes an impact on the debugging phase, which may lead to several possible vulnerabilities unhandled. In smart M-IoT, most common vulnerabilities are identified as the OS level or the application level. The protocol level vulnerabilities are usually known and steps can be determined based on the deployment. However, in several cases, where protocol security is based on credentials, their theft can lead to severe consequences. Some of the key issues causing/leading to vulnerabilities, as discussed by Open Web Application Security Project (OWASP)~\cite{WG1,peguero2018empirical,fang2018deepxss,sagar2018studying}(Fig~\ref{fig11}), for smart M-IoT are listed below:
		\begin{itemize}
			\item Insecure infrastructure: One of the main causes of vulnerabilities in smart M-IoT is the insecure infrastructure that supports transmissions for the involved devices. Architectural layout plays a key role in accessing the network and prioritizing its security. The dominant mode of connections for M-IoT is a cloud, edge, fog architectures, which needs to be prevented from unauthorized access.
			\item Common managing interface: The services which are obtained through a common managing interface are more likely to fall prey to vulnerabilities than the services which are handled by the individual servers. This can be further seen from another dimension. The exploit of vulnerabilities over a common interface may expose the additional services provisioned through it.
			\item Insecure protocols: The protocols mounted for data sharing and authentication in smart M-IoT may be vulnerable to attacks leading to authorization and access control. Thus, the unlimited role of users and non-predetermining the security of the underlying protocol can be other issues causing vulnerabilities in smart M-IoT.
			\item Inefficient transport and data encryption: Usually the broadcasted traffic is not encrypted to avoid performance issues. Thus, vulnerabilities related to access control, such as eavesdropping, is always possible because majority messages are not encrypted.
			\item Cross-site scripting (XSS): Such vulnerabilities are related to insecure web access and are based on access controls such as the same-origin policy, which is applicable to all the devices in M-IoT. Self and mutated XSS are major concerns to be taken care of while dealing with these types of vulnerabilities.
			\item Firmware insecurity: Identification and decision on firmware insecurity is not an easy task. These involve expertise and a common user may easily be fooled to disclosing his/her devices to malicious agents. Such agents exploit the firmware insecurity and lead to several open ports which allow backdoors, worms, trojans, botnet and ransomware to exploit the known/unknown issue on the device.
			\item Process isolation and unauthenticated scans: Several users allow different processes to take control over the device and allow unauthenticated scans. Majority of them are caused by presenting the requirements of an installed application. Non-evaluation of the downloaded application and free access to control the devices leads to several application-level vulnerabilities.
			\item User policies and patching: In the majority of the cases, vulnerabilities are exploited due to limited action from the users. Delays in updating the security settings and unawareness of the released patches lead to the majority of the vulnerabilities. Nowadays, organizations are taking several key steps to force the security updates, still, there is a gap between the user-understandings and update procedures, which lead to several exploits and threats on smart M-IoT.
		\end{itemize}
		Key solutions and possible remedies for preventing the above-discussed vulnerabilities are given below:
		\begin{itemize}
			\item Access control: Limiting the control over device-data and allowing authorized applications to access can help limit the exploits on the known vulnerabilities. Evaluating the content to be accessed and components of shared-data can further elevate the security of devices in smart M-IoT.
			\item Quick identification and release of patches: It is determined that mode and action and time of action play a key role in preventing a device. Thus, quick identification of vulnerability, release of security patches and installing them are major actions that can prevent against severe attacks.
			\item Credential management: For the network-based vulnerability prevention, credential-management, its security, and protection can help to ensure security and privacy for devices. Credential management prevents access to sensitive data and keys which are necessary for encryption as well as securing the communication channels.
			\item Firmware security: It is desired at the developer level to maintain the bug-free release of firmware. Thus, a strong debugging and evaluation against known vulnerabilities must be carried before supplying it to the users or even assemblers.
			\item Device policy compliance: It is necessary that users must comply with the policies laid for a particular device and should not break the codes, which may allow unauthorized applications to take control over a device. Such a vulnerable device may expose the entire network and it is the responsibility of the user to maintain the functionality of the device within the laid guideless.
			\item Script disabling: Majority of developers have shifted their focus on developing applications which do not require client-side scripts. Thus, from futuristic developers, preventing scripts can allow security against vulnerabilities without affecting the services.
			\item Continuous application security: Identification of application security must be followed by the release of the security update or newer versions. Thus, continuous monitoring of applications is required to prevent possible vulnerabilities. Moreover, this is also an effective strategy to prevent the possibilities of zero-day threats and attacks.
		\end{itemize}
		There are several studies that have been dedicated to vulnerabilities in M-IoT and can be followed from~\cite{lancioni2018method,samtani2018identifying,sharma2018behavior,kim2017method,frustaci2018evaluating,kim2017national,stellios2018survey,xie2017vulnerability}. Based on these, it becomes inevitably important to understand the concept, issues, scope and strength of present state of security, privacy and trust for smart M-IoT.}
	
	\subsection{Security, Privacy, and Trust for smart M-IoT}
	Because of a difference in mode of deployment and applicability, security, privacy, and trust of M-IoT devices are of utmost importance. These differences in the characteristics of involved devices raise an alarming factor for securing and isolating each user's operations as the variation in behavior and operations of each device may lead to different kind of threats based on their specifications~\cite{huang2017insight}. Thus, it is important to study all the aspects related to the security, privacy, and trust of smart M-IoT networks. Majority of the threats occur due to inadequate configurations of security properties and some of them are the vulnerabilities that remain undetected over a course of time due to the negligence of their developers~\cite{li2016secure}~\cite{li2017privacy}. Minimizing data acquisition, supporting M2M routing, resolving hidden terminals and encryption can help to secure and privatize each user's data and information.
	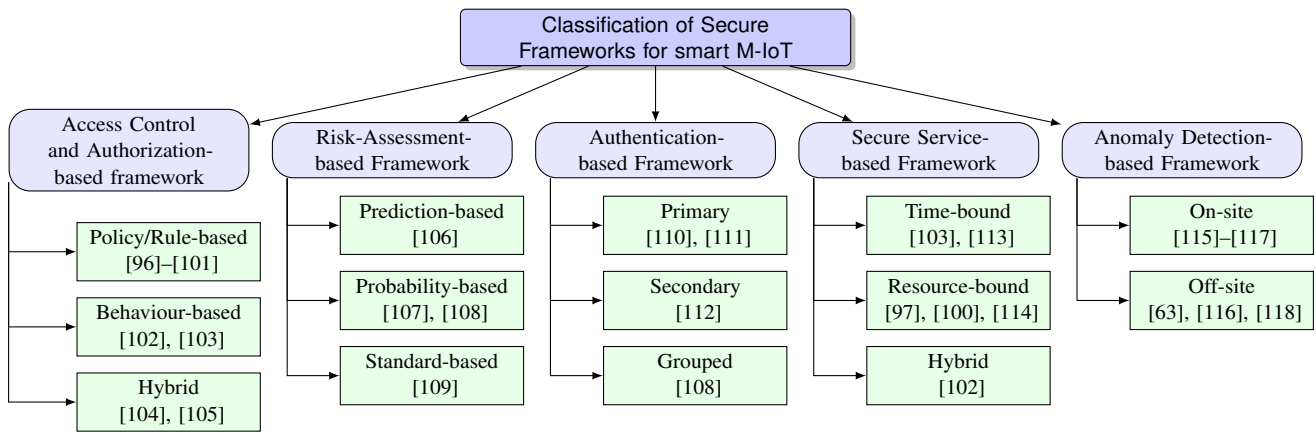
\begin{figure*}[!ht]
		\centering
		\fontsize{8}{10}\selectfont
		\begin{tikzpicture}[
		level 1/.style={sibling distance=35mm},
		edge from parent/.style={->,draw},
		>=latex]
		
		\node[root] {Classification of Secure Frameworks for smart M-IoT}
		child {node[level 2] (c1) {Access Control and Authorization-based framework}}
		child {node[level 2] (c2) {Risk-Assessment-based Framework}}
		child {node[level 2] (c3) {Authentication-based Framework}}
		child {node[level 2] (c4) {Secure Service-based Framework}}
		child {node[level 2] (c5) {Anomaly Detection-based Framework}};
		
		\begin{scope}[every node/.style={level 3}]
		\node [below of = c1, xshift=15pt,yshift=-10pt] (c11) {Policy/Rule-based\\~\cite{dsouza2014policy,cirani2015iot,sahoo2015secured,pacheco2016iota,pereira2014authentication,gonzalez2016sdn}};
		\node [below of = c11] (c12) {Behaviour-based\\~\cite{seitz2013authorization,hernandez2015safir}};
		\node [below of = c12] (c13) {Hybrid\\~\cite{guchhait2018hybrid,dagdee2009credential}};
		
		\node [below of = c2, xshift=15pt] (c21) {Prediction-based\\~\cite{abie2012risk}};
		\node [below of = c21] (c22) {Probability-based\\~\cite{ge2015framework,ray2014scalable}};
		\node [below of = c22] (c23) {Standard-based\\~\cite{kebande2016generic}};
		
		\node [below of = c3, xshift=15pt] (c31) {Primary\\~\cite{wang2014vecure,huang2016seciot}};
		\node [below of = c31] (c32) {Secondary\\~\cite{mclachlan2015adaptive}};
		\node [below of = c32] (c33) {Grouped\\~\cite{ray2014scalable}};

		\node [below of = c4, xshift=15pt] (c41) {Time-bound\\~\cite{tao2018multi,hernandez2015safir}};
		\node [below of = c41] (c42) {Resource-bound\\~\cite{cirani2015iot,pereira2014authentication,rahman2016securing}};
		\node [below of = c42] (c43) {Hybrid\\~\cite{seitz2013authorization}};
		
		\node [below of = c5, xshift=15pt] (c51) {On-site\\~\cite{ahmad2017unsupervised,8322278,toledano2018real}};
		\node [below of = c51] (c52) {Off-site\\~\cite{8322278,ahmed2014network,sharma2017isma}};
		
		\end{scope}
		
		\foreach \value in {1,2,3}
		\draw[->] (c1.195) |- (c1\value.west);
		
		\foreach \value in {1,2,3}
		\draw[->] (c2.195) |- (c2\value.west);
		
		\foreach \value in {1,2,3}
		\draw[->] (c3.195) |- (c3\value.west);
		
		\foreach \value in {1,2,3}
		\draw[->] (c4.195) |- (c4\value.west);
		
		\foreach \value in {1,2}
		\draw[->] (c5.195) |- (c5\value.west);

		\end{tikzpicture}
		\caption{A broad classification of the security framework for smart M-IoT. The security approaches focusing on the frameworks for M-IoT can be broadly classified on the basis of Access control and Authorization, Risk-Assessment features, Authentication, Secure Services, and Anomaly Detection.}\label{fig:tax1}
	\end{figure*}
	
	\subsection{Methodologies for Analyses of Security, Privacy, and Trust in Smart M-IoT}
	An approach is secure for the time being it is not broken, which means security is difficult to analyze as there are no direct simulators and emulators to be used for evaluation of a system for these requirements. Visualization is another big issue for such requirements. Visualization of trust can be obtained as it is comparatively easier to define trust as a metric between the communicating entities; whereas security and privacy are governed by rules and policies which can only be evaluated in an attacker environment. Creation and demonstration of such an environment are difficult as it requires a lot of automation, which is not applicable to most of the available tools. Majority of the solutions are formally analyzed using Burrows–Abadi–Needham (BAN) logic, which is operated on belief theory~\cite{monniaux1999decision}~\cite{cohen2005logical}. Some approaches follow reduction techniques, while others simply rely on evaluating the computational cost of operations. Apart from these, some other methods include formal semantic evaluations, equational theory, etc~\cite{matsuo2010evaluate}. Cryptographic solutions can be evaluated using random oracle model, inductive methods, provable security, etc~\cite{bleumer2011random,mao2003modern,paulson1999inductive}. Model checking and theory of proving are used by some approaches for evaluating the flow of their solution. There are certain tools available which can be used for these evaluations like, Automated Validation of Internet Security Protocols and Applications (AVISPA), A Computational Logic for Applicative Common Lisp (ACL2), ProVerif, Scyther, etc~\cite{armando2005avispa,kaufmann2013computer,kusters2009using,cremers2008scyther}. Irrespective of these evaluations, it is recommended that solutions should conduct certain case studies while presenting outputs of their proposed schemes and should demonstrate the effects on the performance of the system and the network.
	
	\section{Secure Frameworks for smart M-IoT}
	M-IoT networks are vulnerable to a different set of attacks which can be launched due to improper configurations and deployment strategies. It is required that these networks are deployed with ultra-reliable formations, which help to hinder launching of any unknown as well as known attacks. Further, security implications, assessment, and threat modeling can help to identify any such possibilities at a prior, which may support prevention against intruders during the operations of IoT devices~\cite{urien2013llcps}~\cite{park2017security}~\cite{ma2018learning}. Siboni et al.~\cite{siboni2016advanced} highlighted the importance of a framework for securing the content in wearable IoT devices, which are considered as an important part of M-IoT systems. The authors developed an innovative testbed setup for evaluating the security policies of dynamic IoT devices. The need of the hour is to provide such a framework that can be used for supporting the security requirements of Cyber-Physical Systems (CPS) that heavily rely on M-IoT devices for their regular operations~\cite{ning2012cyber}. Authorization, privacy as well as physical security and anonymity should be the core aspect of frameworks, which primarily focus on the security of smart M-IoT networks~\cite{bernabe2014privacy}~\cite{lake2014internet}. Although the existing frameworks provide a base for network formations, these have to be operated with a different set of schemes, protocols, as well as policy-mechanisms for a fully-reliable and secure network establishment.
	
	The smart frameworks should also support the cryptic techniques, that can be built into its system through separate modules~\cite{wang2013system}~\cite{olivier2015new}~\cite{shafagh2014security}. Deployment of M-IoT through SDNs and use of smart IDS are the future aims of the present systems, which tend to facilitate the security of applications operating over low-powered devices~\cite{kasinathan2013ids}~\cite{flauzac2015sdn}. Use of newer concepts, such as fog architecture, Internet of drones, catalytic computing and osmotic computing, can be considered as a base for developing frameworks that can sustain the burden of security as well as the performance at the same time~\cite{sehgal2015smart,gharibi2016internet,sharma2017managing,sharma2017resource,li2018secure}. Based on the security requirements, a taxonomy is presented which classifies the security frameworks for smart M-IoT, as shown in Fig.~\ref{fig:tax1}. The details of these classifications are presented below:
	\subsection{Access Control and Authorization-based Framework}
	The security of devices in M-IoT is subject to the management of accessibility and authorization for using particular services as well as personal data. This type of frameworks helps to limit the control over the usability of network components and provides strong mechanisms for securing the users. The strength of its security lies in the novelty of architecture used for supporting convergence services to M-IoT users. There are some works in this direction, which highlights the main features of access control and management along with user and service authorizations. However, the majority of them operates on general IoT scenario and lacks evidential commitment on their applicability to smart M-IoT scenarios. The access control and authorization-based frameworks can be further classified into three main types as shown below:
	\begin{itemize}
		\item Policy/Rule-based: The main aspects of such type include user authentication, device authentication, resource authorization, Constrained Application Protocol (CoAP) access control and etc. The solutions in this direction focus on acquisition and control over services and user modules to infrastructure security of its network. The main property of this type is the formation of governing conditions, on the basis of which, certain rules and policies are defined for securing the users and services. Solutions in~\cite{dsouza2014policy,cirani2015iot,sahoo2015secured,pacheco2016iota,pereira2014authentication,gonzalez2016sdn} focus on providing frameworks which utilize user and device authentication through policy and rules over device operations in different network setups.
		\item Behaviour-based: This type of access control and authorization depends on the mode of the user's interaction with other users and entities in the network. The operational activity of the users is taken into account for access control and defining conditions of authorization for demanded services. Such types of security frameworks are well suited for modern services such as smart building, smart cities, and smart factory~\cite{seitz2013authorization,hernandez2015safir}.
		\item Hybrid: There are certain solutions for access control and authorization, which form policies or rules by using behavioral aspects of the network entities to ensure its security and continuity in operations. Such types of frameworks are termed as hybrid access control and authorization-based frameworks. Credential-based services and intelligent solutions use such kind of mechanism for ensuring security in a network~\cite{guchhait2018hybrid,dagdee2009credential}.
	\end{itemize}
	\subsection{Risk-Assessment-based Framework}
	Identification of potential conflicting components and users through detection modeling is mainly studied under risk-assessment-based frameworks for security in smart M-IoT. Such kind of frameworks helps to pre-identify any potential risks involved in leveraging services through a particular aspect of the network. These aspects may include situational awareness of every involved entity of the network. Based on the mode of identification, risk-assessment-based frameworks can be further classified into three main types as shown below:
	\begin{itemize}
		\item Prediction-based: The Framework which identifies and manages risk through predictive or estimated evaluations of the network components are termed as prediction-based risk assessment frameworks~\cite{abie2012risk}. Such type of frameworks considers prior and current states to identify the mode of operations and uses decision modeling to arrive at a decision of potential risks in the network.
		\item Probability-based: In probabilistic-based, the network is evaluated for different kind of operations which are executed over a period of time. Then, each process is operated with a probabilistic model which then helps to finalize the probabilistic cost of the networks, while providing knowledge about the factors which dominates the most and can affect the performance as well as security policies. The most common aspect of such frameworks is to identify attack success possibilities in a network while using parameters like false positives, false negatives, accuracy, recall, and precision as considered in~\cite{ge2015framework,ray2014scalable}.
		\item Standard-based: Most of the organizations have a pre-defined set of conditions which are to be fulfilled by every framework which aims to provide a special kind of services to its users. Majority of these conditions are the benchmark and supported by standards organizations such as International Organization for Standardization (ISO), Institute of Electrical and Electronics Engineers (IEEE), International Telecommunications Union (ITU). These organizations provide guidelines for every framework to justify its security considerations for the defined services. One of the examples can be the forensic study of a framework for its applicability to support on-demand services to the mobile users~\cite{kebande2016generic}.
	\end{itemize}
	\subsection{Authentication-based Framework}
	Authentication of the users and devices in smart M-IoT is of utmost importance and highly crucial. It is required that all the services are provided only to the users which authenticates themselves with the security servers usually Authentication, Authorization, and Accounting (AAA) in any network. These security servers ensure the safety of other legitimate and authenticated users by providing a secure mode of communications. One of the crucial aspects of authentication is the positioning of authentication-server along with the number of passes required to reach it. The mode of authentication is quite vast, but for smart M-IoT, it can be classified into primary mode of authentication, the secondary mode of authentication and group-authentication. The choice among each of them depends on the types of device, network architecture and types of services to be supported by the involved entities. The details on each of them are provided below:
	\begin{itemize}
		\item Primary: The authentication which is performed with the core of any network while using the secure channels between the entity and the authentication server is known as primary mode~\cite{wang2014vecure,huang2016seciot}. Such kind of authentication is much secure but often suffers from the consequences of long paths and requirements of route optimizations. Despite its advantage of providing robust security, it often causes additional overheads if each time an entity has to be authenticated through it even in the cases it is always present in the perimeter of the same network. However, the majority of existing solutions prefer a primary mode of authentication because of the ease of deployment and maintenance.
		\item Secondary: Usually, networks which have data to be constrained in a particular periphery or premises opt for the secondary mode of authentication. Such a model is responsible for securing a particular set of nodes which are entitled to communication within the zone of the secondary authentication server~\cite{mclachlan2015adaptive}. Secondary authentication also uses an initial primary authentication for registering its services and users to the core of the network and after initial phases, all the security concerns are managed by it. With the evolution of smart networks, it is preferred to use a hybrid mechanism as it helps to provide a flexible as well as robust security that too with lower overheads.
		\item Grouped: Another mode of authentication can be the group authentication, which entitles similar entities to be authenticated as a group through a common gateway. Group authentication depends on the type of devices involved in a group, and procedure of authentication depends on their type. Some groups with highly crucial devices may involve strong authentication while the ones with limited resources may require light-weight authentications so as to prevent any excessive utilization of their resources~\cite{ray2014scalable}.
	\end{itemize}
	\subsection{Secure Services-based Framework}
	Type of services affects the security of a network. Some services may require light-frameworks which are easy on resources while others may require fast processing frameworks which operate with lesser delays and fewer overheads.
	\onecolumn
	\begin{landscape}
		\begin{center}
			\fontsize{8}{10}\selectfont
			\setlength\LTleft{30pt}            
			\setlength\LTright{0pt}
			\begin{longtable}{@{\extracolsep{\fill}}*{5}{c}}
				\caption{State-of-the-art frameworks applicable to M-IoT security.}\label{Table1} \\
				\hline
				\multicolumn{1}{p{2cm}}{\centering \textbf{Approach/} \\ \textbf{Model} }
				&\multicolumn{1}{p{2cm}}{\centering \textbf{Author} \\ \textbf{(Year)} }
				&\multicolumn{1}{p{2cm}}{\centering \textbf{Ideology}}
				&\multicolumn{1}{p{4cm}}{\centering \textbf{Parameters} \\ \textbf{Focussed}}
				&\multicolumn{1}{p{8cm}}{\centering \textbf{Description}} \\[6pt]
				\hline \\
				
				\endfirsthead
				
				\multicolumn{5}{c}%
				{{\bfseries \tablename\ \thetable{} -- continued from previous page}} \\
				\hline
				\multicolumn{1}{p{2cm}}{\centering \textbf{Approach/} \\ \textbf{Model} }
				&\multicolumn{1}{p{2cm}}{\centering \textbf{Author} \\ \textbf{(Year)} }
				&\multicolumn{1}{p{2cm}}{\centering \textbf{Ideology}}
				&\multicolumn{1}{p{4cm}}{\centering \textbf{Parameters} \\ \textbf{Focussed}}
				&\multicolumn{1}{p{8cm}}{\centering \textbf{Description}} \\[6pt]
				\hline\\\endhead
				
				\hline \multicolumn{5}{l}{{Continued on next page}} \\
				
				\endfoot
				
				\endlastfoot
				
				\multicolumn{1}{p{2cm}}{\centering Policy driven security}
				&\multicolumn{1}{p{2cm}}{\centering [Dsouza \emph{et al.} 2014]~\cite{dsouza2014policy}}
				&\multicolumn{1}{p{2cm}}{\centering Secure collaboration for users in Fog networks}
				&\multicolumn{1}{p{4cm}}{\centering Security resources, \\User authentication, \\Device authentication}
				&\multicolumn{1}{p{8cm}}{This approach focuses on a secure collaboration between the IoT devices by using policy-management of resources through Fog computing architecture. Policy enforcement point is used as a decisive metric, but conflict resolution and anomaly detection are not evaluated in this model. } \\[6pt] \\
				
				\multicolumn{1}{p{2cm}}{\centering IoT-OAS}
				&\multicolumn{1}{p{2cm}}{\centering [Cirani \emph{et al.} 2015]~\cite{cirani2015iot}}
				&\multicolumn{1}{p{2cm}}{\centering Authorization architecture for secure services}
				&\multicolumn{1}{p{4cm}}{\centering Computational overheads, \\Memory utilization, \\Energy consumption, \\Authorization}
				&\multicolumn{1}{p{8cm}}{This approach is based on Open Authorization (OAuth), which is a third party for simple and secure authorization of services. HTTP/CoAP services are targeted for security while maintaining the flexible, dynamic and easily configurable properties of the architecture.} \\[6pt] \\
				
				\multicolumn{1}{p{2cm}}{\centering DFIF-IoT}
				&\multicolumn{1}{p{2cm}}{\centering [Kebande and Ray 2016]~\cite{kebande2016generic}}
				&\multicolumn{1}{p{2cm}}{\centering Digitalized forensic investigation of IoT}
				&\multicolumn{1}{p{4cm}}{\centering Initialization, \\Acquistion, \\Investigation}
				&\multicolumn{1}{p{8cm}}{This framework is capable of supporting digital forensics over IoT infrastructures. The authors focused their framework with standard compliance of ISO. The framework operates by classifying content into digital forensic module through reactive and proactive processing.} \\[6pt] \\
				
				\multicolumn{1}{p{2cm}}{\centering Authorization and access control}
				&\multicolumn{1}{p{2cm}}{\centering [Pereira \emph{et al.} 2014]~\cite{pereira2014authentication}}
				&\multicolumn{1}{p{2cm}}{\centering Secure SOA for IoT}
				&\multicolumn{1}{p{4cm}}{\centering CoAP overheads, \\CoAP access control}
				&\multicolumn{1}{p{8cm}}{This framework supports security of IoT devices by using a service oriented architecture, which uses Constrained Application Protocol (CoAP) for IoT. This approach also provides strategy for tickets and access control for utilizing the features of existing security protocols.} \\[6pt] \\
				
				\multicolumn{1}{p{2cm}}{\centering Security assessment of IoT}
				&\multicolumn{1}{p{2cm}}{\centering [Ge and Kim 2015]~\cite{ge2015framework}}
				&\multicolumn{1}{p{2cm}}{\centering Evaluating security of large scale networks}
				&\multicolumn{1}{p{4cm}}{\centering Reliability, \\ Risk assessment, \\Attack cost, \\Attack success probability}
				&\multicolumn{1}{p{8cm}}{This approach provides methodology and technique for assessing security of large-scale IoT networks. The authors use a set of parameters for analyzing the reliability of the network through risk assessment.} \\[6pt] \\
				
				\multicolumn{1}{p{2cm}}{\centering VeCure}
				&\multicolumn{1}{p{2cm}}{\centering [Wang and Sawhney 2014]~\cite{wang2014vecure}}
				&\multicolumn{1}{p{2cm}}{\centering Resolution of mutual authentication issue for Internet of vehicles}
				&\multicolumn{1}{p{4cm}}{\centering Trust formations, \\Delay evaluations}
				&\multicolumn{1}{p{8cm}}{This approach provides a mechanism for mutual authentication of nodes in Internet of vehicles. The authors illustrated their approach through a verified proof of concept and illustrated lower-delay approach for message evaluations through trust properties.} \\[6pt] \\
				
				\multicolumn{1}{p{2cm}}{\centering SDN-based security framework}
				&\multicolumn{1}{p{2cm}}{\centering [Gonzalez \emph{et al.} 2016]~\cite{gonzalez2016sdn}}
				&\multicolumn{1}{p{2cm}}{\centering Security framework for IoT in grid using SDN}
				&\multicolumn{1}{p{4cm}}{\centering Number of messages, \\OpenFlow modifications}
				&\multicolumn{1}{p{8cm}}{This framework builds a cluster model for IoT devices through SDN. The common controller is employed to form an intrusion detection and prevention system by using predefined rules on the controller.} \\[6pt] \\
				
				\multicolumn{1}{p{2cm}}{\centering Security framework for smart home IoT}
				&\multicolumn{1}{p{2cm}}{\centering [Tao \emph{et al.} 2018]~\cite{tao2018multi}}
				&\multicolumn{1}{p{2cm}}{\centering Multi-layer cloud architecture-based and ontology-based security}
				&\multicolumn{1}{p{4cm}}{\centering Response time, \\ Token assertion}
				&\multicolumn{1}{p{8cm}}{This framework helps in privacy-preservation and maintain security of devices in a multi-hierarchical cloud formation on the basis of ontology, which is formulated over token and encryption assertions.} \\[6pt] \\
				
				\multicolumn{1}{p{2cm}}{\centering Authorization framework}
				&\multicolumn{1}{p{2cm}}{\centering [Seitz \emph{et al.} 2013]~\cite{seitz2013authorization}}
				&\multicolumn{1}{p{2cm}}{\centering Access control and authorization through key management }
				&\multicolumn{1}{p{4cm}}{\centering Request processing time, \\ Accessibility}
				&\multicolumn{1}{p{8cm}}{This framework supports a fine grained and a flexible access control to devices with limited power and memory constraints. The framework is capable of supporting authorization requirements of IoT devices.} \\[6pt] \\
				
				\multicolumn{1}{p{2cm}}{\centering SecIoT}
				&\multicolumn{1}{p{2cm}}{\centering [Huang \emph{et al.} 2016]~\cite{huang2016seciot}}
				&\multicolumn{1}{p{2cm}}{\centering Robust and transparent security for IoT}
				&\multicolumn{1}{p{4cm}}{\centering User authentication, \\Device authentication, \\Authorization, \\Access management}
				&\multicolumn{1}{p{8cm}}{This framework is capable of resolving the basic security requirements such as authentication, authorization, access control and risk assessment. Trust evaluation and availability are yet to be resolved by this framework for IoT networks. } \\[6pt] \\
				
				\multicolumn{1}{p{2cm}}{\centering RFID security framework}
				&\multicolumn{1}{p{2cm}}{\centering [Ray \emph{et al.} 2014]~\cite{ray2014scalable}}
				&\multicolumn{1}{p{2cm}}{\centering Group-based and collaborative approach for scalable security}
				&\multicolumn{1}{p{4cm}}{\centering Computational complexity, \\Payload analysis, \\Hash operations, \\Probability evaluations, \\Scalability tags }
				&\multicolumn{1}{p{8cm}}{This framework emphasizes on the novel identification technique, which is based on a hybrid approach that helps to support security check handoff for RFID systems in an IoT environment.} \\[6pt] \\
				
				\multicolumn{1}{p{2cm}}{\centering SAFIR}
				&\multicolumn{1}{p{2cm}}{\centering [Hernández-Ramos \emph{et al.} 2015]~\cite{hernandez2015safir}}
				&\multicolumn{1}{p{2cm}}{\centering Access framework for smart-buildings IoT networks}
				&\multicolumn{1}{p{4cm}}{\centering Access control, \\ Authentication, \\Evaluation time, \\Discovery time, \\Energy conservation}
				&\multicolumn{1}{p{8cm}}{This framework focuses on the security and privacy of smart building IoT networks. The framework provides security functional components for the establishment of flexible sharing models, context-aware security on IoT scenarios s realized through physical-context awareness.} \\[6pt] \\
				
				\multicolumn{1}{p{2cm}}{\centering Sensor to cloud security}
				&\multicolumn{1}{p{2cm}}{\centering [Rahman \emph{et al.} 2016]~\cite{rahman2016securing}}
				&\multicolumn{1}{p{2cm}}{\centering Cloud-IoT ecosystem security}
				&\multicolumn{1}{p{4cm}}{\centering Security threats Assessment, \\ (Sensor level, Network Level,\\ Cloud level, Data level)}
				&\multicolumn{1}{p{8cm}}{This framework discusses IoT security framework for mitigating threats identified in the sensor to cloud ecosystem. The framework uses layered hierarchy for securing IoT devices.} \\[6pt] \\
				
				\multicolumn{1}{p{2cm}}{\centering SDN framework for IoT}
				&\multicolumn{1}{p{2cm}}{\centering [Sahoo \emph{et al.} 2015]~\cite{sahoo2015secured}}
				&\multicolumn{1}{p{2cm}}{\centering SDN-based security framework}
				&\multicolumn{1}{p{4cm}}{\centering Accessibility, \\Authentication}
				&\multicolumn{1}{p{8cm}}{This framework helps to authenticate devices through policies which are governed by the controller. Also, the policy rules are used for IoT security by managing the node accessibility. } \\[6pt] \\
				
				\multicolumn{1}{p{2cm}}{\centering Trustworthy smart car services}
				&\multicolumn{1}{p{2cm}}{\centering [Pacheco \emph{et al.} 2016]~\cite{pacheco2016iota}}
				&\multicolumn{1}{p{2cm}}{\centering Anomaly behavior analysis}
				&\multicolumn{1}{p{4cm}}{\centering Detection rate, \\Classification Rate}
				&\multicolumn{1}{p{8cm}}{This framework provided IoT security for trustworthy smart car services. This framework
					uses a set of functions and services for securing these services through threat modeling.} \\[6pt] \\
				
				\multicolumn{1}{p{2cm}}{\centering Adaptive security}
				&\multicolumn{1}{p{2cm}}{\centering [Abie and Balasingham 2012]~\cite{abie2012risk}}
				&\multicolumn{1}{p{2cm}}{\centering Risk prediction and assessment}
				&\multicolumn{1}{p{4cm}}{\centering Risk evaluation}
				&\multicolumn{1}{p{8cm}}{This framework focuses on a risk-based adaptive security for e-health applications in IoT. The framework uses game theory and context-awareness techniques for prediction of involved risks and upcoming damages.} \\[6pt] \\
				\hline
			\end{longtable}
		\end{center}
		\twocolumn
	\end{landscape}
	
	Such type of frameworks is usually related to the authentication facilities supported for managing the security of the network as the authentication phase is itself responsible for resource consumption and delays. Based on the requirements of services, these frameworks can be classified into time-bound, resource-bound, and hybrid frameworks as explained below:
	\begin{itemize}
		\item Time-bound: The frameworks which operate with time as a crucial entity in securing the services and the users of a network are studied as a time-bound services-based framework. As studied in~\cite{tao2018multi,hernandez2015safir}, such frameworks are lightweight and highly fast in processing and evaluation of security policies. Usually, such frameworks perform periodic evaluations on the time consumed in authenticating users and allocating communication uplink for data transmissions. Evaluation time, discovery time, and authentication time are the crucial parameters in time-bound security frameworks.
		\item Resource-bound: Most of the devices in smart M-IoT are low on resources and suffer from the threat of average lifetime. Usually, their lifetime is driven by the energy and memory consumed by the services operational on each device and often the mandatory services consume the majority of their services~\cite{cirani2015iot,pereira2014authentication,rahman2016securing}. Thus, it becomes important to develop frameworks which focus on the security while keeping a control on the utilization of M-IoT resources with a limited burden on the operational control and activity of the device. Such type of frameworks uses checkpoint mechanism to manage the resource consumption for the security of M-IoT applications.
		\item Hybrid: Nowadays, the smart applications tend to be time-bound as well as resource-bound. Thus, there is a requirement of frameworks which can apply both these features while forming a hybrid services-based framework that can use both the resource-checkpoints as well as periodic evaluation of security policies for securing activities in smart M-IoT. Accessibility and response time can be considered as mutual parameters for accessing the performance of such frameworks~\cite{seitz2013authorization}.
	\end{itemize}
	\subsection{Anomaly Detection-based Framework}
	Identification of false users, false services and false entities in a network is studied under this category. It is a responsibility of security framework to identify communities and users which pose potential risks to legitimate users of the network. Further, such a classification helps to manage the flow of information as well as limit the accessibility of users with harmful properties and high risks to network services. Anomaly detections are performed by checking the correctness of a device or user against the predefined policies of accurate operations. On a broader side, such frameworks can be classified into on-site and off-site evaluators with the description as given below:
	\begin{itemize}
		\item On-site: The real-time evaluation of the users for legitimate and accurate operations is classified as on-site or real-time anomaly detection. Such type of detections is performed by deploying real-time Intrusion Detection System (IDS) which dedicated sniffs the traffic without breaking its flow and without any excessive overheads~\cite{ahmad2017unsupervised,8322278,toledano2018real}. Majority of evaluations are conducted through sandboxes which do not reveal their identity to the users and prohibits anomaly users from accessing the services across the network.
		\item Off-site: In some cases, real-time evaluations may pose an excessive burden on the network and it is difficult to analyze the high flow of data. Such networks are evaluated off-site at their respective data centers which check for the presence of any abnormal activity for each of its users. Usually, such type is suitable for scenarios which allow delayed transactions without affecting the services such as payment gateways or smart-phone updates~\cite{8322278,ahmed2014network,sharma2017isma}.
	\end{itemize}
	Additionally, anomaly detections can be classified as periodic or continuous depending on the time and procedure of detection.
	\begin{figure*}
		\centering
		\fontsize{8}{10}\selectfont
		\begin{tikzpicture}[
		level 1/.style={sibling distance=35mm},
		edge from parent/.style={->,draw},
		>=latex]
		
		\node[root] {Classification of secure protocols in smart M-IoT}
		child {node[level 2] (c1) {Secure routing}}
		child {node[level 2] (c2) {Authentication mode-based}}
		child {node[level 2] (c3) {Authentication-hierarchy based}}
		child {node[level 2] (c4) {Property-based}};

		\begin{scope}[every node/.style={level 3}]
		\node [below of = c1, xshift=20pt] (c11) {RO-based relaying\\~\cite{shin2017secure}};
		\node [below of = c11,yshift=-10pt] (c12) {Trust-based relaying\\~\cite{sharma2017cooperative}};
		\node [below of = c12,yshift=-10pt] (c13) {Cluster-based relaying\\~\cite{deepa2017hhsrp}};
		\node [below of = c13,yshift=-10pt] (c14) {Secure MAC-based relaying\\~\cite{ullah2014hybrid,yang2016secure}};
		\node [below of = c14,yshift=-10pt] (c15) {Cross-layer-based relaying\\~\cite{li2017intelligent,vinayagam2017adopting,adibi2017novel}};
		\node [below of = c15,yshift=-10pt] (c16) {Topology-based relaying\\~\cite{chze2014secure}};
		\node [below of = c16,yshift=-10pt] (c17) {IP-based relaying\\~\cite{raza2012lightweight,hummen2013tailoring,you2017spfp}};
		\node [below of = c17,yshift=-10pt] (c18) {Mobility-based relaying\\~\cite{sharma2018secure,guan2017extension,shin2017secure}};
		
		\node [below of = c2, xshift=15pt] (c21) {Proactive\\~\cite{xu2014ticket,you2017spfp,sharma2018secure}};
		\node [below of = c21] (c22) {Reactive\\~\cite{yadav2017secure}};
		
		\node [below of = c3, xshift=15pt] (c31) {One-way\\~\cite{kang2017practical}};
		\node [below of = c31] (c32) {Two-way\\~\cite{kothmayr2013dtls,porambage2014two}};
		
		\node [below of = c4, xshift=20pt] (c41) {Freshness-based\\~\cite{amin2018light,gope2017lightweight,kalra2015secure,hummen2013tailoring,porambage2014two,mishra2017efficient,alkuhlani2017lightweight,shin2017secure,sharma2017secure}};
		\node [below of = c41,yshift=-15pt] (c42) {Encryption-based\\~\cite{gope2017lightweight,rahman2017anonpri,ermics2017key,kothmayr2013dtls,kalra2015secure,dhillon2017lightweight,raza2013lithe,hummen2013tailoring,chze2014secure,porambage2014two,mishra2017efficient,alkuhlani2017lightweight,shin2017secure,sharma2017secure}};
		\node [below of = c42,yshift=-15pt] (c43) {Access-based\\~\cite{gope2017lightweight,rahman2017anonpri,ermics2017key,kothmayr2013dtls,kalra2015secure,dhillon2017lightweight,raza2013lithe,porambage2014two,mishra2017efficient,alkuhlani2017lightweight,shin2017secure,sharma2017secure}};
		\node [below of = c43,yshift=-20pt] (c44) {System Integrity-based\\~\cite{gope2017lightweight,kothmayr2013dtls,dhillon2017lightweight,raza2013lithe,hummen2013tailoring,porambage2014two,alkuhlani2017lightweight,siboni2016advanced}};
		
		\end{scope}
		
		\foreach \value in {1,...,8}
		\draw[->] (c1.195) |- (c1\value.west);
		
		\foreach \value in {1,2}
		\draw[->] (c2.195) |- (c2\value.west);
		
		\foreach \value in {1,2}
		\draw[->] (c3.195) |- (c3\value.west);
		
		\foreach \value in {1,2,3,4}
		\draw[->] (c4.195) |- (c4\value.west);
		
		\end{tikzpicture}
		\caption{A broad classification of secure protocols in smart M-IoT. The existing solutions can be classified into secure routing, authentication mode and hierarchy, and property-based secure protocols.}\label{fig:tax_protocols}
	\end{figure*}
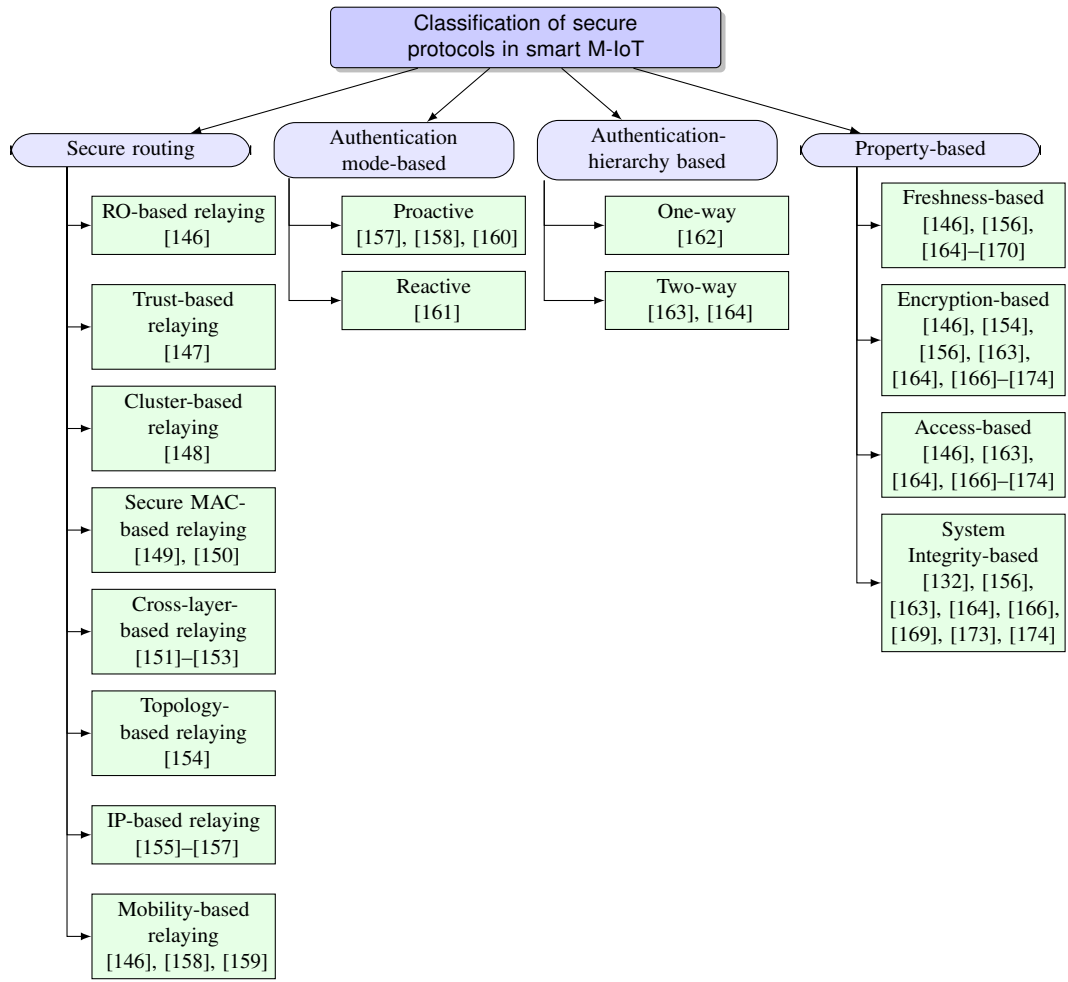
	
	\subsection{Summary and Insights}
	In this section, we have summarized different types of frameworks, which help to secure the operations as well as the network layout of smart M-IoT. The summarized study divides the existing solutions into five broad categories and considering their evaluation metrics and ideologies, a state-of-the-art comparison is presented in Table~\ref{Table1}, which compares the key IoT frameworks with specification on smart mobile network formations. The table helps to understand the reachability of each approach and their primary application of interest. These comparisons can be used for understanding what has been attained so far and what are the directions yet to be focused while securing applications in smart M-IoT networks. Hybridizing network layouts with security policies and involving vulnerability assessments can be used for developing security middleware for smart M-IoT applications. As discussed in~\cite{dsouza2014policy,cirani2015iot}, security resources, user authentication, device authentication, resource utilization and data investigations should be included while developing frameworks for securing smart M-IoT. There are many solutions, which only relies on CoAP, but it is desired to make strategic shifts for enhancing the security of devices and users against known as well as unknown cyber threats.
	
	\section{Security-Aware Protocols for smart M-IoT}
	\textcolor{black}{It is to be considered that with the introduction of new technologies for communications, the links between the M-IoT devices have grown up to many Gigabits, which means the window to perform security operations has further decreased, and it is extremely challenging for the researchers to accommodate existing security policies in such a short timing window. Thus, protocols for M-IoT security are yet to be revolutionized on the basis of their applicability and reachability for M-IoT applications.} \textcolor{black}{Security protocols prevent unauthorized attempts for using resources or data in a defined network~\cite{bonetto2012secure}~\cite{lee2014lightweight}. Communications in M-IoT are usually handled by the dissemination protocols like CoAP,  Advanced Message Queuing Protocol (AMQP), Message Queuing Telemetry Transport (MQTT), Domain Name Server-Service Discovery (DNS-SD), etc~\cite{garcia2016lightweight}, whereas security is supported either by enhancing the features of these protocols; or by using existing security protocols with the routing schemes; or by designing novel security and communication schemes which are usually specific to applications~\cite{heer2011security}~\cite{granjal2015security}. Such solutions may operate well in one scenario and may fall prey to different types of attacks if their application area is changed.} \textcolor{black}{The success of the security protocols is affected by the compliance degree of a user with the recommended settings~\cite{guo2017generic}~\cite{accettura2014optimal}. It is required that security protocols should not affect the performance and their operations (like encryption and decryption) should be completed without many overheads.}
	
	\textcolor{black}{Protection of Peer to Peer (P2P) and Peer to Multi-Peer (P2MP) links is one of the major challenges while designing protocols for the security of M-IoT. Protocols can be protected either by following asymmetric mode or symmetric mode in their key operations. The location of AAA server and its optimized placement are other issues to be resolved in M-IoT. Moreover, Route optimizations are additional concerns which have to be taken care by the security protocols.} \textcolor{black}{Previously known protocols, like Secure File Transfer Protocol (SFTP), Secure Sockets Layer (SSL), Hypertext Transfer Protocol Secure (HTTPS), Session Initiation Protocol (SIP), can be adopted for network security while authentication can be governed by Authentication and Key Agreement (AKA) as it is one of the standard protocols used for security in 3G. Some other crucial protocols include Extensible Authentication Protocol (EAP), Remote Access Dial In User Service (RADIUS), DIAMETER, Protocol for Carrying Authentication for Network Access (PANA),  Lightweight Extensible Authentication Protocol (LEAP), Protected Extensible Authentication Protocol (PEAP), etc. EAP is further facilitated with suitable extensions as AKA, AKA-prime (AKA'), Transport Layer Security (TLS), Message Digest-5 (MD5), Tunneled Transport Layer Security (TTLS), Encrypted Key Exchange (EKE), Generic Token Card (GTC), Pre-Shared Key (PSK), Password (PWD), etc~\cite{syafruddin2010performance}~\cite{nakhjiri2005aaa}. Handover protection is also related to the protocols as these are identified as a crucial part of security in M-IoT. Majority of them use EAP-based authentication (EAP-TLS). Some of the key contributions on handover security include Security Protocol for Fast PMIPv6 (SPFP), Handover Optimized Ticket-based Authentication (HOTA), Ticket-based authentication (TA), Secure Password Authentication Mechanism (SPAM), etc~\cite{you2017spfp}.} These protocols can be classified into routing-based, authentication-mode based, authentication-hierarchy-based, and property-based mechanisms, as shown in Fig.~\ref{fig:tax_protocols}.
	\subsection{Secure routing-based}
	There are a plethora of solutions, which focus on providing secure routing for smart M-IoT applications. The routing schemes which are available for general networks, be it reactive or proactive, holds true for smart M-IoT setups. Existing routing mechanisms can be used while leveraging the security guidelines to secure the communications between the M-IoT users. From a broader point of view, the secure routing-based protocols can be classified into following types:
	\begin{itemize}
		\item Route Optimization (RO)-based relaying: Finding shortest paths and reducing the path of authentication can be attained through optimized relaying in the networks. Such RO-based relaying often removes dependencies from the intermediate entities to provide low-overhead based solutions for security~\cite{shin2017secure}.
		\item Trust-based relaying: Finding nodes on the basis of trust calculations and using them for transmissions are another kind of security protocols. Such protocols use trust as a weighted metric for calculating paths between the users in M-IoT~\cite{sharma2017cooperative}.
		\item Cluster-based relaying: In some scenarios, network entities operate in a group while depending on a core entity which acts as their head leading to the formation of multiple clusters in the network. There are certain routing protocols which aim to support the security of communication between the cluster heads allowing secure relaying between the nodes with lesser overheads and computational complexity~\cite{deepa2017hhsrp}. Clustering is effective in case the protocols depend on group-based authentication, however, in primary and secondary modes of authentication, it may cause excessive overheads.
		\item Secure-Medium Access Control (MAC) based relaying: Control over timing policies and accessibility of user operations lead to the requirements of a secure MAC based relaying for users in smart M-IoT. Such relaying protocols use command over congestion window and packet forwarding policies to control the flow of packets as well as uses cryptographic solutions for securing its relaying procedures~\cite{ullah2014hybrid,yang2016secure}.
		\item Cross-layer-based relaying: Secure routing can be obtained over the network layer while obtaining properties from other layers such as the transport layer or the MAC layer. The protocols on the network layer use parameters like Received Signal Strength Indicator (RSSI) and use it as a weighted condition to select nodes in smart M-IoT~\cite{li2017intelligent,vinayagam2017adopting,adibi2017novel}. Such a relaying can be effective in scenarios where the resources are limited and the lifetime of the network is of utmost importance.
		\item Topology-based relaying: Identification of nodes on the basis of their location and checking the path of authentication before transmissions lead to the formation of secure topology-based routing~\cite{chze2014secure}. Such protocols are effective where the dynamic nature of nodes is crucial and often changes. However, it is difficult to control such scenario and topology-aware relaying is often combined with mobility-management procedures for attaining a secure and fast relaying.
		\item IP-based relaying: This is the core relaying mechanism for the majority of the mobile applications as it uses Mobile IPv6 (MIPv6) and Fast Mobile IPv6 (FMIPv6) procedures to support the selection of nodes. Further, the security in such protocols is provided by proxy-mechanisms and can be seen in various proxy-based protocols such as Proxy Mobile IPv6 (PMIPv6) and F-PMIPv6~\cite{raza2012lightweight,hummen2013tailoring,you2017spfp}. Such relaying solutions can be combined with media independent schemes to form Media Independent Handover (MIH)-PMIPv6 relaying with specific implementation over smart M-IoT applications.
		\item Mobility-based relaying: Mobility management is often studied as a part of handovers; however, existing routing schemes can be classified on the basis of mobility management. Such schemes are responsible for securing the path of the nodes when they are moving in an intra- or inter-mode of a given authentication server. Mobility management schemes can be studied as distributed, centralized, semi-distributed or even hierarchical~\cite{sharma2018secure,guan2017extension,shin2017secure}.
	\end{itemize}
	\subsection{Authentication mode-based}
	Similar to authentication-based frameworks, authentication protocols allow identification of legitimate users which can interact with each other for acquiring particular services over the network. Authentication protocols help to validate the users for transmissions in M-IoT. The vulnerability and importance of M-IoT demand the employment of exceedingly reliable methods in the design of secure systems. Authentication protocols are one of the most important design parameters. These protocols help to achieve a reliable trust and security for exchange information. On the basis of mode of operations, the authentication protocols can be classified into following two types:
	\begin{itemize}
		\item Proactive: Authentication protocols which focus on pre-verification of the users before beginning the transmissions are termed as proactive authentication~\cite{xu2014ticket,you2017spfp,sharma2018secure}. Such schemes are highly reliable but sometimes slower in operations. Thus, these are often the primary preference of setups that focus on the services over smart M-IoT.
		\item Reactive: Authentication protocols which focus on the on-demand verification of the users and support a direct linking between the network users are termed as reactive authentication~\cite{yadav2017secure}. Reactive authentication is fast in operations, but is usually, vulnerable to a lot of network attacks which raises a question about their secure usability for smart M-IoT. However, with modern solutions like crowdsourcing~\cite{irazabal2019blockchain} and blockchains~\cite{8734799,8743548,sharma2018block}, reactive protocols can easily be extended and secure for their usability in smart M-IoT setups.
	\end{itemize}
	
	\subsection{Authentication hierarchy-based}
	Authentication involves multiple entities which secure themselves by verifying each other either directly or through an Authentication Server (AS). On the basis of operations and hierarchy, authentication protocols can be classified into one-way or two-way authentication based protocols.
	
	\onecolumn
	\begin{landscape}
		\begin{center}
			\fontsize{8}{11}\selectfont
			\setlength\LTleft{30pt}            
			\setlength\LTright{0pt}
			\begin{longtable}{@{\extracolsep{\fill}}*{11}{c}}
				\caption{State-of-the-art protocols for M-IoT security.}\label{Table2} \\
				\hline
				\multicolumn{1}{p{1.5cm}}{\centering \textbf{Approach}}
				&\multicolumn{1}{p{1.5cm}}{\centering \textbf{Author} \\ \textbf{(Year)} }
				&\multicolumn{1}{p{1.5cm}}{\centering \textbf{Ideology}}
				&\multicolumn{1}{p{1.2cm}}{\centering \textbf{System}\\ \textbf{Integrity} }
				&\multicolumn{1}{p{1cm}}{\centering \textbf{Freshness}}
				&\multicolumn{1}{p{1.2cm}}{\centering \textbf{Confidentiality}}
				&\multicolumn{1}{p{1.2cm}}{\centering \textbf{Mutual}\\ \textbf{Authentication}}
				&\multicolumn{1}{p{1cm}}{\centering \textbf{Access}\\ \textbf{Control}}
				&\multicolumn{1}{p{1.2cm}}{\centering \textbf{Overheads}}
				&\multicolumn{1}{p{1.2cm}}{\centering \textbf{Encryption}}
				&\multicolumn{1}{p{1.2cm}}{\centering \textbf{Non-repudiation}} \\[6pt]\\
				\hline \\
				
				\endfirsthead
				
				\multicolumn{11}{c}%
				{{\bfseries \tablename\ \thetable{} -- continued from previous page}} \\
				\hline
				
				\multicolumn{1}{p{1.5cm}}{\centering \textbf{Approach}}
				&\multicolumn{1}{p{1.5cm}}{\centering \textbf{Author} \\ \textbf{(Year)} }
				&\multicolumn{1}{p{1.5cm}}{\centering \textbf{Ideology}}
				&\multicolumn{1}{p{1.2cm}}{\centering \textbf{System}\\ \textbf{Integrity} }
				&\multicolumn{1}{p{1cm}}{\centering \textbf{Freshness}}
				&\multicolumn{1}{p{1.2cm}}{\centering \textbf{Confidentiality}}
				&\multicolumn{1}{p{1.2cm}}{\centering \textbf{Mutual}\\ \textbf{Authentication}}
				&\multicolumn{1}{p{1cm}}{\centering \textbf{Access}\\ \textbf{Control}}
				&\multicolumn{1}{p{1.2cm}}{\centering \textbf{Overheads}}
				&\multicolumn{1}{p{1.2cm}}{\centering \textbf{Encryption}}
				&\multicolumn{1}{p{1.2cm}}{\centering \textbf{Non-repudiation}} \\[6pt]\\
				\hline\\\endhead
				
				\hline \multicolumn{11}{l}{{Continued on next page}} \\
				
				\endfoot
				
				\endlastfoot
				
				\multicolumn{1}{p{1.5cm}}{\centering  Light weight authentication}
				&\multicolumn{1}{p{1.5cm}}{\centering [Amin \emph{et al.} 2018]\cite{amin2018light}}
				&\multicolumn{1}{p{1.5cm}}{\centering Distributed cloud computing environment}
				&\multicolumn{1}{p{1.2cm}}{\centering \xmark}
				&\multicolumn{1}{p{1cm}}{\centering \cmark}
				&\multicolumn{1}{p{1.2cm}}{\centering \cmark}
				&\multicolumn{1}{p{1.2cm}}{\centering \cmark}
				&\multicolumn{1}{p{1cm}}{\centering -}
				&\multicolumn{1}{p{1.2cm}}{\centering Medium}
				&\multicolumn{1}{p{1.2cm}}{\centering \xmark}
				&\multicolumn{1}{p{1.2cm}}{\centering \xmark} \\[6pt]\\
				
				\multicolumn{1}{p{1.5cm}}{\centering RFID authentication scheme }
				&\multicolumn{1}{p{1.5cm}}{\centering [Gope \emph{et al.} 2017]\cite{gope2017lightweight}}
				&\multicolumn{1}{p{1.5cm}}{\centering Distributed IoT infrastructure}
				&\multicolumn{1}{p{1.2cm}}{\centering \cmark}
				&\multicolumn{1}{p{1cm}}{\centering \cmark}
				&\multicolumn{1}{p{1.2cm}}{\centering \cmark}
				&\multicolumn{1}{p{1.2cm}}{\centering \cmark}
				&\multicolumn{1}{p{1cm}}{\centering \cmark}
				&\multicolumn{1}{p{1.2cm}}{\centering High}
				&\multicolumn{1}{p{1.2cm}}{\centering \cmark}
				&\multicolumn{1}{p{1.2cm}}{\centering \xmark} \\[6pt]\\
				
				\multicolumn{1}{p{1.5cm}}{\centering Anonymous private authentication}
				&\multicolumn{1}{p{1.5cm}}{\centering [Rahman \emph{et al.} 2017]\cite{rahman2017anonpri}}
				&\multicolumn{1}{p{1.5cm}}{\centering Security of RFID systems}
				&\multicolumn{1}{p{1.2cm}}{\centering -}
				&\multicolumn{1}{p{1cm}}{\centering \xmark}
				&\multicolumn{1}{p{1.2cm}}{\centering \cmark}
				&\multicolumn{1}{p{1.2cm}}{\centering \xmark}
				&\multicolumn{1}{p{1cm}}{\centering \cmark}
				&\multicolumn{1}{p{1.2cm}}{\centering High}
				&\multicolumn{1}{p{1.2cm}}{\centering \cmark}
				&\multicolumn{1}{p{1.2cm}}{\centering \xmark} \\[6pt]\\
				
				\multicolumn{1}{p{1.5cm}}{\centering  Key agreement mechanism}
				&\multicolumn{1}{p{1.5cm}}{\centering [Ermis \emph{et al.}  2017]\cite{ermics2017key}}
				&\multicolumn{1}{p{1.5cm}}{\centering Partial backward confidentiality}
				&\multicolumn{1}{p{1.2cm}}{\centering -}
				&\multicolumn{1}{p{1cm}}{\centering \xmark}
				&\multicolumn{1}{p{1.2cm}}{\centering \cmark}
				&\multicolumn{1}{p{1.2cm}}{\centering \xmark}
				&\multicolumn{1}{p{1cm}}{\centering \cmark}
				&\multicolumn{1}{p{1.2cm}}{\centering Medium}
				&\multicolumn{1}{p{1.2cm}}{\centering \cmark}
				&\multicolumn{1}{p{1.2cm}}{\centering \xmark} \\[6pt]\\
				
				\multicolumn{1}{p{1.5cm}}{\centering two-way authentication}
				&\multicolumn{1}{p{1.5cm}}{\centering [Kothmayr \emph{et al.}  2013]\cite{kothmayr2013dtls}}
				&\multicolumn{1}{p{1.5cm}}{\centering DTLS based security}
				&\multicolumn{1}{p{1.2cm}}{\centering \cmark}
				&\multicolumn{1}{p{1cm}}{\centering \xmark}
				&\multicolumn{1}{p{1.2cm}}{\centering \cmark}
				&\multicolumn{1}{p{1.2cm}}{\centering \xmark}
				&\multicolumn{1}{p{1cm}}{\centering \cmark}
				&\multicolumn{1}{p{1.2cm}}{\centering Medium}
				&\multicolumn{1}{p{1.2cm}}{\centering \cmark}
				&\multicolumn{1}{p{1.2cm}}{\centering \xmark} \\[6pt]\\
				
				\multicolumn{1}{p{1.5cm}}{\centering  Secure authentication scheme}
				&\multicolumn{1}{p{1.5cm}}{\centering [Kalra and Sood 2015]\cite{kalra2015secure}}
				&\multicolumn{1}{p{1.5cm}}{\centering Authentication of IoT and cloud servers}
				&\multicolumn{1}{p{1.2cm}}{\centering -}
				&\multicolumn{1}{p{1cm}}{\centering \cmark}
				&\multicolumn{1}{p{1.2cm}}{\centering \cmark}
				&\multicolumn{1}{p{1.2cm}}{\centering \cmark}
				&\multicolumn{1}{p{1cm}}{\centering \cmark}
				&\multicolumn{1}{p{1.2cm}}{\centering Medium}
				&\multicolumn{1}{p{1.2cm}}{\centering \cmark}
				&\multicolumn{1}{p{1.2cm}}{\centering \xmark} \\[6pt]\\
				
				\multicolumn{1}{p{1.5cm}}{\centering  User authentication scheme}
				&\multicolumn{1}{p{1.5cm}}{\centering [Dhillon and Kalra 2017]\cite{dhillon2017lightweight}}
				&\multicolumn{1}{p{1.5cm}}{\centering Lightweight biometrics based}
				&\multicolumn{1}{p{1.2cm}}{\centering \cmark}
				&\multicolumn{1}{p{1cm}}{\centering \xmark}
				&\multicolumn{1}{p{1.2cm}}{\centering \cmark}
				&\multicolumn{1}{p{1.2cm}}{\centering \cmark}
				&\multicolumn{1}{p{1cm}}{\centering \cmark}
				&\multicolumn{1}{p{1.2cm}}{\centering Medium}
				&\multicolumn{1}{p{1.2cm}}{\centering \cmark}
				&\multicolumn{1}{p{1.2cm}}{\centering \xmark} \\[6pt]\\
				
				\multicolumn{1}{p{1.5cm}}{\centering Key management protocol}
				&\multicolumn{1}{p{1.5cm}}{\centering [Raza \emph{et al.}  2012]\cite{raza2012lightweight}}
				&\multicolumn{1}{p{1.5cm}}{\centering Lightweight IKEv2}
				&\multicolumn{1}{p{1.2cm}}{\centering \xmark}
				&\multicolumn{1}{p{1cm}}{\centering \xmark}
				&\multicolumn{1}{p{1.2cm}}{\centering \xmark}
				&\multicolumn{1}{p{1.2cm}}{\centering \xmark}
				&\multicolumn{1}{p{1cm}}{\centering -}
				&\multicolumn{1}{p{1.2cm}}{\centering Medium}
				&\multicolumn{1}{p{1.2cm}}{\centering \xmark}
				&\multicolumn{1}{p{1.2cm}}{\centering \xmark} \\[6pt]\\
				
				\multicolumn{1}{p{1.5cm}}{\centering  Constrained application protocol}
				&\multicolumn{1}{p{1.5cm}}{\centering [Raza \emph{et al.} 2013]\cite{raza2013lithe}}
				&\multicolumn{1}{p{1.5cm}}{\centering DTLS based security}
				&\multicolumn{1}{p{1.2cm}}{\centering \cmark}
				&\multicolumn{1}{p{1cm}}{\centering \xmark}
				&\multicolumn{1}{p{1.2cm}}{\centering \cmark}
				&\multicolumn{1}{p{1.2cm}}{\centering -}
				&\multicolumn{1}{p{1cm}}{\centering \cmark}
				&\multicolumn{1}{p{1.2cm}}{\centering Low}
				&\multicolumn{1}{p{1.2cm}}{\centering \cmark}
				&\multicolumn{1}{p{1.2cm}}{\centering \xmark} \\[6pt]\\
				
				\multicolumn{1}{p{1.5cm}}{\centering End-to-End IP security}
				&\multicolumn{1}{p{1.5cm}}{\centering [Hummen \emph{et al.}  2013]\cite{hummen2013tailoring}}
				&\multicolumn{1}{p{1.5cm}}{\centering Common protocol functionality-based IP security}
				&\multicolumn{1}{p{1.2cm}}{\centering \cmark}
				&\multicolumn{1}{p{1cm}}{\centering \cmark}
				&\multicolumn{1}{p{1.2cm}}{\centering -}
				&\multicolumn{1}{p{1.2cm}}{\centering \cmark}
				&\multicolumn{1}{p{1cm}}{\centering -}
				&\multicolumn{1}{p{1.2cm}}{\centering Medium}
				&\multicolumn{1}{p{1.2cm}}{\centering \cmark}
				&\multicolumn{1}{p{1.2cm}}{\centering \cmark} \\[6pt]\\
				
				\multicolumn{1}{p{1.5cm}}{\centering  Secure multi-hop routing}
				&\multicolumn{1}{p{1.5cm}}{\centering [Chze and Leong 2014]\cite{chze2014secure}}
				&\multicolumn{1}{p{1.5cm}}{\centering Secure IoT Communication}
				&\multicolumn{1}{p{1.2cm}}{\centering \xmark}
				&\multicolumn{1}{p{1cm}}{\centering \xmark}
				&\multicolumn{1}{p{1.2cm}}{\centering -}
				&\multicolumn{1}{p{1.2cm}}{\centering \xmark}
				&\multicolumn{1}{p{1cm}}{\centering -}
				&\multicolumn{1}{p{1.2cm}}{\centering Medium}
				&\multicolumn{1}{p{1.2cm}}{\centering \cmark}
				&\multicolumn{1}{p{1.2cm}}{\centering \xmark} \\[6pt]\\
				
				\multicolumn{1}{p{1.5cm}}{\centering  Two-phase authentication}
				&\multicolumn{1}{p{1.5cm}}{\centering [Porambage \emph{et al.} 2014]\cite{porambage2014two}}
				&\multicolumn{1}{p{1.5cm}}{\centering Authentication of distributed IoT applications}
				&\multicolumn{1}{p{1.2cm}}{\centering \cmark}
				&\multicolumn{1}{p{1cm}}{\centering \cmark}
				&\multicolumn{1}{p{1.2cm}}{\centering -}
				&\multicolumn{1}{p{1.2cm}}{\centering \cmark}
				&\multicolumn{1}{p{1cm}}{\centering \cmark}
				&\multicolumn{1}{p{1.2cm}}{\centering Low}
				&\multicolumn{1}{p{1.2cm}}{\centering \cmark}
				&\multicolumn{1}{p{1.2cm}}{\centering \xmark} \\[6pt]\\
				
				\multicolumn{1}{p{1.5cm}}{\centering  Authentication protocol for multimedia}
				&\multicolumn{1}{p{1.5cm}}{\centering [Mishra \emph{et al.} 2017]\cite{mishra2017efficient}}
				&\multicolumn{1}{p{1.5cm}}{\centering Secure multimedia communications}
				&\multicolumn{1}{p{1.2cm}}{\centering \xmark}
				&\multicolumn{1}{p{1cm}}{\centering \cmark}
				&\multicolumn{1}{p{1.2cm}}{\centering \cmark}
				&\multicolumn{1}{p{1.2cm}}{\centering \cmark}
				&\multicolumn{1}{p{1cm}}{\centering \cmark}
				&\multicolumn{1}{p{1.2cm}}{\centering Low}
				&\multicolumn{1}{p{1.2cm}}{\centering \cmark}
				&\multicolumn{1}{p{1.2cm}}{\centering \xmark} \\[6pt]\\
				
				\multicolumn{1}{p{1.5cm}}{\centering  Authentication and key Agreement}
				&\multicolumn{1}{p{1.5cm}}{\centering [Alkuhlani and Thorat 2017]\cite{alkuhlani2017lightweight}}
				&\multicolumn{1}{p{1.5cm}}{\centering Anonymity-preserving agreement}
				&\multicolumn{1}{p{1.2cm}}{\centering \cmark}
				&\multicolumn{1}{p{1cm}}{\centering \cmark}
				&\multicolumn{1}{p{1.2cm}}{\centering -}
				&\multicolumn{1}{p{1.2cm}}{\centering \cmark}
				&\multicolumn{1}{p{1cm}}{\centering \cmark}
				&\multicolumn{1}{p{1.2cm}}{\centering -}
				&\multicolumn{1}{p{1.2cm}}{\centering \cmark}
				&\multicolumn{1}{p{1.2cm}}{\centering \xmark} \\[6pt]\\
				
				\multicolumn{1}{p{1.5cm}}{\centering  Secure route optimization}
				&\multicolumn{1}{p{1.5cm}}{\centering [Shin \emph{et al.} 2017]\cite{shin2017secure}}
				&\multicolumn{1}{p{1.5cm}}{\centering Security of PMIPv6-smart home IoT}
				&\multicolumn{1}{p{1.2cm}}{\centering \cmark}
				&\multicolumn{1}{p{1cm}}{\centering \cmark}
				&\multicolumn{1}{p{1.2cm}}{\centering \cmark}
				&\multicolumn{1}{p{1.2cm}}{\centering \cmark}
				&\multicolumn{1}{p{1cm}}{\centering \cmark}
				&\multicolumn{1}{p{1.2cm}}{\centering Low}
				&\multicolumn{1}{p{1.2cm}}{\centering \cmark}
				&\multicolumn{1}{p{1.2cm}}{\centering \xmark} \\[6pt]\\
				
				\multicolumn{1}{p{1.5cm}}{\centering  Remote user authentication scheme}
				&\multicolumn{1}{p{1.5cm}}{\centering [Sharma and Kalra 2017]\cite{sharma2017secure}}
				&\multicolumn{1}{p{1.5cm}}{\centering Authentication for e-governance applications }
				&\multicolumn{1}{p{1.2cm}}{\centering -}
				&\multicolumn{1}{p{1cm}}{\centering \cmark}
				&\multicolumn{1}{p{1.2cm}}{\centering -}
				&\multicolumn{1}{p{1.2cm}}{\centering \cmark}
				&\multicolumn{1}{p{1cm}}{\centering \cmark}
				&\multicolumn{1}{p{1.2cm}}{\centering -}
				&\multicolumn{1}{p{1.2cm}}{\centering \cmark}
				&\multicolumn{1}{p{1.2cm}}{\centering \xmark} \\[6pt]\\
				
				\hline
			\end{longtable}
		\end{center}
		\twocolumn
	\end{landscape}
	\begin{itemize}
		\item One-way authentication: One-way authentication involves user-side verification with respect to the rules provided by the governing server (AAA or AS)~\cite{kang2017practical}. The genuineness of the users is proved by the properties which are only shared by the user itself.
		\item Two-way authentication: Two-way authentication involves both user-sides as well as server-side verifications~\cite{kothmayr2013dtls,porambage2014two}. The genuineness of the users, as well as the servers, is proved through their respective properties which are shared amongst them. Two-way authentication can further be extended into different modes of handshakes depending on the level of security to be verified before beginning the transmissions.
	\end{itemize}
	\subsection{Property-based}
	Security protocols can also be classified on the basis of properties which are used for securing the transmissions between the nodes. Based on some key requirements, the security protocols can be categorized on the basis of following properties:
	\begin{itemize}
		\item Freshness-based: Freshness means that messages exchanged in a session are generated specifically for a particular session. The attacker cannot use the previous session for messages. Freshness based protocols are used for communication between the two parties by establishing a secure channel on the basis of the freshness of sessions. The receiver believes that the obtained information is fresh and authenticated. Freshness is achieved by updating keys and sessions through consistent changes in parameters like seeds, nonce and sequence numbers of involved entities in smart M-IoT. Approaches based on freshness of keys and sessions are discussed in~\cite{amin2018light,gope2017lightweight,kalra2015secure,hummen2013tailoring,porambage2014two,mishra2017efficient,alkuhlani2017lightweight,shin2017secure,sharma2017secure}.
		\item Encryption-based: Encryption is an interesting piece of technology that works by scrambling data or information so it is unreadable by attackers. Encryption is a key-based approach to combine confidentiality and integrity, and provides a secure mechanism against external threats such as chosen plaintext and chosen ciphertext attacks. Encryption based protocol ensure the confidentiality of sharing information between the users in smart M-IoT~\cite{gope2017lightweight,rahman2017anonpri,ermics2017key,kothmayr2013dtls,kalra2015secure,dhillon2017lightweight,raza2013lithe,hummen2013tailoring,chze2014secure,porambage2014two,mishra2017efficient,alkuhlani2017lightweight,shin2017secure,sharma2017secure}.
		\item Access-based: Limiting the users from accessing a particular service is one of the key requirements of smart M-IoT applications. Protocols which can help to define role to every user and control their activity are classified into access-based security protocols. There are a lot of existing solutions, which aim at enhancing the security of the mobile network by limiting the user operations while using the policies for information flow, management, and control~\cite{gope2017lightweight,rahman2017anonpri,ermics2017key,kothmayr2013dtls,kalra2015secure,dhillon2017lightweight,raza2013lithe,porambage2014two,mishra2017efficient,alkuhlani2017lightweight,shin2017secure,sharma2017secure}. A highly stabilized access control protocol can prevent misleading or eavesdropper from gaining access to crucial information in smart M-IoT.
		\item System Integrity-based: System integrity protection is a necessary step to ensure a high level of security. As discussed in~\cite{gope2017lightweight,kothmayr2013dtls,dhillon2017lightweight,raza2013lithe,hummen2013tailoring,porambage2014two,alkuhlani2017lightweight,siboni2016advanced}, development of system integrity protection protocols can help to manage information disturbances and prevent attacks. The involved parties in smart M-IoT setups want to assure that all the remote data they receive is from systems that satisfy the users' integrity requirements. Therefore, it is important that system integrity based protocols can protect the information results from being polluted by attackers.
	\end{itemize}
	\begin{figure*}
		\centering
		\fontsize{8}{10}\selectfont
		\begin{tikzpicture}[
		level 1/.style={sibling distance=35mm},
		edge from parent/.style={->,draw},
		>=latex]
		
		\node[root] {Classification of privacy-preservation approaches in smart M-IoT}
		child {node[level 2] (c1) {Encryption-based}}
		child {node[level 2] (c2) {Architecture-based}}
		child {node[level 2] (c3) {Protocol-based\\~\cite{rahman2017anonpri,alcaide2013anonymous}}}
		child {node[level 2] (c4) {Tool-based\\~\cite{evans2012efficient,ukil2012negotiation,sharma2017cooperative}}};
		
		\begin{scope}[every node/.style={level 3}]
		\node [below of = c1, xshift=20pt] (c11) {Symmetric-key\\~\cite{perez2017towards}};
		\node [below of = c11,yshift=-10pt] (c12) {Asymmetric-key\\~\cite{jayaraman2017privacy,belguith2018phoabe,bamasag2015lightweight,hu2011identity,doukas2012enabling,wang2014performance,li2014p3,bao2016lightweight}};
		\node [below of = c12,yshift=-10pt] (c13) {Homomorphic\\~\cite{bao2015new,gong2015medical,lu2017lightweight}};
		
		\node [below of = c2, xshift=20pt] (c21) {Centralized\\~\cite{mascetti2009longitude}};
		\node [below of = c21] (c22) {Distributed\\~\cite{alcaide2013anonymous}~\cite{addo2014reference}};
		
		\end{scope}
		
		\foreach \value in {1,2,3}
		\draw[->] (c1.195) |- (c1\value.west);
		
		\foreach \value in {1,2}
		\draw[->] (c2.195) |- (c2\value.west);
		
		\end{tikzpicture}
		\caption{A broad classification of privacy-preservation schemes for smart M-IoT. The existing solutions can be studied by classifying them into encryption-based, architecture-based, protocol-based, and tool-based mechanisms for privacy preservation.}\label{fig:tax_privacy}
	\end{figure*}
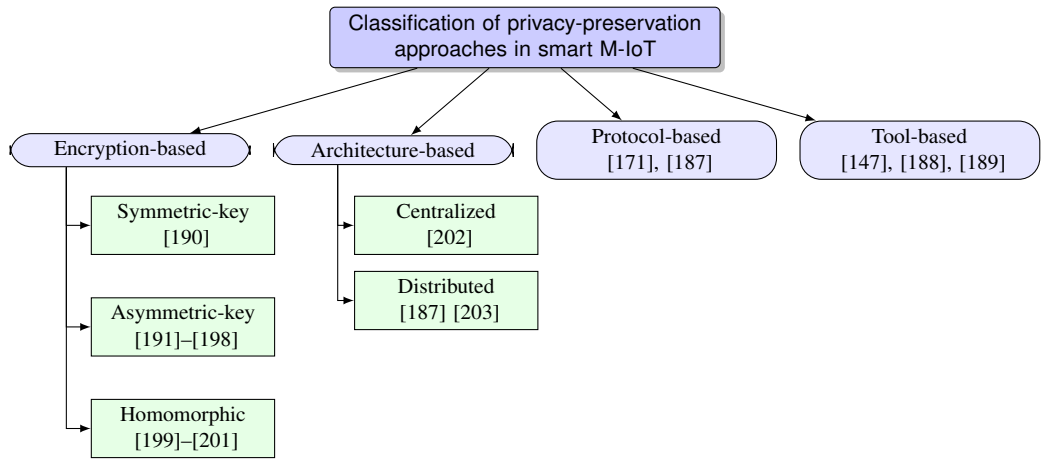
	
	\subsection{Summary and Insights}
	In this section, we presented a detailed study on security-aware protocols for smart M-IoT. Following the above-discussed classification, some major contributions to security protocols which are applicable to M-IoT are highlighted in Table~\ref{Table2}. The existing protocols are evaluated on the basis of system integrity, freshness, confidentiality, mutual authentication, access control, overheads, encryption, and non-repudiation. Apart from these, several schemes can be followed from SPORE~\cite{SG5}, which is a repository of security protocols.. Over the last decade, protocols have been improvised by utilizing security as a crucial metric to decide a path; however, with the evolution of new CPS, dynamic nodes, and energy-constraint mobile devices, this direction of research remains open and require protocols, which can operate beyond authentication.
	
	Research can be extended towards the designing of a secure protocol stack, which can include channel as well as message protection without compromising the QoS to its users. Network patrolling, perimeter evaluation, and deployment of intrusion detection protocols can help to further secure the operations of smart M-IoT.
	
	\section{Privacy Preservation Approaches for smart M-IoT}
	Data in M-IoT is highly crucial as well as sensitive and any eavesdropping may result in leakage of users' personal information~\cite{vasilomanolakis2015security}~\cite{alqassem2014taxonomy}. With data processing reaching a fine granularity level, it becomes tedious to privatize the content as new issues arise because of many dependencies on the platform used for transmissions. The collection and control of data are two of the main reasons that increase threat-level for data privacy in M-IoT~\cite{patil2014big,bertino2016data,perera2015big}.
	
	With the difference in architectural deployment, smart M-IoT possesses large-scale implications for removing issues which may leak the entire information of the networks. Most of the approaches fail to support access control and authorization while deploying applications for smart M-IoT networks. Reducing the reachability of every user and keeping a watch on the amount and level of contents accessed by an individual can help in privacy-preservation~\cite{hwang2015iot,medaglia2010overview,zhang2014iot}.
	
	Encryption of data for every link can further help this cause, however, with the networks attaining a high-speed property, it becomes necessary to support both encryption and decryption at a rapid pace~\cite{jiang2017efficient}. Majority of the intermediate procedures should be done on the cipher itself, as this will help to prevent any unauthorized decryption of the text being shared between M-IoT devices. Further, approaches can use customized identifiers for creating policies for maintaining the anonymity of access between its users. Along with these, prevention of hidden terminals is another major requirement for privacy-preservation~\cite{sadeghi2015security,skarmeta2014decentralized,lin2015insecurity,schurgot2015experiments}.
	
	Practical problems like network partitioning and isolations increase the risks of leakage of data and it is necessary to formulate approaches that can help to identify such issues before-hand and with low-complexity~\cite{campan2009data,cheng2010k,zou2009k,abomhara2014security,weber2010internet}. Data privacy can be guaranteed by using solutions, which prevent sniffing and do not yield any information even if discovered by intermediates~\cite{kozlov2012security}~\cite{ukil2015privacy}~\cite{ukil2014iot}. This can be further enhanced by using non-store approach, which refers to the immediate forwarding of the data without consuming the excessive time stamp as well as keeping the freshness of the keys. Privacy can further be assured by preventing third party-based evaluations as these may disclose the encryption mechanisms of the entire route as well as of the traffic~\cite{dorri2017blockchain}.
	
	Distribution of incoming traffic not only prevents DoS or DDoS but also helps to make sure about the identification of any eavesdropper that may be listening to the incoming or outgoing traffic~\cite{abhishta2017comparing}~\cite{kotenko2017parallel}. Updating security policies, maintenance of logs and refining network architecture at regular intervals for the detected traffic and anomalies can further help in privacy-preservation of M-IoT networks~\cite{zhou2014security}.
	
	Some of the major contributions on data privacy in IoT, which are applicable to M-IoT architecture, are discussed in Table~\ref{Table3}. These schemes can be further classified into four major types, encryption-based, architecture-based, protocol-based and tool-based privacy preservation, as shown in Fig.~\ref{fig:tax_privacy}. The details of each of these are provided below:
	\subsection{Encryption-based}
	Privacy is mainly the protection of personal information of users and devices in smart M-IoT. Disclosure of information can be protected through encryption of data which prohibits any eavesdropper from obtaining any knowledge even if he or she is able to capture the majority of its parts. Encryption-based schemes are not different from usual encryption algorithms. Thus, the existing solutions can be classified into traditional encryption schemes on the basis of algorithm or mechanism used by them for protecting the data. These types are as follows:\\\\
	\begin{itemize}
		\item Symmetric encryption: The symmetric key encryption relies on the same key for encryption and decryption i.e. the key used for the encryption and the decryption should be same at both the parties. Symmetric-key encryption is essentially the same as a secret code that each of the two entities must know in order to encrypt and decrypt information. The symmetric key encryption has the major problem of exchange overheads of keys between the two parties, especially with maintaining trust when encryption is used for authentication and integrity checking~\cite{perez2017towards}.
		\item Asymmetric encryption: Asymmetrical encryption is also known as public key cryptography, uses two keys to encrypt or decrypt of a plain text. The secret keys are exchanged over the Internet or a large network. The message encrypted by a public key can only be decrypted using a private key and similarly, data encrypted using a private key can only be decrypted using a public key~\cite{jayaraman2017privacy,belguith2018phoabe,bamasag2015lightweight,hu2011identity,doukas2012enabling,wang2014performance,li2014p3,bao2016lightweight}. Asymmetric encryption is far better in ensuring the security of information transmitted during communication.
		\item Homomorphic encryptions: Homomorphic encryptions allow complex mathematical operations to be performed on encrypted data without compromising the encryption. The encrypted data set is transformed into another data set by preserving relationships between elements in both sets. Studies conducted on the topic of Homomorphic encryption in~\cite{bao2015new,gong2015medical,lu2017lightweight} highlight their applicability over the smart M-IoT.\\
	\end{itemize}
	
	\subsection{Architecture-based}
	Privacy preservation schemes can also be marked on the basis of architecture used for deployment and operations. Generally, the existing solutions depend on a centralized mechanism, but with solutions like blockchain which primarily uses public key operations, the architectural deployments become distributed.
	
	\onecolumn
	\begin{landscape}
		\begin{center}
			\fontsize{7}{10}\selectfont
			\setlength\LTleft{30pt}            
			\setlength\LTright{0pt}
			\begin{longtable}{@{\extracolsep{\fill}}*{10}{c}}
				\caption{State-of-the-art approaches for data privacy in M-IoT.}\label{Table3} \\
				\hline
				\multicolumn{1}{p{1.5cm}}{\centering \textbf{Approach}}
				&\multicolumn{1}{p{2.5cm}}{\centering \textbf{Author} \\ \textbf{(Year)} }
				&\multicolumn{1}{p{2.5cm}}{\centering \textbf{Ideology}}
				&\multicolumn{1}{p{1.2cm}}{\centering \textbf{Application}\\ \textbf{Area} }
				&\multicolumn{1}{p{1cm}}{\centering \textbf{Encryption}}
				&\multicolumn{1}{p{1.2cm}}{\centering \textbf{End to End}\\ \textbf{Security}}
				&\multicolumn{1}{p{1cm}}{\centering \textbf{Perfect Forward}\\ \textbf{Secrecy}}
				&\multicolumn{1}{p{1.2cm}}{\centering \textbf{Persistence against}\\ \textbf{Replay attack}}
				&\multicolumn{1}{p{1.2cm}}{\centering \textbf{Password Protection}}
				&\multicolumn{1}{p{1.2cm}}{\centering \textbf{Trust-Assessment}} \\[6pt]\\
				\hline \\
				
				\endfirsthead
				
				\multicolumn{10}{c}%
				{{\bfseries \tablename\ \thetable{} -- continued from previous page}} \\
				\hline
				
				\multicolumn{1}{p{1.5cm}}{\centering \textbf{Approach}}
				&\multicolumn{1}{p{2.5cm}}{\centering \textbf{Author} \\ \textbf{(Year)} }
				&\multicolumn{1}{p{2.5cm}}{\centering \textbf{Ideology}}
				&\multicolumn{1}{p{1.2cm}}{\centering \textbf{Application}\\ \textbf{Area} }
				&\multicolumn{1}{p{1cm}}{\centering \textbf{Encryption}}
				&\multicolumn{1}{p{1.2cm}}{\centering \textbf{End to End}\\ \textbf{Security}}
				&\multicolumn{1}{p{1cm}}{\centering \textbf{Perfect Forward}\\ \textbf{Secrecy}}
				&\multicolumn{1}{p{1.2cm}}{\centering \textbf{Persistence against}\\ \textbf{Replay attack}}
				&\multicolumn{1}{p{1.2cm}}{\centering \textbf{Password Protection}}
				&\multicolumn{1}{p{1.2cm}}{\centering \textbf{Trust-Assessment}} \\[6pt]\\
				\hline\\\endhead
				
				\hline \multicolumn{10}{l}{{Continued on next page}} \\
				
				\endfoot
				
				\endlastfoot
				
				\multicolumn{1}{p{1.5cm}}{\centering  End to end privacy}
				&\multicolumn{1}{p{2.5cm}}{\centering [Jayaraman \emph{et al.} 2017]\cite{jayaraman2017privacy}}
				&\multicolumn{1}{p{2.5cm}}{\centering Privacy preserving IoT architecture}
				&\multicolumn{1}{p{1.2cm}}{\centering IoT}
				&\multicolumn{1}{p{1cm}}{\centering \cmark}
				&\multicolumn{1}{p{1.2cm}}{\centering \cmark}
				&\multicolumn{1}{p{1cm}}{\centering \xmark}
				&\multicolumn{1}{p{1.2cm}}{\centering \cmark}
				&\multicolumn{1}{p{1.2cm}}{\centering \xmark}
				&\multicolumn{1}{p{1.2cm}}{\centering \cmark} \\[6pt]\\
				
				\multicolumn{1}{p{1.5cm}}{\centering  Decentralized anonymous authentication}
				&\multicolumn{1}{p{2.5cm}}{\centering [Alcaide \emph{et al.} 2013]\cite{alcaide2013anonymous}}
				&\multicolumn{1}{p{2.5cm}}{\centering Privacy preserving protocol}
				&\multicolumn{1}{p{1.2cm}}{\centering IoT target-driven applications}
				&\multicolumn{1}{p{1cm}}{\centering \cmark}
				&\multicolumn{1}{p{1.2cm}}{\centering \xmark}
				&\multicolumn{1}{p{1cm}}{\centering \xmark}
				&\multicolumn{1}{p{1.2cm}}{\centering \cmark}
				&\multicolumn{1}{p{1.2cm}}{\centering \xmark}
				&\multicolumn{1}{p{1.2cm}}{\centering \cmark} \\[6pt]\\
				
				\multicolumn{1}{p{1.5cm}}{\centering Multi-authority attribute based encryption}
				&\multicolumn{1}{p{2.5cm}}{\centering [Belguith \emph{et al.} 2018]\cite{belguith2018phoabe}}
				&\multicolumn{1}{p{2.5cm}}{\centering PHOABE (Policy-Hidden Outsourced ABE scheme)}
				&\multicolumn{1}{p{1.2cm}}{\centering Cloud-assisted IoT}
				&\multicolumn{1}{p{1cm}}{\centering \cmark}
				&\multicolumn{1}{p{1.2cm}}{\centering -}
				&\multicolumn{1}{p{1cm}}{\centering \cmark}
				&\multicolumn{1}{p{1.2cm}}{\centering \cmark}
				&\multicolumn{1}{p{1.2cm}}{\centering \xmark}
				&\multicolumn{1}{p{1.2cm}}{\centering \cmark} \\[6pt]\\
				
				\multicolumn{1}{p{1.5cm}}{\centering Privacy and integrity preservation}
				&\multicolumn{1}{p{2.5cm}}{\centering [Bamasag 2015]\cite{bamasag2015lightweight}}
				&\multicolumn{1}{p{2.5cm}}{\centering ID-based signcryption scheme}
				&\multicolumn{1}{p{1.2cm}}{\centering Smart grid}
				&\multicolumn{1}{p{1cm}}{\centering \cmark}
				&\multicolumn{1}{p{1.2cm}}{\centering \xmark}
				&\multicolumn{1}{p{1cm}}{\centering -}
				&\multicolumn{1}{p{1.2cm}}{\centering \cmark}
				&\multicolumn{1}{p{1.2cm}}{\centering \xmark}
				&\multicolumn{1}{p{1.2cm}}{\centering \cmark} \\[6pt]\\
				
				\multicolumn{1}{p{1.5cm}}{\centering Identity-based personal location system}
				&\multicolumn{1}{p{2.5cm}}{\centering [Hu \emph{et al.} 2011]\cite{hu2011identity}}
				&\multicolumn{1}{p{2.5cm}}{\centering Identity-based privacy preservation}
				&\multicolumn{1}{p{1.2cm}}{\centering IoT}
				&\multicolumn{1}{p{1cm}}{\centering \cmark}
				&\multicolumn{1}{p{1.2cm}}{\centering -}
				&\multicolumn{1}{p{1cm}}{\centering \xmark}
				&\multicolumn{1}{p{1.2cm}}{\centering \xmark}
				&\multicolumn{1}{p{1.2cm}}{\centering \xmark}
				&\multicolumn{1}{p{1.2cm}}{\centering \cmark} \\[6pt]\\
				
				\multicolumn{1}{p{1.5cm}}{\centering Data protection mechanism}
				&\multicolumn{1}{p{2.5cm}}{\centering [Doukas \emph{et al.} 2012]\cite{doukas2012enabling}}
				&\multicolumn{1}{p{2.5cm}}{\centering PKI encryption}
				&\multicolumn{1}{p{1.2cm}}{\centering IoT m-Health devices}
				&\multicolumn{1}{p{1cm}}{\centering \cmark}
				&\multicolumn{1}{p{1.2cm}}{\centering \cmark}
				&\multicolumn{1}{p{1cm}}{\centering \xmark}
				&\multicolumn{1}{p{1.2cm}}{\centering \xmark}
				&\multicolumn{1}{p{1.2cm}}{\centering \xmark}
				&\multicolumn{1}{p{1.2cm}}{\centering \cmark} \\[6pt]\\
				
				\multicolumn{1}{p{1.5cm}}{\centering Privacy management mechanism}
				&\multicolumn{1}{p{2.5cm}}{\centering [Evans \emph{et al.} 2012]\cite{evans2012efficient}}
				&\multicolumn{1}{p{2.5cm}}{\centering Efficient data tagging}
				&\multicolumn{1}{p{1.2cm}}{\centering IoT}
				&\multicolumn{1}{p{1cm}}{\centering \xmark}
				&\multicolumn{1}{p{1.2cm}}{\centering -}
				&\multicolumn{1}{p{1cm}}{\centering \xmark}
				&\multicolumn{1}{p{1.2cm}}{\centering -}
				&\multicolumn{1}{p{1.2cm}}{\centering \xmark}
				&\multicolumn{1}{p{1.2cm}}{\centering \cmark} \\[6pt]\\
				
				\multicolumn{1}{p{1.5cm}}{\centering Attribute-based encryption}
				&\multicolumn{1}{p{2.5cm}}{\centering [Wang \emph{et al.} 2014]\cite{wang2014performance}}
				&\multicolumn{1}{p{2.5cm}}{\centering public key encryption}
				&\multicolumn{1}{p{1.2cm}}{\centering IoT}
				&\multicolumn{1}{p{1cm}}{\centering \cmark}
				&\multicolumn{1}{p{1.2cm}}{\centering -}
				&\multicolumn{1}{p{1cm}}{\centering -}
				&\multicolumn{1}{p{1.2cm}}{\centering -}
				&\multicolumn{1}{p{1.2cm}}{\centering \xmark}
				&\multicolumn{1}{p{1.2cm}}{\centering \cmark} \\[6pt]\\
				
				\multicolumn{1}{p{1.5cm}}{\centering Privacy preservation protocol}
				&\multicolumn{1}{p{2.5cm}}{\centering [Li \emph{et al.} 2014]\cite{li2014p3}}
				&\multicolumn{1}{p{2.5cm}}{\centering Attribute-based encryption (ABE) key management}
				&\multicolumn{1}{p{1.2cm}}{\centering Smart grid}
				&\multicolumn{1}{p{1cm}}{\centering \cmark}
				&\multicolumn{1}{p{1.2cm}}{\centering -}
				&\multicolumn{1}{p{1cm}}{\centering \xmark}
				&\multicolumn{1}{p{1.2cm}}{\centering -}
				&\multicolumn{1}{p{1.2cm}}{\centering \xmark}
				&\multicolumn{1}{p{1.2cm}}{\centering \cmark} \\[6pt]\\
				
				\multicolumn{1}{p{1.5cm}}{\centering Reference software architecture }
				&\multicolumn{1}{p{2.5cm}}{\centering [Addo \emph{et al.} 2014]\cite{addo2014reference}}
				&\multicolumn{1}{p{2.5cm}}{\centering Collaborative pervasive systems}
				&\multicolumn{1}{p{1.2cm}}{\centering Cloud enabled IoT applications}
				&\multicolumn{1}{p{1cm}}{\centering \cmark}
				&\multicolumn{1}{p{1.2cm}}{\centering -}
				&\multicolumn{1}{p{1cm}}{\centering \xmark}
				&\multicolumn{1}{p{1.2cm}}{\centering -}
				&\multicolumn{1}{p{1.2cm}}{\centering \xmark}
				&\multicolumn{1}{p{1.2cm}}{\centering \cmark} \\[6pt]\\
				
				\multicolumn{1}{p{1.5cm}}{\centering RERUM:Reliable, Resilient and secure IoT for smart city applications }
				&\multicolumn{1}{p{2.5cm}}{\centering [Pohls \emph{et al.} 2014]\cite{pohls2014rerum}}
				&\multicolumn{1}{p{2.5cm}}{\centering Smart object (SO) hardware prototypes}
				&\multicolumn{1}{p{1.2cm}}{\centering Smart city IoT}
				&\multicolumn{1}{p{1cm}}{\centering \cmark}
				&\multicolumn{1}{p{1.2cm}}{\centering \cmark}
				&\multicolumn{1}{p{1cm}}{\centering -}
				&\multicolumn{1}{p{1.2cm}}{\centering \cmark}
				&\multicolumn{1}{p{1.2cm}}{\centering \xmark}
				&\multicolumn{1}{p{1.2cm}}{\centering \cmark} \\[6pt]\\
				
				\multicolumn{1}{p{1.5cm}}{\centering Private data aggregation with fault tolerance (DPAFT)}
				&\multicolumn{1}{p{2.5cm}}{\centering [Bao and Lu 2015]\cite{bao2015new}}
				&\multicolumn{1}{p{2.5cm}}{\centering Boneh-Goh-Nissim cryptosystem}
				&\multicolumn{1}{p{1.2cm}}{\centering Smart grid}
				&\multicolumn{1}{p{1cm}}{\centering \cmark}
				&\multicolumn{1}{p{1.2cm}}{\centering -}
				&\multicolumn{1}{p{1cm}}{\centering -}
				&\multicolumn{1}{p{1.2cm}}{\centering \cmark}
				&\multicolumn{1}{p{1.2cm}}{\centering \xmark}
				&\multicolumn{1}{p{1.2cm}}{\centering \cmark} \\[6pt]\\
				
				\multicolumn{1}{p{1.5cm}}{\centering Network-level security and privacy}
				&\multicolumn{1}{p{2.5cm}}{\centering [Sivaraman \emph{et al.} 2015]\cite{sivaraman2015network}}
				&\multicolumn{1}{p{2.5cm}}{\centering SDN-based approach}
				&\multicolumn{1}{p{1.2cm}}{\centering Smart-home IoT}
				&\multicolumn{1}{p{1cm}}{\centering \cmark}
				&\multicolumn{1}{p{1.2cm}}{\centering -}
				&\multicolumn{1}{p{1cm}}{\centering -}
				&\multicolumn{1}{p{1.2cm}}{\centering \cmark}
				&\multicolumn{1}{p{1.2cm}}{\centering \xmark}
				&\multicolumn{1}{p{1.2cm}}{\centering \cmark} \\[6pt]\\
				
				\multicolumn{1}{p{1.5cm}}{\centering Privacy protection mechanism }
				&\multicolumn{1}{p{2.5cm}}{\centering [Gong \emph{et al.} 2015]\cite{gong2015medical}}
				&\multicolumn{1}{p{2.5cm}}{\centering Lightweight private homomorphism algorithm and encryption algorithm}
				&\multicolumn{1}{p{1.2cm}}{\centering Medical IoT }
				&\multicolumn{1}{p{1cm}}{\centering \cmark}
				&\multicolumn{1}{p{1.2cm}}{\centering -}
				&\multicolumn{1}{p{1cm}}{\centering -}
				&\multicolumn{1}{p{1.2cm}}{\centering \cmark}
				&\multicolumn{1}{p{1.2cm}}{\centering \xmark}
				&\multicolumn{1}{p{1.2cm}}{\centering \cmark} \\[6pt]\\
				
				\multicolumn{1}{p{1.5cm}}{\centering Data-centric security }
				&\multicolumn{1}{p{2.5cm}}{\centering [Wrona \emph{et al.} 2015]\cite{wrona2016data}}
				&\multicolumn{1}{p{2.5cm}}{\centering End to End security solution}
				&\multicolumn{1}{p{1.2cm}}{\centering Military applications}
				&\multicolumn{1}{p{1cm}}{\centering \cmark}
				&\multicolumn{1}{p{1.2cm}}{\centering \cmark}
				&\multicolumn{1}{p{1cm}}{\centering -}
				&\multicolumn{1}{p{1.2cm}}{\centering -}
				&\multicolumn{1}{p{1.2cm}}{\centering \xmark}
				&\multicolumn{1}{p{1.2cm}}{\centering \cmark} \\[6pt]\\
				
				\multicolumn{1}{p{1.5cm}}{\centering Lightweight privacy-preserving data aggregation (LPDA)}
				&\multicolumn{1}{p{2.5cm}}{\centering [Lu \emph{et al.} 2017]\cite{lu2017lightweight}}
				&\multicolumn{1}{p{2.5cm}}{\centering Homomorphic Paillier encryption, Chinese Remainder theorem}
				&\multicolumn{1}{p{1.2cm}}{\centering Hybrid IoTs}
				&\multicolumn{1}{p{1cm}}{\centering \cmark}
				&\multicolumn{1}{p{1.2cm}}{\centering -}
				&\multicolumn{1}{p{1cm}}{\centering -}
				&\multicolumn{1}{p{1.2cm}}{\centering \cmark}
				&\multicolumn{1}{p{1.2cm}}{\centering \xmark}
				&\multicolumn{1}{p{1.2cm}}{\centering \cmark} \\[6pt]\\
				
				\multicolumn{1}{p{1.5cm}}{\centering Lightweight data report scheme}
				&\multicolumn{1}{p{2.5cm}}{\centering [Bao and Chen 2016]\cite{bao2016lightweight}}
				&\multicolumn{1}{p{2.5cm}}{\centering Pseudonym identity-based privacy-preserving}
				&\multicolumn{1}{p{1.2cm}}{\centering Smart grid}
				&\multicolumn{1}{p{1cm}}{\centering \cmark}
				&\multicolumn{1}{p{1.2cm}}{\centering -}
				&\multicolumn{1}{p{1cm}}{\centering -}
				&\multicolumn{1}{p{1.2cm}}{\centering \cmark}
				&\multicolumn{1}{p{1.2cm}}{\centering \xmark}
				&\multicolumn{1}{p{1.2cm}}{\centering \cmark} \\[6pt]\\
				
				\multicolumn{1}{p{1.5cm}}{\centering Negotiation-based privacy preservation scheme}
				&\multicolumn{1}{p{2.5cm}}{\centering [Ukil \emph{et al.} 2012]\cite{ukil2012negotiation}}
				&\multicolumn{1}{p{2.5cm}}{\centering Data masking tool}
				&\multicolumn{1}{p{1.2cm}}{\centering IoT}
				&\multicolumn{1}{p{1cm}}{\centering \xmark}
				&\multicolumn{1}{p{1.2cm}}{\centering \xmark}
				&\multicolumn{1}{p{1cm}}{\centering \xmark}
				&\multicolumn{1}{p{1.2cm}}{\centering \cmark}
				&\multicolumn{1}{p{1.2cm}}{\centering \xmark}
				&\multicolumn{1}{p{1.2cm}}{\centering \cmark} \\[6pt]\\
				
				\multicolumn{1}{p{1.5cm}}{\centering CP-ABE Application}
				&\multicolumn{1}{p{2.5cm}}{\centering [Perez \emph{et al.} 2017]\cite{perez2017towards}}
				&\multicolumn{1}{p{2.5cm}}{\centering Symmetric key encryption techniques}
				&\multicolumn{1}{p{1.2cm}}{\centering IoT}
				&\multicolumn{1}{p{1cm}}{\centering \cmark}
				&\multicolumn{1}{p{1.2cm}}{\centering -}
				&\multicolumn{1}{p{1cm}}{\centering \cmark}
				&\multicolumn{1}{p{1.2cm}}{\centering \cmark}
				&\multicolumn{1}{p{1.2cm}}{\centering \xmark}
				&\multicolumn{1}{p{1.2cm}}{\centering \cmark} \\[6pt]\\
				\hline
			\end{longtable}
		\end{center}
		\twocolumn
	\end{landscape}
	
	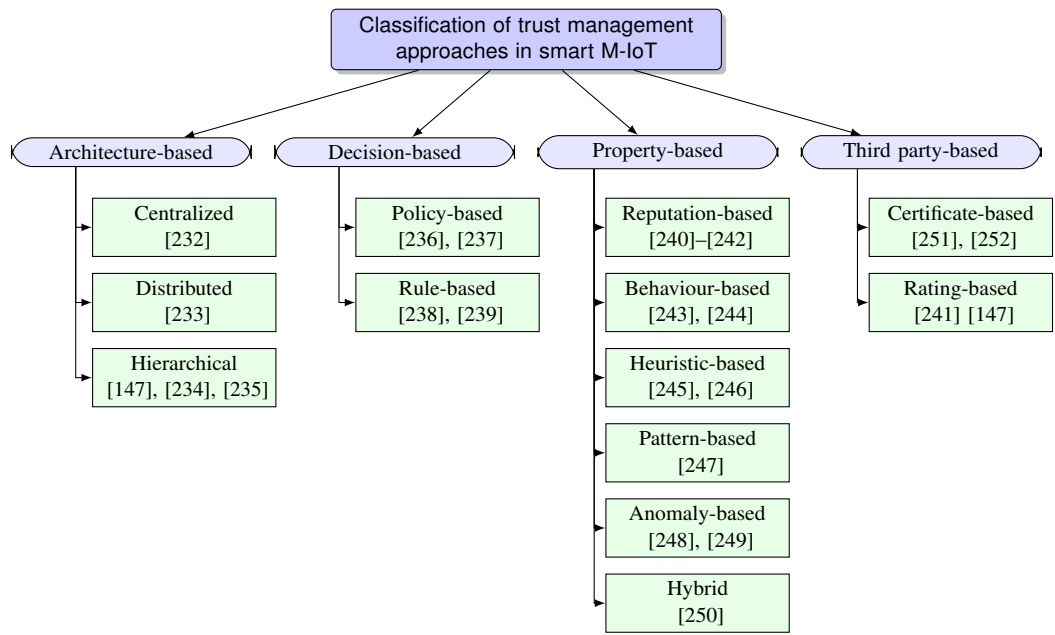
\begin{figure*}[!ht]
		\centering
		\fontsize{8}{10}\selectfont
		\begin{tikzpicture}[
		level 1/.style={sibling distance=35mm},
		edge from parent/.style={->,draw},
		>=latex]
		
		\node[root] {Classification of trust management approaches in smart M-IoT}
		child {node[level 2] (c1) {Architecture-based}}
		child {node[level 2] (c2) {Decision-based}}
		child {node[level 2] (c3) {Property-based}}
		child {node[level 2] (c4) {Third party-based}};
		
		\begin{scope}[every node/.style={level 3}]
		\node [below of = c1, xshift=20pt] (c11) {Centralized\\~\cite{alshehri2017centralized}};
		\node [below of = c11] (c12) {Distributed\\~\cite{chen2016trust}};
		\node [below of = c12] (c13) {Hierarchical\\~\cite{wu2017multi,guo2017mobile,sharma2017cooperative}};
		
		\node [below of = c2, xshift=20pt] (c21) {Policy-based\\~\cite{peshwe2017algorithm,ziegler2017anastacia}};
		\node [below of = c21] (c22) {Rule-based\\~\cite{chahal2017fuzzy,mahalle2013fuzzy}};
		
		\node [below of = c3, xshift=15pt] (c31) {Reputation-based\\~\cite{son2017adaptive,sharma2017computational,chen2011trm}};
		\node [below of = c31] (c32) {Behaviour-based\\~\cite{khan2017trusta,kravari2017ordain}};
		\node [below of = c32] (c33) {Heuristic-based\\~\cite{santos2013opcode,zhou2008application}};
		\node [below of = c33] (c34) {Pattern-based\\~\cite{niki2009drive}};
		\node [below of = c34] (c35) {Anomaly-based\\~\cite{gai2017multidimensional,wang2018trust}};
		\node [below of = c35] (c36) {Hybrid\\~\cite{ozcelik2017hybrid}};
		
		\node [below of = c4, xshift=15pt] (c41) {Certificate-based\\~\cite{hinarejos2018risklaine,obaidi2017persona}};
		\node [below of = c41] (c42) {Rating-based\\~\cite{sharma2017computational}~\cite{sharma2017cooperative}};
		
		\end{scope}
		
		\foreach \value in {1,2,3}
		\draw[->] (c1.195) |- (c1\value.west);
		
		\foreach \value in {1,2}
		\draw[->] (c2.195) |- (c2\value.west);
		
		\foreach \value in {1,...,6}
		\draw[->] (c3.195) |- (c3\value.west);
		
		\foreach \value in {1,2}
		\draw[->] (c4.195) |- (c4\value.west);
		\end{tikzpicture}
		\caption{A broad classification of trust-management schemes for smart M-IoT. The existing solutions can be studied by classifying them into architecture-based, property-based, decision-based, and IDS-based trust management in smart M-IoT.}\label{fig:tax_trust}
	\end{figure*}
	
	\begin{itemize}
		\item Centralized: Approaches which use a controller or centralized entity as a key enabler for privacy-preservation are studied in this type. Centralized solutions are effective from the monitoring perspective, but these pose a threat to a single point of failure which is difficult to sustain for any network~\cite{mascetti2009longitude}. Especially, in smart M-IoT, if all the traffic is regulated by the centralized authority, it becomes necessary to develop schemes which will define the policies of load management as well as prevent excessive utilization of resources for the traffic coming from a single source.
		\item Distributed: Such schemes depend on the distributed and flat nature of architectures and prevent a common point of failure as privacy preservation is initiated by the user or a node which are abstracted from other components of the network. In some scenarios, multiple nodes are used for defining policies for privacy preservation. However, the success of such approaches depends on their compliance degree and synergy in supporting common algorithms for a large set of nodes~\cite{alcaide2013anonymous}~\cite{addo2014reference}.
	\end{itemize}
	
	\subsection{Protocol-based}
	As discussed in the protocol section, privacy can be supported by defining rules which are operated as a part of conditions and help to decide on the sharing of information between the users. Protocol-based privacy is easier to achieve and an efficient way for networks that operate in close proximity to each other~\cite{rahman2017anonpri,alcaide2013anonymous}. Such schemes are extremely useful for networks using crowdsourcing and can be used as broadcast mechanisms for blockchain-based distributed solutions for privacy preservations.
	
	\subsection{Tool-based}
	Such an approach is easier to manage as it only involves process like masking, tagging or user-controlled policies~\cite{evans2012efficient,ukil2012negotiation,sharma2017cooperative}. Tool-based privacy is governed by the properties and services offered by the application platforms running for smart M-IoT. However, the correctness of such solutions is dependent on the legitimacy of the service providers and their honesty which cannot be measured through any tool and depends on the level of commitment to their users.
	
	\subsection{Summary and Insights}
	In this section, we summarized the privacy-preservation approaches for smart M-IoT on the basis of encryption, architecture, protocols, and tools. Data privacy is achievable through message protection and protocols can be used for authorizing applications and users before accessing personalized data of the smart M-IoT owner. Privacy can be attained through novel security protocols as well as positioning of AAA that can ensure the end to end data privacy.
	
	Policy-based, identity-based, ID-based, attribute-based encryptions and Public Key Infrastructure (PKI) can be the major enables for privacy preservation. Solutions, like blockchain and tangle (directed acyclic graph), can be used for preserving privacy through transactions between smart M-IoT users. The choice of encryption plays a key role as it affects the policies of session management between end to end devices based on the factors like freshness, integrity and perfect forward secrecy, which are attainable through secure key operations. More advances are expected in tool-based privacy preservation as well as personalized management as users are becoming much aware and demand personalized settings for each operation.
	\section{Trust Management Approaches for smart M-IoT}
	M-IoT aims at maintaining a secure relationship between the entities involved in service provisioning as well as data dissemination~\cite{namal2015autonomic}~\cite{nitti2012subjective}. Most of the trust-enabled networks establish a reputation system based on a centralized entity that can help to check whether a particular node in the network can be relied upon or not. Such evaluations of reliability are an integral part of trust management systems~\cite{mhetre2016trust}. With a billion of devices, the complexity of maintaining trust increases and it becomes relatively difficult to handle such an enormous number of devices, which leads the network into attacks by false reputation enhancement of an intruder~\cite{nitti2014trustworthiness}.
	
	Most of the trust management systems are governed by policies which are decided on the basis of the configurations of the network as well as the types of services supported by the M-IoT devices~\cite{lize2014trust}. Trust management depends heavily on the distributed computations as slow computations may cause excessive overheads which are a hazard for secure systems. Crowdsourcing, computational offloading, dividing of service accessibility, distributed policy formations, distributed trust maintenance,  and D2D computations, help in reducing the overheads and complexities associated with the building of trust-relaying systems for M-IoT~\cite{duan2014energy,chen2014trust,wang2013distributed,saied2013trust}. Trust-based solutions for smart M-IoT can be classified into following types, as shown in Fig.~\ref{fig:tax_trust}:
	
	\subsection{Architecture-based}
	Trust in smart M-IoT is attainable through a unique implementation of architecture while placing each entity in such a way that it provides a pathway for believing each other before communications. On the basis of architectural setup, trust management approaches can be classified into following three types:\\\\
	\begin{itemize}
		\item Centralized: It constitutes an entity which is present at the center of a given network and is responsible for handling trust computations for the entire network~\cite{alshehri2017centralized}. The problem with such a deployment is the risk of single point of failure.
		\item Distributed: It constitutes trust evaluation through distributed entities which prevent a single point of failure. Distributed trusts are usually operated as P2P or P2MP, but not peer to all~\cite{chen2016trust}.
		\item Hierarchical: It constitutes calculations by using a layered architecture which focuses on evaluating trust for entities on each layer~\cite{wu2017multi,guo2017mobile,sharma2017cooperative}. This allows selection of accurate nodes in the next order of hierarchy.
	\end{itemize}
	\subsection{Decision-based}
	Trust is a decision-based entity, which in some cases is marked by following certain principles of communications. Node management and selection of next hop are two of the examples of decision-based trust management. On the basis of ideology, decision-based trust management can be categorized into following two types:\\
	\begin{itemize}
		\item Policy-based: Using conditions to take a decision on the situation of entities is treated as a policy-based solution. The policy-based approach often results in a centralized or hierarchical solution as a governing body is required to form the policies for evaluating trust of the involved entities in smart M-IoT~\cite{peshwe2017algorithm,ziegler2017anastacia}.
		\item Rule-based: Using conditions to evaluate given information for generating relevant knowledge regarding the trust of an entity is treated as a rule-based solution. The rule-based approach utilizes any type of architecture; however, it always has dominance for deciding rules or a consensus model for arriving at a common decision while formulating principles of trust evaluations~\cite{chahal2017fuzzy,mahalle2013fuzzy}.
	\end{itemize}
	
	\subsection{Property-based}
	Trust is itself a property of a device in smart M-IoT. However, this core property can be classified into sub-categories through which trust can be ensured in any type of network as explained below:\\
	\begin{itemize}
		\item Reputation-based: Reputation is a fundamental concept in several situations which can be involved in the interaction between mutually distrusting parties~\cite{son2017adaptive,sharma2017computational}. Reputation-based trust relies on a ``soft computational" approach to formulate the problem of trust. The trust systems rely on the basic idea of analyses and combination of paths and networks of trust relationships. Trust and reputation systems play a significant role in decision support for Internet-mediated service provisioning. Reputation-based trust management helps to mitigate the security complications of smart M-IoT~\cite{chen2011trm}.
		\item Behaviour-based: Behaviour-based trust models include a fixed evaluation scheme. The scheme uses the knowledge of behaviour in previous interactions and derives the trustworthiness of an entity~\cite{khan2017trusta,kravari2017ordain}. The behaviour-specific knowledge can be obtained from the feedbacks and recommendations.
		\item Heuristics based: Heuristics are used to aid the decision or estimation process by evaluating indirect trust of an agent into the direct trust estimation. The decision formulation is handled with the estimation through metrics~\cite{santos2013opcode,zhou2008application}.
		\item Pattern-based: A set of design patterns are used for designing systems with the explicit intention of increasing trust between entities. The behavioural patterns are followed to achieve a sustainable trust. Patterns are used to solve recurring problems in trust-based communications for smart M-IoT. Patterns have been developed in a range of disciplines for a variety of domains to make a trust model. The patterns can be obtained by behaviour, rules, policy, flow, etc~\cite{niki2009drive}.
		\item Anomaly detection based: The anomalies are abnormal behavior which is intended to affect the systems. Anomalies can be detected based on their own signatures and settings. The rules and threat modeling can be done with the help of system behaviors and signatures. Anomalies are inspected over the high malicious network traffic to improve the detection accuracy of trust model~\cite{gai2017multidimensional,wang2018trust}. Signature-based IDS are the well-known anomaly detection systems in smart M-IoT networks.
		
		\onecolumn
		\begin{landscape}
			\begin{center}
				\fontsize{7}{10}\selectfont
				\setlength\LTleft{30pt}            
				\setlength\LTright{0pt}
				\begin{longtable}{@{\extracolsep{\fill}}*{9}{c}}
					\caption{State-of-the-art approaches applicable for trust management in M-IoT.}\label{Table7} \\
					\hline
					\multicolumn{1}{p{1.5cm}}{\centering \textbf{Approach}}
					&\multicolumn{1}{p{1.5cm}}{\centering \textbf{Author} \\ \textbf{(Year)} }
					&\multicolumn{1}{p{2.5cm}}{\centering \textbf{Ideology}}
					&\multicolumn{1}{p{1.2cm}}{\centering \textbf{Application}\\ \textbf{Area} }
					&\multicolumn{1}{p{2.5cm}}{\centering \textbf{Parameters}\\\textbf{focused}}
					&\multicolumn{1}{p{1.2cm}}{\centering \textbf{Computational}\\ \textbf{Offloading}}
					&\multicolumn{1}{p{1cm}}{\centering \textbf{Visualization}}
					&\multicolumn{1}{p{1.2cm}}{\centering \textbf{Reliability}}
					&\multicolumn{1}{p{1.2cm}}{\centering \textbf{Security}\\\textbf{Constraints}} \\[6pt]\\
					\hline \\
					
					\endfirsthead
					
					\multicolumn{9}{c}%
					{{\bfseries \tablename\ \thetable{} -- continued from previous page}} \\
					\hline
					
					\multicolumn{1}{p{1.5cm}}{\centering \textbf{Approach}}
					&\multicolumn{1}{p{1.5cm}}{\centering \textbf{Author} \\ \textbf{(Year)} }
					&\multicolumn{1}{p{2.5cm}}{\centering \textbf{Ideology}}
					&\multicolumn{1}{p{1.2cm}}{\centering \textbf{Application}\\ \textbf{Area} }
					&\multicolumn{1}{p{2.5cm}}{\centering \textbf{Parameters}\\\textbf{focused}}
					&\multicolumn{1}{p{1.2cm}}{\centering \textbf{Computational}\\ \textbf{Offloading}}
					&\multicolumn{1}{p{1cm}}{\centering \textbf{Visualization}}
					&\multicolumn{1}{p{1.2cm}}{\centering \textbf{Reliability}}
					&\multicolumn{1}{p{1.2cm}}{\centering \textbf{Security}\\\textbf{Constraints}} \\[6pt]\\
					\hline\\\endhead
					
					\hline \multicolumn{9}{l}{{Continued on next page}} \\
					
					\endfoot
					
					\endlastfoot
					
					\multicolumn{1}{p{1.5cm}}{\centering ORDAIN}
					&\multicolumn{1}{p{1.5cm}}{\centering [Kravari and Bassiliades 2017]\cite{kravari2017ordain}}
					&\multicolumn{1}{p{2.5cm}}{\centering Identification of social and non-social metrics}
					&\multicolumn{1}{p{1.2cm}}{\centering IoT}
					&\multicolumn{1}{p{2.5cm}}{\centering Social and Non-social features}
					&\multicolumn{1}{p{1.2cm}}{\centering -}
					&\multicolumn{1}{p{1cm}}{\centering \xmark}
					&\multicolumn{1}{p{1.2cm}}{\centering \xmark}
					&\multicolumn{1}{p{1.2cm}}{\centering \xmark } \\[6pt]\\
					
					\multicolumn{1}{p{1.5cm}}{\centering Multi-domain trust management}
					&\multicolumn{1}{p{1.5cm}}{\centering [Wu and Li 2017]\cite{wu2017multi}}
					&\multicolumn{1}{p{2.5cm}}{\centering Hierarchical trust management framework}
					&\multicolumn{1}{p{1.2cm}}{\centering RFID}
					&\multicolumn{1}{p{2.5cm}}{\centering Convergence speed, Malicious event detection rate, Mobility }
					&\multicolumn{1}{p{1.2cm}}{\centering -}
					&\multicolumn{1}{p{1cm}}{\centering \xmark}
					&\multicolumn{1}{p{1.2cm}}{\centering -}
					&\multicolumn{1}{p{1.2cm}}{\centering \cmark } \\[6pt]\\
					
					\multicolumn{1}{p{1.5cm}}{\centering Multidimensional trust-based anomaly detection}
					&\multicolumn{1}{p{1.5cm}}{\centering [Gai \emph{et al.} 2017]\cite{gai2017multidimensional}}
					&\multicolumn{1}{p{2.5cm}}{\centering QoS and social relationship}
					&\multicolumn{1}{p{1.2cm}}{\centering IoT}
					&\multicolumn{1}{p{2.5cm}}{\centering Trust level evolvement, False alarm rate, Malicious nodes percentage}
					&\multicolumn{1}{p{1.2cm}}{\centering -}
					&\multicolumn{1}{p{1cm}}{\centering \xmark}
					&\multicolumn{1}{p{1.2cm}}{\centering \cmark}
					&\multicolumn{1}{p{1.2cm}}{\centering \cmark } \\[6pt]\\
					
					\multicolumn{1}{p{1.5cm}}{\centering CTM-IoT}
					&\multicolumn{1}{p{1.5cm}}{\centering [Alshehri and Hussain 2017]\cite{alshehri2017centralized}}
					&\multicolumn{1}{p{2.5cm}}{\centering Centralized trust management}
					&\multicolumn{1}{p{1.2cm}}{\centering IoT}
					&\multicolumn{1}{p{2.5cm}}{\centering Trust management module, Communication module}
					&\multicolumn{1}{p{1.2cm}}{\centering -}
					&\multicolumn{1}{p{1cm}}{\centering \xmark}
					&\multicolumn{1}{p{1.2cm}}{\centering -}
					&\multicolumn{1}{p{1.2cm}}{\centering -} \\[6pt]\\
					
					\multicolumn{1}{p{1.5cm}}{\centering Resilient routing mechanism}
					&\multicolumn{1}{p{1.5cm}}{\centering [Khan \emph{et al.} 2017]\cite{khan2017trusta}}
					&\multicolumn{1}{p{2.5cm}}{\centering Routing protocol for low power and lossy networks}
					&\multicolumn{1}{p{1.2cm}}{\centering IoT}
					&\multicolumn{1}{p{2.5cm}}{\centering Delivery ratio, Path length, Bad paths, Network reliability}
					&\multicolumn{1}{p{1.2cm}}{\centering -}
					&\multicolumn{1}{p{1cm}}{\centering \xmark}
					&\multicolumn{1}{p{1.2cm}}{\centering \cmark}
					&\multicolumn{1}{p{1.2cm}}{\centering \cmark} \\[6pt]\\
					
					\multicolumn{1}{p{1.5cm}}{\centering Trust-based policy hidden communication}
					&\multicolumn{1}{p{1.5cm}}{\centering [Peshwe and Das 2017]\cite{peshwe2017algorithm}}
					&\multicolumn{1}{p{2.5cm}}{\centering SIGMA-I - policy hiding prefix based encryption}
					&\multicolumn{1}{p{1.2cm}}{\centering IoT}
					&\multicolumn{1}{p{2.5cm}}{\centering Private service discovery, SIGMA-I, RSSI}
					&\multicolumn{1}{p{1.2cm}}{\centering -}
					&\multicolumn{1}{p{1cm}}{\centering \xmark}
					&\multicolumn{1}{p{1.2cm}}{\centering -}
					&\multicolumn{1}{p{1.2cm}}{\centering \cmark} \\[6pt]\\
					
					\multicolumn{1}{p{1.5cm}}{\centering Timely trust establishment}
					&\multicolumn{1}{p{1.5cm}}{\centering [Yusof \emph{et al.} 2017]\cite{yusof2017timely}}
					&\multicolumn{1}{p{2.5cm}}{\centering Swift trust formation}
					&\multicolumn{1}{p{1.2cm}}{\centering IoT}
					&\multicolumn{1}{p{2.5cm}}{\centering Swift trust, Global virtual teams}
					&\multicolumn{1}{p{1.2cm}}{\centering -}
					&\multicolumn{1}{p{1cm}}{\centering \xmark}
					&\multicolumn{1}{p{1.2cm}}{\centering \cmark}
					&\multicolumn{1}{p{1.2cm}}{\centering \cmark} \\[6pt]\\
					
					\multicolumn{1}{p{1.5cm}}{\centering Hybrid trust-based IDS}
					&\multicolumn{1}{p{1.5cm}}{\centering [Ozcelik \emph{et al.} 2017]\cite{ozcelik2017hybrid}}
					&\multicolumn{1}{p{2.5cm}}{\centering Functional reputation and misuse}
					&\multicolumn{1}{p{1.2cm}}{\centering WSNs}
					&\multicolumn{1}{p{2.5cm}}{\centering Energy consumption, Network lifetime}
					&\multicolumn{1}{p{1.2cm}}{\centering \xmark}
					&\multicolumn{1}{p{1cm}}{\centering \cmark}
					&\multicolumn{1}{p{1.2cm}}{\centering \cmark}
					&\multicolumn{1}{p{1.2cm}}{\centering \cmark} \\[6pt]\\
					
					\multicolumn{1}{p{1.5cm}}{\centering ANASTACIA}
					&\multicolumn{1}{p{1.5cm}}{\centering [Ziegler \emph{et al.} 2017]\cite{ziegler2017anastacia}}
					&\multicolumn{1}{p{2.5cm}}{\centering Trustworthy-by-design autonomic}
					&\multicolumn{1}{p{1.2cm}}{\centering CPS-IoT }
					&\multicolumn{1}{p{2.5cm}}{\centering Policy-based access control, Smart security planning}
					&\multicolumn{1}{p{1.2cm}}{\centering -}
					&\multicolumn{1}{p{1cm}}{\centering \cmark}
					&\multicolumn{1}{p{1.2cm}}{\centering \cmark}
					&\multicolumn{1}{p{1.2cm}}{\centering \cmark} \\[6pt]\\
					
					\multicolumn{1}{p{1.5cm}}{\centering Trust-based decision making}
					&\multicolumn{1}{p{1.5cm}}{\centering [Al-Hamadi and Chen 2017]\cite{al2017trust}}
					&\multicolumn{1}{p{2.5cm}}{\centering Trust-based information sharing}
					&\multicolumn{1}{p{1.2cm}}{\centering Health-IoT }
					&\multicolumn{1}{p{2.5cm}}{\centering Correct decision ratio (CDR), Malicious nodes}
					&\multicolumn{1}{p{1.2cm}}{\centering \cmark}
					&\multicolumn{1}{p{1cm}}{\centering -}
					&\multicolumn{1}{p{1.2cm}}{\centering \cmark}
					&\multicolumn{1}{p{1.2cm}}{\centering \cmark} \\[6pt]\\
					
					\multicolumn{1}{p{1.5cm}}{\centering TMCoI-SIoT}
					&\multicolumn{1}{p{1.5cm}}{\centering [Abderrahim \emph{et al.} 2017]\cite{abderrahim2017tmcoi}}
					&\multicolumn{1}{p{2.5cm}}{\centering Trust management in SIoT}
					&\multicolumn{1}{p{1.2cm}}{\centering SIoT }
					&\multicolumn{1}{p{2.5cm}}{\centering Trust evaluation, Trust prediction}
					&\multicolumn{1}{p{1.2cm}}{\centering -}
					&\multicolumn{1}{p{1cm}}{\centering -}
					&\multicolumn{1}{p{1.2cm}}{\centering \cmark}
					&\multicolumn{1}{p{1.2cm}}{\centering \cmark} \\[6pt]\\
					
					\multicolumn{1}{p{1.5cm}}{\centering Trust estimation scheme}
					&\multicolumn{1}{p{1.5cm}}{\centering [Son \emph{et al.} 2017]\cite{son2017adaptive}}
					&\multicolumn{1}{p{2.5cm}}{\centering Interaction history and stereotypical reputation}
					&\multicolumn{1}{p{1.2cm}}{\centering IoT }
					&\multicolumn{1}{p{2.5cm}}{\centering Trust value}
					&\multicolumn{1}{p{1.2cm}}{\centering -}
					&\multicolumn{1}{p{1cm}}{\centering \xmark}
					&\multicolumn{1}{p{1.2cm}}{\centering -}
					&\multicolumn{1}{p{1.2cm}}{\centering \cmark} \\[6pt]\\
					
					\multicolumn{1}{p{1.5cm}}{\centering Hierarchical trust management}
					&\multicolumn{1}{p{1.5cm}}{\centering [Guo and Chen 2017]\cite{guo2017mobile}}
					&\multicolumn{1}{p{2.5cm}}{\centering Hierarchical trust management in mobile cloud}
					&\multicolumn{1}{p{1.2cm}}{\centering IoT }
					&\multicolumn{1}{p{2.5cm}}{\centering Trust value}
					&\multicolumn{1}{p{1.2cm}}{\centering \cmark}
					&\multicolumn{1}{p{1cm}}{\centering \xmark}
					&\multicolumn{1}{p{1.2cm}}{\centering \cmark}
					&\multicolumn{1}{p{1.2cm}}{\centering \cmark} \\[6pt]\\
					
					\multicolumn{1}{p{1.5cm}}{\centering Trust-based distributed intrusion detection}
					&\multicolumn{1}{p{1.5cm}}{\centering [Khan and Herrmann 2017]\cite{khan2017trust}}
					&\multicolumn{1}{p{2.5cm}}{\centering Distributed IDS}
					&\multicolumn{1}{p{1.2cm}}{\centering IoT }
					&\multicolumn{1}{p{2.5cm}}{\centering False positives, False negatives}
					&\multicolumn{1}{p{1.2cm}}{\centering \xmark}
					&\multicolumn{1}{p{1cm}}{\centering \xmark}
					&\multicolumn{1}{p{1.2cm}}{\centering \xmark}
					&\multicolumn{1}{p{1.2cm}}{\centering -} \\[6pt]\\
					
					\multicolumn{1}{p{1.5cm}}{\centering Computational offloading for efficient trust management}
					&\multicolumn{1}{p{1.5cm}}{\centering [Sharma \emph{et al.} 2017]\cite{sharma2017computational}}
					&\multicolumn{1}{p{2.5cm}}{\centering Osmotic computing}
					&\multicolumn{1}{p{1.2cm}}{\centering POSNs}
					&\multicolumn{1}{p{2.5cm}}{\centering Trust visualization, Monitoring cost, Average osmosis time}
					&\multicolumn{1}{p{1.2cm}}{\centering \cmark}
					&\multicolumn{1}{p{1cm}}{\centering \cmark}
					&\multicolumn{1}{p{1.2cm}}{\centering \cmark}
					&\multicolumn{1}{p{1.2cm}}{\centering \xmark} \\[6pt]\\
					
					\multicolumn{1}{p{1.5cm}}{\centering Trust management via SOA}
					&\multicolumn{1}{p{1.5cm}}{\centering [Chen \emph{et al.} 2016]\cite{chen2016trust}}
					&\multicolumn{1}{p{2.5cm}}{\centering Distributed collaborative filtering}
					&\multicolumn{1}{p{1.2cm}}{\centering SOA-based IoT}
					&\multicolumn{1}{p{2.5cm}}{\centering Trust value, Decay parameter, Trust convergence}
					&\multicolumn{1}{p{1.2cm}}{\centering \xmark}
					&\multicolumn{1}{p{1cm}}{\centering -}
					&\multicolumn{1}{p{1.2cm}}{\centering -}
					&\multicolumn{1}{p{1.2cm}}{\centering -} \\[6pt]\\
					
					\multicolumn{1}{p{1.5cm}}{\centering Trust-based access control}
					&\multicolumn{1}{p{1.5cm}}{\centering [Mahalle \emph{et al.} 2013]\cite{mahalle2013fuzzy}}
					&\multicolumn{1}{p{2.5cm}}{\centering Fuzzy-based approach}
					&\multicolumn{1}{p{1.2cm}}{\centering IoT}
					&\multicolumn{1}{p{2.5cm}}{\centering Energy consumption, Residual energy}
					&\multicolumn{1}{p{1.2cm}}{\centering \xmark}
					&\multicolumn{1}{p{1cm}}{\centering \xmark}
					&\multicolumn{1}{p{1.2cm}}{\centering -}
					&\multicolumn{1}{p{1.2cm}}{\centering -} \\[6pt]\\
					
					\multicolumn{1}{p{1.5cm}}{\centering TRM-IoT}
					&\multicolumn{1}{p{1.5cm}}{\centering [Chen \emph{et al.} 2011]\cite{chen2011trm}}
					&\multicolumn{1}{p{2.5cm}}{\centering Fuzzy-reputation}
					&\multicolumn{1}{p{1.2cm}}{\centering IoT}
					&\multicolumn{1}{p{2.5cm}}{\centering End-to-end packet forwarding ratio (EPFR), Energy consumption, Convergence speed, Detection probability}
					&\multicolumn{1}{p{1.2cm}}{\centering \xmark}
					&\multicolumn{1}{p{1cm}}{\centering \xmark}
					&\multicolumn{1}{p{1.2cm}}{\centering \xmark}
					&\multicolumn{1}{p{1.2cm}}{\centering \xmark} \\[6pt]\\
					
					\multicolumn{1}{p{1.5cm}}{\centering Cooperative spectrum sensing data fusion}
					&\multicolumn{1}{p{1.5cm}}{\centering [Wang \emph{et al.} 2018]\cite{wang2018trust}}
					&\multicolumn{1}{p{2.5cm}}{\centering Mechanism design theory}
					&\multicolumn{1}{p{1.2cm}}{\centering Cognitive radio networks}
					&\multicolumn{1}{p{2.5cm}}{\centering Malicious nodes percentage, Decision rate, Trust threshold}
					&\multicolumn{1}{p{1.2cm}}{\centering \xmark}
					&\multicolumn{1}{p{1cm}}{\centering \xmark}
					&\multicolumn{1}{p{1.2cm}}{\centering \xmark}
					&\multicolumn{1}{p{1.2cm}}{\centering \xmark} \\[6pt]\\
					
					\multicolumn{1}{p{1.5cm}}{\centering Cooperative trust relaying and privacy preservation}
					&\multicolumn{1}{p{1.5cm}}{\centering [Sharma \emph{et al.} 2017]\cite{sharma2017cooperative}}
					&\multicolumn{1}{p{2.5cm}}{\centering Edge-crowdsourcing via fission computing}
					&\multicolumn{1}{p{1.2cm}}{\centering SIoT}
					&\multicolumn{1}{p{2.5cm}}{\centering Fission time, Combined entropy, Integration cost, Per node relaying time}
					&\multicolumn{1}{p{1.2cm}}{\centering \cmark}
					&\multicolumn{1}{p{1cm}}{\centering \cmark}
					&\multicolumn{1}{p{1.2cm}}{\centering \xmark}
					&\multicolumn{1}{p{1.2cm}}{\centering \xmark} \\[6pt]\\

					\hline
				\end{longtable}
			\end{center}
			\twocolumn
		\end{landscape}
		
		\item Hybrid: Such a trust management system which combines all the above-discussed solutions as a single mechanism is a part of hybrid trust management in smart M-IoT. Hybrid approaches use all the existing property-based approaches and choose the one which suits best to the given conditions and configurations~\cite{ozcelik2017hybrid}.
	\end{itemize}
	\subsection{Third party-based}
	Depending on external mode for calculating trust is one of the prominent solutions of modern day networks. Such a solution uses mechanisms like deep learning, data analytic, neural networks or AI for evaluating the trust of communicating entities. Based on the outputs from third-party evaluations, there can be two main types:
	\begin{itemize}
		\item Certificate-based: Providing certificate of assurance on successful evaluation of required trust is easier and a comprehensive solution, which is also capable of providing a detailed report on the operations of a device~\cite{hinarejos2018risklaine,obaidi2017persona}. Third parties use certain policies, cookies, and cached entries to ensure trust while generating certificates for the required device in a smart M-IoT.
		\item Rating-based: In certain scenarios, third parties are involved in giving ranking or ratings to each individual involved in the formation of the network. Such an approach is termed as rating-based trust management. A threshold is marked on the basis of some predetermined score and each entity is evaluated against this threshold value~\cite{sharma2017computational}~\cite{sharma2017cooperative}.
	\end{itemize}
	\subsection{Summary and Insights}
	In this section, we provided a detailed classification of trust management approaches for smart M-IoT. Trust relationships not only secure the M-IoT but also help in building reliable CPS. Evaluation of trust by using a limited set of metrics is a challenge for M-IoT, however, such a system offers huge scalability and can be operated with less management and better control~\cite{chang2012survey}. Incorporation of software security, privacy control, and security constraints further strengthen the trust modeling in M-IoT. Along with these, trust-based solutions can be modeled into secure communication systems through security protocols, which use encryption policies for defining new security schemes by using a similar model of trust-relaying systems~\cite{bao2012trust}~\cite{yan2014survey}~\cite{bao2012dynamic}.
	
	To summarize, a detailed state-of-the-art comparative study on various trust management schemes is presented in Table~\ref{Table7}, which can be extended for their use in the smart M-IoT environment. The table helps to understand the key features and parameters focused by most of the existing solutions along with their core ideology for maintaining trust between the IoT entities.
	\begin{figure}
		\centering
		\fontsize{8}{10}\selectfont
		\begin{tikzpicture}[
		level 1/.style={sibling distance=35mm},
		edge from parent/.style={->,draw},
		>=latex]
		
		\node[root] {Classification of physical layer security approaches in smart M-IoT}
		child {node[level 2] (c1) {Service-based}}
		child {node[level 2] (c2) {Channel-based}};
		
		\begin{scope}[every node/.style={level 3}]
		\node [below of = c1, xshift=25pt] (c11) {Cryptographic\\~\cite{liu2010securing,zaman2017polarization}};
		\node [below of = c11,yshift=-10pt] (c12) {Access Control and Trasmissions\\~\cite{xu2016security,chen2016securing,zhang2016secure,hu2017secure,islam2017secured}};
		\node [below of = c12,yshift=-10pt] (c13) {Jamming\\~\cite{choi2017physical,hu2017cooperative,li2016worst}};
		
		\node [below of = c2, xshift=25pt] (c21) {Modulation-based\\~\cite{wei2016polarization}~\cite{gao2016physical}};
		\node [below of = c21] (c22) {Encoding-based\\~\cite{li2015compressed}~\cite{limmanee2010secure}};
		
		\end{scope}
		
		\foreach \value in {1,2,3}
		\draw[->] (c1.195) |- (c1\value.west);
		
		\foreach \value in {1,2}
		\draw[->] (c2.195) |- (c2\value.west);
		\end{tikzpicture}
		\caption{A broad classification of secure physical layer schemes for smart M-IoT. The existing solutions can be classified into service-based and channel based mechanisms.}\label{fig:tax_physical}
	\end{figure}
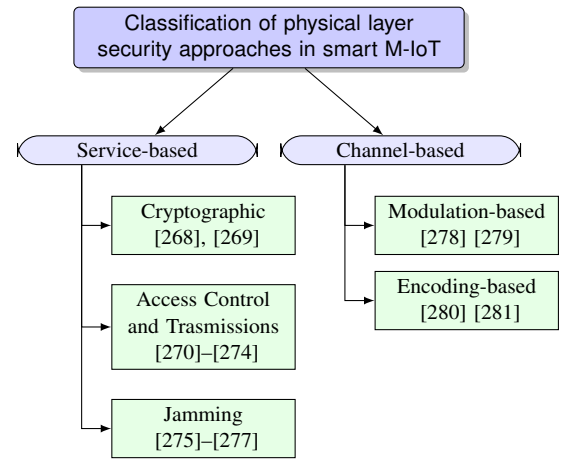
	
	\section{Physical Layer Security for Smart M-IoT} 
	Unlike traditional security solutions, which focus on the logical aspects of the networks, physical security is hardest and difficult to follow because of a difference in the type and make of an M-IoT device. With each device following a different set of parameters and configurations, it becomes difficult to provide a common solution which can withstand the Channel State Information (CSI) requirements of the entire network while securing the physical transmission of the network~\cite{trappe2015challenges}~\cite{mukherjee2015physical}~\cite{pecorella2016role}. Network coding and multiplexing approaches usually rely on cryptographic solutions only to reduce the complexity of physical layer; however, this makes the system vulnerable to different types of attacks that can be launched over the used mechanisms.  With devices being operated on battery, physical layer security becomes far more challenging and should be attained with lesser overheads as well as a lesser number of computations. A highly burdened operation may deplete the energy resources and an operational M-IoT network becomes of no use. The types of technology, 3G, 4G/LTE or upcoming 5G, play a crucial role in selecting an approach that can fit into the physical configurations as well as can support the load at a dedicated frame size~\cite{mukherjee2015physical}~\cite{zhang2017securing}~\cite{kitana2016impact}.
	
	Designing of security schemes on the physical layer may seem to be difficult, but it provides all set of new opportunities for improving the QoS as well as QoE for the end-users. The strength of the physical layer security depends on the adversary model which is used for evaluating the developed solution~\cite{trappe2015challenges}~\cite{zeng2015physical}. Such solutions are usually driven by the assumptions of the CSI as well as device type and may or may not stand once new vulnerabilities are discovered over a course of time~\cite{altolini2013low}~\cite{lee2014securing}~\cite{brilli2016physical}. The existing solutions can be broadly classified into two main types, service-based physical layer security, and channel-based physical layer security, as shown in Fig.~\ref{fig:tax_physical}. The details on both of these are presented below:
	
	\begin{figure*}[!ht]
		\centering
		\fontsize{8}{10}\selectfont
		\begin{tikzpicture}[
		level 1/.style={sibling distance=35mm},
		edge from parent/.style={->,draw},
		>=latex]
		
		\node[root] {Proactive or Reactive Handover Authentication}
		child {node[level 2] (c1) {Initiation-based}}
		child {node[level 2] (c2) {Architecture-based}}
		child {node[level 2] (c3) {Property-based}};
		
		\begin{scope}[every node/.style={level 3}]
		\node [below of = c1, xshift=20pt] (c11) {Host-initiated\\~\cite{lee2012host}};
		\node [below of = c11] (c12) {Network-initiated\\~\cite{lee2013distributed}};
		
		\node [below of = c2, xshift=20pt] (c21) {Centralized\\~\cite{ghahfarokhi2012context}};
		\node [below of = c21] (c22) {Distributed\\~\cite{lee2013distributed,sharma2018block}};
		
		\node [below of = c3, xshift=20pt] (c31) {IP-based\\~\cite{chai2017security,chai2015enhanced,shin2017secure,you2017spfp}};
		\node [below of = c31,yshift=-6pt] (c32) {Reliabilty-based\\~\cite{sharma2018secure,ndibanje2017secure}};
		\node [below of = c32,yshift=-6pt] (c33) {Encryption-based\\~\cite{ndibanje2017secure,saxena2016authentication,shin2017secure}};
		\node [below of = c33,yshift=-6pt] (c34) {Uniform\\~\cite{cao2012uniform,haddad2016secure}};
		\node [below of = c34,yshift=-6pt] (c35) {Media-Independent\\~\cite{chiang2017forward,ameur2017enhanced}};
		\end{scope}
		
		\foreach \value in {1,2}
		\draw[->] (c1.195) |- (c1\value.west);
		
		\foreach \value in {1,2}
		\draw[->] (c2.195) |- (c2\value.west);
		
		\foreach \value in {1,...,5}
		\draw[->] (c3.195) |- (c3\value.west);
		
		\end{tikzpicture}
		\caption{A broad classification of secure handover schemes for smart M-IoT.}\label{fig:tax_handover}
	\end{figure*}
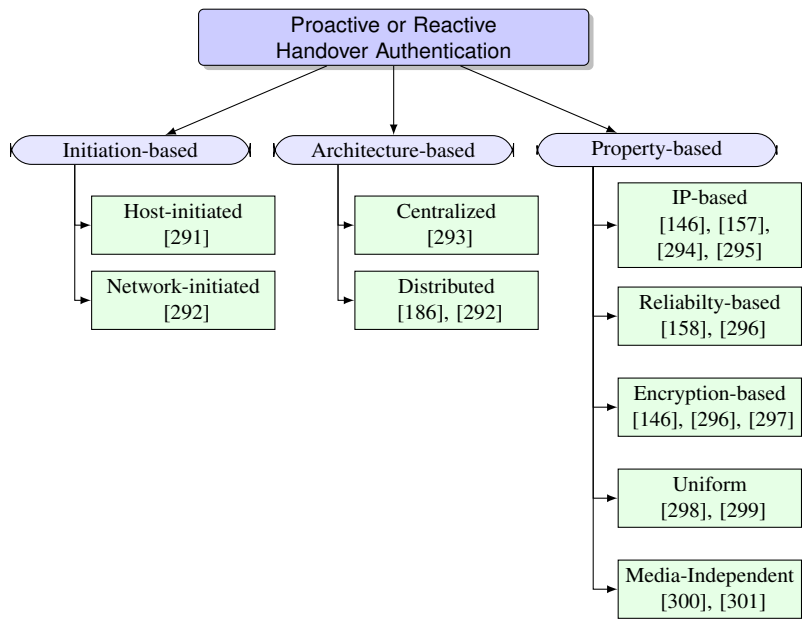
	
	\subsection{Service-based}
	Physical layer security in smart M-IoT can be obtained through service management, control over interference issues and performing accessibility management. Based on the services supported by the smart M-IoT, physical layer security can be studied in three parts:
	\begin{itemize}
		\item Cryptographic: The solutions, which use cryptographic mechanisms for preventing any eavesdropping, are studied in this type. As discussed in~\cite{liu2010securing,zaman2017polarization}, these systems combine physical layer properties with cryptographic mechanisms to ensure the safety of communication between the devices in M-IoT. Such a security is complex to attain but powerful in applicability.
		\item Access Control and Transmissions: The solutions, which control the signal possessions by the users as analyzed in~\cite{xu2016security,chen2016securing,zhang2016secure,hu2017secure,islam2017secured}, are studied in this type. Access control and transmission based solutions are generally low complex and focus on interference management along with control over secrecy probability.
		\item Jamming: There are certain solutions as provided in~\cite{choi2017physical,hu2017cooperative,li2016worst}, which prohibit users from unintentional uplink or downlink in a specified zone. These approaches are responsible for energy-efficient security at the physical layer.
	\end{itemize}
	\subsection{Channel-based}
	Physical layer solutions which emphasize the security of channel used for communications are dependent on the signal alterations and induction of bit codes into the transmission medium. Such solutions should operate with a low-complexity and their operations must be completed in few nanoseconds. The success of these solutions depends on the type of communication setup used for transmissions and the approaches used for securing the bits. Based on the mode of operations, these can be classified into modulation-based and encoding-based solutions:
	\begin{itemize}
		\item Modulation-based: Such schemes changes the signal properties (Amplitude, Phase, or Frequency) for preventing any eavesdropping on the transmitted data. In general, secure-spectrums can help to attain modulation-based channel security in smart M-IoT. These solutions are performed by using carrier waves~\cite{wei2016polarization}~\cite{gao2016physical}.
		\item Encoding-based: Using different codes for the security algorithms at the physical layer helps to secure the traffic and such an approach is classified into encoding based solution. These are performed through binary codes~\cite{li2015compressed}~\cite{limmanee2010secure}.
	\end{itemize}
	
	\subsection{Summary and Insights}
	In this section, we summarized the existing studies into two main categories of physical-layer approaches namely, service-based and channel-based solutions. These solutions were further studies by classifying them on the basis of cryptographic mechanism, access control and transmission policies, jamming facilities, modulation, and encoding. From the study, it is evident that channel estimation, M2M modeling, fading losses, noisy models, energy-constraints are some of the crucial aspects to be taken care of while deploying security solutions for physical layer in M-IoT~\cite{raza2014secure,swetina2014toward,rajaram2017opportunistic,centenaro2016long}.
	
	Physical security of the M-IoT network is also impacted by the burden of devices and interference-management, which are driven by the density of the network. Most of the physical layer security solutions are driven by Signal-to-Interference Ratio (SIR), Signal-to-Interference-plus-Noise Ratio (SINR), secrecy, outage policies, and transmit energies. Despite a plethora of approaches for IoT's physical layer security, there are only a few solutions which can withstand the requirements of M-IoT; thus, a comparison study is presented in Table~\ref{Table4}, which helps to understand the reach and level of security provided by the existing solutions.
	
	\section{Handover Security for Smart M-IoT} 
	Handovers can be hard, soft, horizontal, vertical, terminal and network controlled, and terminal and network initiated, as shown in Fig.~\ref{fig:tax_handover}.
	\onecolumn
	\begin{landscape}
		\begin{center}
			\fontsize{7}{10}\selectfont
			\setlength\LTleft{30pt}            
			\setlength\LTright{0pt}
			\begin{longtable}{@{\extracolsep{\fill}}*{10}{c}}
				\caption{State-of-the-art approaches for physical layer security in M-IoT.}\label{Table4} \\
				\hline
				\multicolumn{1}{p{1.5cm}}{\centering \textbf{Approach}}
				&\multicolumn{1}{p{1.5cm}}{\centering \textbf{Author} \\ \textbf{(Year)} }
				&\multicolumn{1}{p{1.5cm}}{\centering \textbf{Ideology}}
				&\multicolumn{1}{p{1.2cm}}{\centering \textbf{Application}\\ \textbf{Area} }
				&\multicolumn{1}{p{2.5cm}}{\centering \textbf{Parameters}\\\textbf{focused}}
				&\multicolumn{1}{p{1.2cm}}{\centering \textbf{Computational}\\ \textbf{Complexity}}
				&\multicolumn{1}{p{1cm}}{\centering \textbf{Memory}\\ \textbf{Consumption}}
				&\multicolumn{1}{p{1.2cm}}{\centering \textbf{Energy}\\ \textbf{Efficiency}}
				&\multicolumn{1}{p{1.2cm}}{\centering \textbf{Scalability}}
				&\multicolumn{1}{p{1.2cm}}{\centering \textbf{Secrecy}} \\[6pt]\\
				\hline \\
				
				\endfirsthead
				
				\multicolumn{10}{c}%
				{{\bfseries \tablename\ \thetable{} -- continued from previous page}} \\
				\hline
				
				\multicolumn{1}{p{1.5cm}}{\centering \textbf{Approach}}
				&\multicolumn{1}{p{1.5cm}}{\centering \textbf{Author} \\ \textbf{(Year)} }
				&\multicolumn{1}{p{1.5cm}}{\centering \textbf{Ideology}}
				&\multicolumn{1}{p{1.2cm}}{\centering \textbf{Application}\\ \textbf{Area} }
				&\multicolumn{1}{p{2.5cm}}{\centering \textbf{Parameters}\\\textbf{focused}}
				&\multicolumn{1}{p{1.2cm}}{\centering \textbf{Computational}\\ \textbf{Complexity}}
				&\multicolumn{1}{p{1cm}}{\centering \textbf{Memory}\\ \textbf{Consumption}}
				&\multicolumn{1}{p{1.2cm}}{\centering \textbf{Energy}\\ \textbf{Efficiency}}
				&\multicolumn{1}{p{1.2cm}}{\centering \textbf{Scalability}}
				&\multicolumn{1}{p{1.2cm}}{\centering \textbf{Secrecy}} \\[6pt]\\
				\hline\\\endhead
				
				\hline \multicolumn{10}{l}{{Continued on next page}} \\
				
				\endfoot
				
				\endlastfoot

				\multicolumn{1}{p{1.5cm}}{\centering  Security enhancement against eavesdropping }
				&\multicolumn{1}{p{1.5cm}}{\centering [Xu \emph{et al.} 2016]\cite{xu2016security}}
				&\multicolumn{1}{p{1.5cm}}{\centering Secure relay communications in IoT}
				&\multicolumn{1}{p{1.2cm}}{\centering IoT}
				&\multicolumn{1}{p{2.5cm}}{\centering Secrecy outage probability, Secrecy rate, Achievable communication distance }
				&\multicolumn{1}{p{1.2cm}}{\centering -}
				&\multicolumn{1}{p{1cm}}{\centering - }
				&\multicolumn{1}{p{1.2cm}}{\centering - }
				&\multicolumn{1}{p{1.2cm}}{\centering High}
				&\multicolumn{1}{p{1.2cm}}{\centering \cmark } \\[6pt]\\
				
				\multicolumn{1}{p{1.5cm}}{\centering  Securing uplink transmissions }
				&\multicolumn{1}{p{1.5cm}}{\centering [Chen \emph{et al.} 2016]\cite{chen2016securing}}
				&\multicolumn{1}{p{1.5cm}}{\centering Light-weight mechanism for preventing eavesdropping }
				&\multicolumn{1}{p{1.2cm}}{\centering MIMO-IoT}
				&\multicolumn{1}{p{2.5cm}}{\centering Symbol error rate}
				&\multicolumn{1}{p{1.2cm}}{\centering Low}
				&\multicolumn{1}{p{1cm}}{\centering - }
				&\multicolumn{1}{p{1.2cm}}{\centering - }
				&\multicolumn{1}{p{1.2cm}}{\centering High}
				&\multicolumn{1}{p{1.2cm}}{\centering \cmark } \\[6pt]\\
				
				\multicolumn{1}{p{1.5cm}}{\centering  Securing cyber-physical communications }
				&\multicolumn{1}{p{1.5cm}}{\centering [Xu \emph{et al.} 2017]\cite{xu2017security}}
				&\multicolumn{1}{p{1.5cm}}{\centering Security aware wave formations}
				&\multicolumn{1}{p{1.2cm}}{\centering CPS}
				&\multicolumn{1}{p{2.5cm}}{\centering Secrecy rate, Transmit energy}
				&\multicolumn{1}{p{1.2cm}}{\centering Low}
				&\multicolumn{1}{p{1cm}}{\centering - }
				&\multicolumn{1}{p{1.2cm}}{\centering High }
				&\multicolumn{1}{p{1.2cm}}{\centering High}
				&\multicolumn{1}{p{1.2cm}}{\centering \cmark } \\[6pt]\\
				
				\multicolumn{1}{p{1.5cm}}{\centering  Channel aware security }
				&\multicolumn{1}{p{1.5cm}}{\centering [Choi 2017]\cite{choi2017physical}}
				&\multicolumn{1}{p{1.5cm}}{\centering Opportunistic jamming }
				&\multicolumn{1}{p{1.2cm}}{\centering IoT}
				&\multicolumn{1}{p{2.5cm}}{\centering Secrecy rate, Achievable rate, Outage Probability }
				&\multicolumn{1}{p{1.2cm}}{\centering -}
				&\multicolumn{1}{p{1cm}}{\centering - }
				&\multicolumn{1}{p{1.2cm}}{\centering High}
				&\multicolumn{1}{p{1.2cm}}{\centering High}
				&\multicolumn{1}{p{1.2cm}}{\centering \cmark } \\[6pt]\\
				
				\multicolumn{1}{p{1.5cm}}{\centering  Physical layer security}
				&\multicolumn{1}{p{1.5cm}}{\centering [Hu \emph{et al.} 2017]\cite{hu2017cooperative}}
				&\multicolumn{1}{p{1.5cm}}{\centering Cooperative jamming in IoT}
				&\multicolumn{1}{p{1.2cm}}{\centering IoT}
				&\multicolumn{1}{p{2.5cm}}{\centering Secrecy outage probability, Power allocation ratio, }
				&\multicolumn{1}{p{1.2cm}}{\centering -}
				&\multicolumn{1}{p{1cm}}{\centering - }
				&\multicolumn{1}{p{1.2cm}}{\centering High }
				&\multicolumn{1}{p{1.2cm}}{\centering High}
				&\multicolumn{1}{p{1.2cm}}{\centering \cmark } \\[6pt]\\
				
				\multicolumn{1}{p{1.5cm}}{\centering  Secure communications in C-IoT }
				&\multicolumn{1}{p{1.5cm}}{\centering [Li \emph{et al.} 2016]\cite{li2016worst}}
				&\multicolumn{1}{p{1.5cm}}{\centering Worst case channel jamming for spectrum leasing}
				&\multicolumn{1}{p{1.2cm}}{\centering Cognitive-IoT}
				&\multicolumn{1}{p{2.5cm}}{\centering SINR, Energy-harvesting, Channel Uncertainty }
				&\multicolumn{1}{p{1.2cm}}{\centering High}
				&\multicolumn{1}{p{1cm}}{\centering - }
				&\multicolumn{1}{p{1.2cm}}{\centering High}
				&\multicolumn{1}{p{1.2cm}}{\centering -}
				&\multicolumn{1}{p{1.2cm}}{\centering \cmark } \\[6pt]\\
				
				\multicolumn{1}{p{1.5cm}}{\centering  Secure distributed Detection in IoT}
				&\multicolumn{1}{p{1.5cm}}{\centering [Zhang and Sun 2016]\cite{zhang2016secure}}
				&\multicolumn{1}{p{1.5cm}}{\centering Security in energy-constrained IoT networks }
				&\multicolumn{1}{p{1.2cm}}{\centering IoT}
				&\multicolumn{1}{p{2.5cm}}{\centering Error probability, SNR}
				&\multicolumn{1}{p{1.2cm}}{\centering Low}
				&\multicolumn{1}{p{1cm}}{\centering - }
				&\multicolumn{1}{p{1.2cm}}{\centering High }
				&\multicolumn{1}{p{1.2cm}}{\centering High}
				&\multicolumn{1}{p{1.2cm}}{\centering \cmark } \\[6pt]\\
				
				\multicolumn{1}{p{1.5cm}}{\centering  Secure communications in CIoT}
				&\multicolumn{1}{p{1.5cm}}{\centering [Hu \emph{et al.} 2017]\cite{hu2017secure}}
				&\multicolumn{1}{p{1.5cm}}{\centering Secure energy-efficient relay communications}
				&\multicolumn{1}{p{1.2cm}}{\centering Cognitive-IoT}
				&\multicolumn{1}{p{2.5cm}}{\centering Secure transmission rate, SNR }
				&\multicolumn{1}{p{1.2cm}}{\centering -}
				&\multicolumn{1}{p{1cm}}{\centering - }
				&\multicolumn{1}{p{1.2cm}}{\centering High}
				&\multicolumn{1}{p{1.2cm}}{\centering -}
				&\multicolumn{1}{p{1.2cm}}{\centering \cmark } \\[6pt]\\
				
				\multicolumn{1}{p{1.5cm}}{\centering  Security enhancement during inteference }
				&\multicolumn{1}{p{1.5cm}}{\centering [Islam \emph{et al.} 2017]\cite{islam2017secured}}
				&\multicolumn{1}{p{1.5cm}}{\centering Confidential transmission in IoT-relays}
				&\multicolumn{1}{p{1.2cm}}{\centering IoT}
				&\multicolumn{1}{p{2.5cm}}{\centering Mean square error, SNR}
				&\multicolumn{1}{p{1.2cm}}{\centering -}
				&\multicolumn{1}{p{1cm}}{\centering - }
				&\multicolumn{1}{p{1.2cm}}{\centering - }
				&\multicolumn{1}{p{1.2cm}}{\centering High}
				&\multicolumn{1}{p{1.2cm}}{\centering \cmark } \\[6pt]\\

				\hline
			\end{longtable}
		\end{center}
		\twocolumn
	\end{landscape}
	
	The handovers allow the shifting of radios between the same or different media in a network. M-IoT devices undergo handoffs once they leave their service-space and enter an area governed by a different entity. Most of the handovers in M-IoT are vertical that require efficient security measures for the protection of links during their switching~\cite{khan2017enabling,sharma2018secure,ju2015efficient}. There is a huge requirement of trust as well as seamless shifting of services across the terminals while performing handoffs and mobility management in the network~\cite{luzuriaga2015handling}~\cite{valera2010architecture}. Usually, the M-IoT networks focus on using an Access Point (AP), M-IoT device, AS, and core terminals for shifting services across the network~\cite{you2017spfp}. Most of the networks require seamless services and faster authentication which can be obtained through proactive mechanisms~\cite{gaur2017iot}. These proactive approaches define pre-determined system model over which the authentication is performed and verified against the attacker models. Most of the approaches are selected on the basis of handoff latency, and time consumed in laying off their services onto the next terminals along with their cost of operations~\cite{baek2017spatially}. SDNs, media-independent technologies, network slicing and the inclusion of PMIPv6-based solutions can enhance the performance of security solutions that aim at securing the handovers in M-IoT~\cite{guan2017extension,li2017sat,lee2013comparative,shin2017secure,sharma2017saca,khan2014sensor}. The proactive and reactive handover authentication approaches can be further classified into initiation-based, architecture-based, and property-based schemes for security in smart M-IoT.
	
	Note that all of the handover authentication solutions may either use primary, secondary or group mode for authentication irrespective of the classification. The details on each of these classifications are as follows:
	\subsection{Initiation-based}
	Handovers are operated through a governing entity which initiates the procedures of attachment and detachment of a node in the network. Based on the initiation, the handovers authentication procedures can be divided into following two types:
	\begin{itemize}
		\item Host-initiated: When the service consuming entity starts the procedures of handovers, the type of handovers is marked as host-initiated. Host-initiated handovers consume much signaling and might have a weak security because of a failure in the identification of requests which may come from an anomaly node~\cite{lee2012host}.
		\item Network-initiated: When the service providing entity starts the procedures of handovers, the type of handovers is marked as network-initiated. This type of handover is low complex and more secure in because of control by a centralized authority~\cite{lee2013distributed}. However, security layouts and architectural complexity can affect the performance of such handovers.
	\end{itemize}
	\subsection{Architecture-based}
	As discussed earlier, the handovers authentication procedures can also be studied from the architectural point of view and can be distinguished into following two types:
	\begin{itemize}
		\item Centralized: This includes the authentication procedures, which are driven by a centralized authority. SDN-based or topology-based authentications are usually centralized in nature and consequently pose a risk of single point of failure~\cite{ghahfarokhi2012context}. Further, centralized layout increases the security path, which requires RO approaches for increasing the performance.
		\item Distributed: This includes solutions like blockchain-DMM, P2P, P2MP and crowdsourcing like authentications which can help to remove the dependencies on a single entity in smart M-IoT~\cite{lee2013distributed,sharma2018block}.
	\end{itemize}
	
	\textcolor{black}{Moreover, location privacy is another factor to be considered for mobility of M-IoT. It helps to maintain the anonymity of user location and its specifications. Considering the inclusion of location-based services in M-IoT, use of location-privacy solutions helps to protect the system at the network as well as the user's end~\cite{chen2017robustness}~\cite{ni2018location}~\cite{han2018caslp}. M-IoT can also be facilitated by using location-privacy through obfuscation~\cite{zakhary2018location}. This will also allow the extension of M-IoT to opportunistic scenarios. Liao et al.~\cite{liao2017framework} developed a trajectory-protecting solution, which supports location-based service privacy for IoT-cloud systems. The authors rely on K-Anonymity Trajectory (KAT) algorithm, which shows low complex simulated results. Location-privacy can also be considered as an additional metric for trust evaluation~\cite{mirzamohammadi2017ditio}~\cite{mao2018privacy}. Such solutions are facilitated by hybrid security architectures and uses different algorithms for different modules of the architecture.}
	
	\textcolor{black}{With the involvement of crowdsources in M-IoT, location-privacy is a dominant metric to be considered for protecting location-based threats and prevent issues related to backward broadcasting or tunneling~\cite{wang2018truthful}~\cite{ullah2018esot}~\cite{sun2017efficient}. Especially, for the inclusion of such solutions to M-IoT, it is desired to developed novel key distribution and credential management system that can elongate the efforts for location-based privacy preservation.}
	
	\subsection{Property-based}
	Handover authentication mechanisms can be classified on the basis of property which governs their security aspects. These include
	\begin{itemize}
		\item IP-based: This includes authentication mechanisms followed by the majority of mobile applications as it uses proxy procedures to support the security of nodes in smart M-IoT. PMIPv6 and F-PMIPv6 are among the popular solutions for secure and seamless handovers~\cite{chai2017security,chai2015enhanced,shin2017secure,you2017spfp}.
		\item Reliability-based: Approaches like~\cite{sharma2018secure,ndibanje2017secure}, which not only provides strong authentication but also supports the reliability of connections, are studied under this category. Such approaches help to sustain the connections for longer durations without affecting compromising the security considerations of the network.
		\item Encryption-based: Authentication solutions, which focus on using encryption-based solutions for security, are studied under this type. Encryption based handovers help to protect the user data as well as the control information which is passed between the entities laying off from a zone of one entity and moving into the zone of other entity~\cite{ndibanje2017secure,saxena2016authentication,shin2017secure}.
		
		\onecolumn
		\begin{landscape}
			\begin{center}
				\fontsize{7}{10}\selectfont
				\setlength\LTleft{30pt}            
				\setlength\LTright{0pt}
				\begin{longtable}{@{\extracolsep{\fill}}*{9}{c}}
					\caption{Proactive authentication mechanisms for secure handovers.}\label{Table5} \\
					\hline
					\multicolumn{1}{p{2cm}}{\centering \textbf{Approach}}
					&\multicolumn{1}{p{2.2cm}}{\centering \textbf{Author} \\ \textbf{(Year)} }
					&\multicolumn{1}{p{2.5cm}}{\centering \textbf{Ideology}}
					&\multicolumn{1}{p{1cm}}{\centering \textbf{Scalable}}
					&\multicolumn{1}{p{1cm}}{\centering \textbf{Latency}}
					&\multicolumn{1}{p{1cm}}{\centering \textbf{Bandwidth}}
					&\multicolumn{1}{p{1cm}}{\centering \textbf{Handoff Time}}
					&\multicolumn{1}{p{1.5cm}}{\centering \textbf{Mutual }\\ \textbf{Authentication}}
					&\multicolumn{1}{p{1.2cm}}{\centering \textbf{Location}\textbf{ Privacy}}\\[6pt]
					\hline \\
					
					\endfirsthead
					
					\multicolumn{9}{c}%
					{{\bfseries \tablename\ \thetable{} -- continued from previous page}} \\
					\hline
					\multicolumn{1}{p{2cm}}{\centering \textbf{Approach}}
					&\multicolumn{1}{p{2.2cm}}{\centering \textbf{Author} \\ \textbf{(Year)} }
					&\multicolumn{1}{p{2.5cm}}{\centering \textbf{Ideology}}
					&\multicolumn{1}{p{1cm}}{\centering \textbf{Scalable}}
					&\multicolumn{1}{p{1cm}}{\centering \textbf{Latency}}
					&\multicolumn{1}{p{1cm}}{\centering \textbf{Bandwidth}}
					&\multicolumn{1}{p{1cm}}{\centering \textbf{Handoff Time}}
					&\multicolumn{1}{p{1.5cm}}{\centering \textbf{Mutual }\\ \textbf{Authentication}}
					&\multicolumn{1}{p{1.2cm}}{\centering \textbf{Location}\textbf{ Privacy}}\\[6pt]
					\hline\\\endhead
					
					\hline \multicolumn{9}{l}{{Continued on next page}} \\
					
					\endfoot
					
					\endlastfoot

					\multicolumn{1}{p{2cm}}{\centering Delay optimization handoffs}
					&\multicolumn{1}{p{2.2cm}}{\centering [Lopez \emph{et al.} 2007]\cite{lopez2007network}}
					&\multicolumn{1}{p{2.5cm}}{\centering Network layer authentication }
					&\multicolumn{1}{p{1cm}}{\centering -}
					&\multicolumn{1}{p{1cm}}{\centering -}
					&\multicolumn{1}{p{1cm}}{\centering -}
					&\multicolumn{1}{p{1cm}}{\centering Low}
					&\multicolumn{1}{p{1.5cm}}{\centering \xmark}
					&\multicolumn{1}{p{1.2cm}}{\centering \xmark}\\[8pt]
					
					\multicolumn{1}{p{2cm}}{\centering BASH}
					&\multicolumn{1}{p{2.2cm}}{\centering [He and Perkins 2008]\cite{he2008bash}}
					&\multicolumn{1}{p{2.5cm}}{\centering Backhaul-aided seamless handovers}
					&\multicolumn{1}{p{1cm}}{\centering \cmark}
					&\multicolumn{1}{p{1cm}}{\centering Low}
					&\multicolumn{1}{p{1cm}}{\centering High}
					&\multicolumn{1}{p{1cm}}{\centering Low}
					&\multicolumn{1}{p{1.5cm}}{\centering \xmark}
					&\multicolumn{1}{p{1.2cm}}{\centering \xmark}\\[8pt]
					
					\multicolumn{1}{p{2cm}}{\centering Efficient handover authentication}
					&\multicolumn{1}{p{2.2cm}}{\centering [Fu \emph{et al.} 2012]\cite{fu2012efficient}}
					&\multicolumn{1}{p{2.5cm}}{\centering Privacy preservation for 802.16m}
					&\multicolumn{1}{p{1cm}}{\centering \cmark}
					&\multicolumn{1}{p{1cm}}{\centering -}
					&\multicolumn{1}{p{1cm}}{\centering -}
					&\multicolumn{1}{p{1cm}}{\centering High}
					&\multicolumn{1}{p{1.5cm}}{\centering \cmark}
					&\multicolumn{1}{p{1.2cm}}{\centering \cmark}\\[8pt]
					
					\multicolumn{1}{p{2cm}}{\centering Ticket-based handoffs}
					&\multicolumn{1}{p{2.2cm}}{\centering [Xu \emph{et al.} 2014]\cite{xu2014ticket}}
					&\multicolumn{1}{p{2.5cm}}{\centering Handoff authentication for mesh networks}
					&\multicolumn{1}{p{1cm}}{\centering -}
					&\multicolumn{1}{p{1cm}}{\centering -}
					&\multicolumn{1}{p{1cm}}{\centering -}
					&\multicolumn{1}{p{1cm}}{\centering -}
					&\multicolumn{1}{p{1.5cm}}{\centering \cmark}
					&\multicolumn{1}{p{1.2cm}}{\centering \cmark} \\[8pt]
					
					\multicolumn{1}{p{2cm}}{\centering Secure and efficient handovers}
					&\multicolumn{1}{p{2.2cm}}{\centering [Zhang \emph{et al.} 2014]\cite{zhang2014generic}}
					&\multicolumn{1}{p{2.5cm}}{\centering Authentication using EAP in wireless networks}
					&\multicolumn{1}{p{1cm}}{\centering \cmark}
					&\multicolumn{1}{p{1cm}}{\centering -}
					&\multicolumn{1}{p{1cm}}{\centering High}
					&\multicolumn{1}{p{1cm}}{\centering Medium}
					&\multicolumn{1}{p{1.5cm}}{\centering \cmark}
					&\multicolumn{1}{p{1.2cm}}{\centering \xmark}\\[8pt]
					
					\multicolumn{1}{p{2cm}}{\centering Fast pre-hand authentication}
					&\multicolumn{1}{p{2.2cm}}{\centering [Chien \emph{et al.} 2008]\cite{chien2008fast}}
					&\multicolumn{1}{p{2.5cm}}{\centering Minimized overhead and high security}
					&\multicolumn{1}{p{1cm}}{\centering -}
					&\multicolumn{1}{p{1cm}}{\centering Low}
					&\multicolumn{1}{p{1cm}}{\centering -}
					&\multicolumn{1}{p{1cm}}{\centering Low}
					&\multicolumn{1}{p{1.5cm}}{\centering \cmark}
					&\multicolumn{1}{p{1.2cm}}{\centering \xmark}\\[8pt]

					\multicolumn{1}{p{2cm}}{\centering Handover authentication}
					&\multicolumn{1}{p{2.2cm}}{\centering [Choi and Jung 2010]\cite{choi2010handover}}
					&\multicolumn{1}{p{2.5cm}}{\centering Backhaul-aided seamless handovers}
					&\multicolumn{1}{p{1cm}}{\centering \cmark}
					&\multicolumn{1}{p{1cm}}{\centering Low}
					&\multicolumn{1}{p{1cm}}{\centering -}
					&\multicolumn{1}{p{1cm}}{\centering Medium}
					&\multicolumn{1}{p{1.5cm}}{\centering \cmark}
					&\multicolumn{1}{p{1.2cm}}{\centering \cmark} \\[8pt]
					
					\multicolumn{1}{p{2cm}}{\centering Re-authentication for 3GPP}
					&\multicolumn{1}{p{2.2cm}}{\centering [Shidhani and Leung 2011]\cite{al2011fast}}
					&\multicolumn{1}{p{2.5cm}}{\centering Mutual re-authentication}
					&\multicolumn{1}{p{1cm}}{\centering -}
					&\multicolumn{1}{p{1cm}}{\centering Low}
					&\multicolumn{1}{p{1cm}}{\centering -}
					&\multicolumn{1}{p{1cm}}{\centering Low}
					&\multicolumn{1}{p{1.5cm}}{\centering \cmark}
					&\multicolumn{1}{p{1.2cm}}{\centering \xmark} \\[8pt]
					
					\multicolumn{1}{p{2cm}}{\centering Secure continuous handovers }
					&\multicolumn{1}{p{2.2cm}}{\centering [Kalong \emph{et al.} 2010]\cite{kalong2010dynamic}}
					&\multicolumn{1}{p{2.5cm}}{\centering Dynamic key management}
					&\multicolumn{1}{p{1cm}}{\centering \cmark}
					&\multicolumn{1}{p{1cm}}{\centering Low}
					&\multicolumn{1}{p{1cm}}{\centering High}
					&\multicolumn{1}{p{1cm}}{\centering -}
					&\multicolumn{1}{p{1.5cm}}{\centering \cmark}
					&\multicolumn{1}{p{1.2cm}}{\centering \xmark}\\[8pt]
					
					\multicolumn{1}{p{2cm}}{\centering Handover for seamless multimedia transmissions }
					&\multicolumn{1}{p{2.2cm}}{\centering [Saxena and Roy 2011]\cite{saxena2011novel}}
					&\multicolumn{1}{p{2.5cm}}{\centering Proactive authentication over 802.11}
					&\multicolumn{1}{p{1cm}}{\centering \cmark}
					&\multicolumn{1}{p{1cm}}{\centering Low}
					&\multicolumn{1}{p{1cm}}{\centering High}
					&\multicolumn{1}{p{1cm}}{\centering -}
					&\multicolumn{1}{p{1.5cm}}{\centering \xmark}
					&\multicolumn{1}{p{1.2cm}}{\centering \xmark}\\[8pt]
					
					\multicolumn{1}{p{2cm}}{\centering Privacy preserving handover}
					&\multicolumn{1}{p{2.2cm}}{\centering [Jing \emph{et al.} 2011]\cite{jing2011privacy}}
					&\multicolumn{1}{p{2.5cm}}{\centering EAP-based wireless networks}
					&\multicolumn{1}{p{1cm}}{\centering \cmark}
					&\multicolumn{1}{p{1cm}}{\centering Low}
					&\multicolumn{1}{p{1cm}}{\centering -}
					&\multicolumn{1}{p{1cm}}{\centering Low}
					&\multicolumn{1}{p{1.5cm}}{\centering \cmark}
					&\multicolumn{1}{p{1.2cm}}{\centering \xmark}\\[8pt]
					
					\multicolumn{1}{p{2cm}}{\centering Secure inter-ASN handovers}
					&\multicolumn{1}{p{2.2cm}}{\centering [Nguyen and Ma 2012]\cite{nguyen2012enhanced}}
					&\multicolumn{1}{p{2.5cm}}{\centering EAP-based pre-authentication}
					&\multicolumn{1}{p{1cm}}{\centering \cmark}
					&\multicolumn{1}{p{1cm}}{\centering Low}
					&\multicolumn{1}{p{1cm}}{\centering -}
					&\multicolumn{1}{p{1cm}}{\centering Medium}
					&\multicolumn{1}{p{1.5cm}}{\centering \cmark}
					&\multicolumn{1}{p{1.2cm}}{\centering \xmark}\\[8pt]
					
					\multicolumn{1}{p{2cm}}{\centering Mechanism for E-UTRAN and Non-3GPP}
					&\multicolumn{1}{p{2.2cm}}{\centering [Cao \emph{et al.} 2012]\cite{cao2012uniform}}
					&\multicolumn{1}{p{2.5cm}}{\centering Uniform handover authentication}
					&\multicolumn{1}{p{1cm}}{\centering \cmark}
					&\multicolumn{1}{p{1cm}}{\centering Low}
					&\multicolumn{1}{p{1cm}}{\centering -}
					&\multicolumn{1}{p{1cm}}{\centering -}
					&\multicolumn{1}{p{1.5cm}}{\centering \cmark}
					&\multicolumn{1}{p{1.2cm}}{\centering \xmark}\\[8pt]
					
					\multicolumn{1}{p{2cm}}{\centering CPAL}
					&\multicolumn{1}{p{2.2cm}}{\centering [Lai \emph{et al.} 2014]\cite{lai2014cpal}}
					&\multicolumn{1}{p{2.5cm}}{\centering Privacy-preserving authentication with access linkability}
					&\multicolumn{1}{p{1cm}}{\centering -}
					&\multicolumn{1}{p{1cm}}{\centering Low}
					&\multicolumn{1}{p{1cm}}{\centering -}
					&\multicolumn{1}{p{1cm}}{\centering High}
					&\multicolumn{1}{p{1.5cm}}{\centering \cmark}
					&\multicolumn{1}{p{1.2cm}}{\centering \xmark}\\[8pt]
					
					\multicolumn{1}{p{2cm}}{\centering Secure fast WLAN handoff}
					&\multicolumn{1}{p{2.2cm}}{\centering [Chien and Hsu 2009]\cite{chien2009secure}}
					&\multicolumn{1}{p{2.5cm}}{\centering Time-bound delegated authentication}
					&\multicolumn{1}{p{1cm}}{\centering \cmark}
					&\multicolumn{1}{p{1cm}}{\centering Low}
					&\multicolumn{1}{p{1cm}}{\centering -}
					&\multicolumn{1}{p{1cm}}{\centering -}
					&\multicolumn{1}{p{1.5cm}}{\centering \cmark}
					&\multicolumn{1}{p{1.2cm}}{\centering \xmark}\\[8pt]
					
					\multicolumn{1}{p{2cm}}{\centering Re-authentication scheme for handovers}
					&\multicolumn{1}{p{2.2cm}}{\centering [Ma \emph{et al.} 2013]\cite{ma2013proxy}}
					&\multicolumn{1}{p{2.5cm}}{\centering Proxy signature approach}
					&\multicolumn{1}{p{1cm}}{\centering \cmark}
					&\multicolumn{1}{p{1cm}}{\centering Low}
					&\multicolumn{1}{p{1cm}}{\centering High}
					&\multicolumn{1}{p{1cm}}{\centering -}
					&\multicolumn{1}{p{1.5cm}}{\centering \cmark}
					&\multicolumn{1}{p{1.2cm}}{\centering \xmark}\\[8pt]
					
					\multicolumn{1}{p{2cm}}{\centering Handauth}
					&\multicolumn{1}{p{2.2cm}}{\centering [He \emph{et al.} 2013]\cite{he2013handauth}}
					&\multicolumn{1}{p{2.5cm}}{\centering Authentication with conditional privacy}
					&\multicolumn{1}{p{1cm}}{\centering \cmark}
					&\multicolumn{1}{p{1cm}}{\centering Low}
					&\multicolumn{1}{p{1cm}}{\centering -}
					&\multicolumn{1}{p{1cm}}{\centering -}
					&\multicolumn{1}{p{1.5cm}}{\centering \cmark}
					&\multicolumn{1}{p{1.2cm}}{\centering \xmark}\\[8pt]
					
					\multicolumn{1}{p{2cm}}{\centering EAP-based pre-authentication}
					&\multicolumn{1}{p{2.2cm}}{\centering [Wang \emph{et al.} 2017]\cite{wang2017efficient}}
					&\multicolumn{1}{p{2.5cm}}{\centering Inter-WRAN Handover authentication}
					&\multicolumn{1}{p{1cm}}{\centering \cmark}
					&\multicolumn{1}{p{1cm}}{\centering Low}
					&\multicolumn{1}{p{1cm}}{\centering -}
					&\multicolumn{1}{p{1cm}}{\centering Low}
					&\multicolumn{1}{p{1.5cm}}{\centering \cmark}
					&\multicolumn{1}{p{1.2cm}}{\centering \xmark}\\[8pt]
					
					\multicolumn{1}{p{2cm}}{\centering Fast handovers in 5G Xhaul}
					&\multicolumn{1}{p{2.2cm}}{\centering [Sharma \emph{et al.} 2018]\cite{sharma2018secure}}
					&\multicolumn{1}{p{2.5cm}}{\centering Secure and fast handoffs in 5G-Xhaul and IoT}
					&\multicolumn{1}{p{1cm}}{\centering \cmark}
					&\multicolumn{1}{p{1cm}}{\centering Low}
					&\multicolumn{1}{p{1cm}}{\centering -}
					&\multicolumn{1}{p{1cm}}{\centering Low}
					&\multicolumn{1}{p{1.5cm}}{\centering \cmark}
					&\multicolumn{1}{p{1.2cm}}{\centering \cmark}\\[8pt]
					
					\hline
				\end{longtable}
			\end{center}
		\end{landscape}
		\begin{landscape}
			\begin{center}
				\fontsize{7}{10}\selectfont
				\setlength\LTleft{30pt}            
				\setlength\LTright{0pt}
				\begin{longtable}{@{\extracolsep{\fill}}*{10}{c}}
					\caption{Approaches for secure handovers in M-IoT.}\label{Table6} \\
					\hline
					\multicolumn{1}{p{1.5cm}}{\centering \textbf{Approach}}
					&\multicolumn{1}{p{1.5cm}}{\centering \textbf{Author} \\ \textbf{(Year)} }
					&\multicolumn{1}{p{1.5cm}}{\centering \textbf{Ideology}}
					&\multicolumn{1}{p{1.2cm}}{\centering \textbf{Application}\\ \textbf{Area} }
					&\multicolumn{1}{p{2.5cm}}{\centering \textbf{Parameters}\\\textbf{focused}}
					&\multicolumn{1}{p{1.2cm}}{\centering \textbf{Mutual}\\ \textbf{Authentication}}
					&\multicolumn{1}{p{1cm}}{\centering \textbf{Handoff}\\ \textbf{Time}}
					&\multicolumn{1}{p{1.2cm}}{\centering \textbf{Latency}}
					&\multicolumn{1}{p{1.2cm}}{\centering \textbf{Reliability}}
					&\multicolumn{1}{p{1.2cm}}{\centering \textbf{Security}\\\textbf{Constraints}} \\[6pt]\\
					\hline \\
					
					\endfirsthead
					
					\multicolumn{10}{c}%
					{{\bfseries \tablename\ \thetable{} -- continued from previous page}} \\
					\hline
					
					\multicolumn{1}{p{1.5cm}}{\centering \textbf{Approach}}
					&\multicolumn{1}{p{1.5cm}}{\centering \textbf{Author} \\ \textbf{(Year)} }
					&\multicolumn{1}{p{1.5cm}}{\centering \textbf{Ideology}}
					&\multicolumn{1}{p{1.2cm}}{\centering \textbf{Application}\\ \textbf{Area} }
					&\multicolumn{1}{p{2.5cm}}{\centering \textbf{Parameters}\\\textbf{focused}}
					&\multicolumn{1}{p{1.2cm}}{\centering \textbf{Mutual}\\ \textbf{Authentication}}
					&\multicolumn{1}{p{1cm}}{\centering \textbf{Handoff}\\ \textbf{Time}}
					&\multicolumn{1}{p{1.2cm}}{\centering \textbf{Latency}}
					&\multicolumn{1}{p{1.2cm}}{\centering \textbf{Reliability}}
					&\multicolumn{1}{p{1.2cm}}{\centering \textbf{Security}\\\textbf{Constraints}} \\[6pt]\\
					\hline\\\endhead
					
					\hline \multicolumn{10}{l}{{Continued on next page}} \\
					
					\endfoot
					
					\endlastfoot
					
					\multicolumn{1}{p{1.5cm}}{\centering  Inter-LMA domain handover}
					&\multicolumn{1}{p{1.5cm}}{\centering [Chai \emph{et al.} 2017]\cite{chai2017security}}
					&\multicolumn{1}{p{1.5cm}}{\centering Proxy-based FPMIPv6}
					&\multicolumn{1}{p{1.2cm}}{\centering Mobile IPv6 networks}
					&\multicolumn{1}{p{2.5cm}}{\centering Handover latency}
					&\multicolumn{1}{p{1.2cm}}{\centering \cmark}
					&\multicolumn{1}{p{1cm}}{\centering Low }
					&\multicolumn{1}{p{1.2cm}}{\centering Low }
					&\multicolumn{1}{p{1.2cm}}{\centering -}
					&\multicolumn{1}{p{1.2cm}}{\centering \cmark } \\[6pt]\\
					
					\multicolumn{1}{p{1.5cm}}{\centering  Mobility management scheme}
					&\multicolumn{1}{p{1.5cm}}{\centering [Chai \emph{et al.} 2015]\cite{chai2015enhanced}}
					&\multicolumn{1}{p{1.5cm}}{\centering Proxy-based FPMIPv6}
					&\multicolumn{1}{p{1.2cm}}{\centering IoT}
					&\multicolumn{1}{p{2.5cm}}{\centering Handover latency, Inter-domain movement}
					&\multicolumn{1}{p{1.2cm}}{\centering -}
					&\multicolumn{1}{p{1cm}}{\centering Low }
					&\multicolumn{1}{p{1.2cm}}{\centering Low }
					&\multicolumn{1}{p{1.2cm}}{\centering -}
					&\multicolumn{1}{p{1.2cm}}{\centering \cmark } \\[6pt]\\
					
					\multicolumn{1}{p{1.5cm}}{\centering  Uniform handover}
					&\multicolumn{1}{p{1.5cm}}{\centering [Cao \emph{et al.} 2012]\cite{cao2015ugha}}
					&\multicolumn{1}{p{1.5cm}}{\centering E-UTRAN}
					&\multicolumn{1}{p{1.2cm}}{\centering LTE-A Networks}
					&\multicolumn{1}{p{2.5cm}}{\centering Signaling messages,Computational cost}
					&\multicolumn{1}{p{1.2cm}}{\centering -}
					&\multicolumn{1}{p{1cm}}{\centering -}
					&\multicolumn{1}{p{1.2cm}}{\centering - }
					&\multicolumn{1}{p{1.2cm}}{\centering -}
					&\multicolumn{1}{p{1.2cm}}{\centering \cmark } \\[6pt]\\
					
					\multicolumn{1}{p{1.5cm}}{\centering  Uniform handover}
					&\multicolumn{1}{p{1.5cm}}{\centering [Haddad \emph{et al.} 2016]\cite{haddad2016secure}}
					&\multicolumn{1}{p{1.5cm}}{\centering Authentication and registration with HSS}
					&\multicolumn{1}{p{1.2cm}}{\centering LTE-A networks}
					&\multicolumn{1}{p{2.5cm}}{\centering Computational delay, Communication overhead, Storage cost}
					&\multicolumn{1}{p{1.2cm}}{\centering \cmark}
					&\multicolumn{1}{p{1cm}}{\centering Low }
					&\multicolumn{1}{p{1.2cm}}{\centering Low }
					&\multicolumn{1}{p{1.2cm}}{\centering -}
					&\multicolumn{1}{p{1.2cm}}{\centering \cmark } \\[6pt]\\
					
					\multicolumn{1}{p{1.5cm}}{\centering  Session key management}
					&\multicolumn{1}{p{1.5cm}}{\centering [Kong \emph{et al.} 2017]\cite{kong2017achieve}}
					&\multicolumn{1}{p{1.5cm}}{\centering Mobile relaying-based session management}
					&\multicolumn{1}{p{1.2cm}}{\centering LTE-A networks}
					&\multicolumn{1}{p{2.5cm}}{\centering Computational delay, Communication overhead, Storage cost}
					&\multicolumn{1}{p{1.2cm}}{\centering \cmark}
					&\multicolumn{1}{p{1cm}}{\centering Low }
					&\multicolumn{1}{p{1.2cm}}{\centering Low }
					&\multicolumn{1}{p{1.2cm}}{\centering -}
					&\multicolumn{1}{p{1.2cm}}{\centering \cmark } \\[6pt]\\
					
					\multicolumn{1}{p{1.5cm}}{\centering  Secure and efficient protocol}
					&\multicolumn{1}{p{1.5cm}}{\centering [Sharma \emph{et al.} 2018]\cite{sharma2018secure}}
					&\multicolumn{1}{p{1.5cm}}{\centering Key exchange and authentication}
					&\multicolumn{1}{p{1.2cm}}{\centering 5G-Xhaul}
					&\multicolumn{1}{p{2.5cm}}{\centering Handover latency, Failure factor, Signaling overheads}
					&\multicolumn{1}{p{1.2cm}}{\centering \cmark}
					&\multicolumn{1}{p{1cm}}{\centering Low }
					&\multicolumn{1}{p{1.2cm}}{\centering Low }
					&\multicolumn{1}{p{1.2cm}}{\centering \cmark}
					&\multicolumn{1}{p{1.2cm}}{\centering \cmark } \\[6pt]\\
					
					\multicolumn{1}{p{1.5cm}}{\centering  Route optimization}
					&\multicolumn{1}{p{1.5cm}}{\centering [Shin \emph{et al.} 2017]\cite{shin2017secure}}
					&\multicolumn{1}{p{1.5cm}}{\centering PMIPv6-based RO}
					&\multicolumn{1}{p{1.2cm}}{\centering Smart home IoT networks}
					&\multicolumn{1}{p{2.5cm}}{\centering Transmission rate, Packet loss, Network throughput}
					&\multicolumn{1}{p{1.2cm}}{\centering \cmark}
					&\multicolumn{1}{p{1cm}}{\centering Low }
					&\multicolumn{1}{p{1.2cm}}{\centering Low }
					&\multicolumn{1}{p{1.2cm}}{\centering -}
					&\multicolumn{1}{p{1.2cm}}{\centering \cmark } \\[6pt]\\
					
					\multicolumn{1}{p{1.5cm}}{\centering  Seamless handover}
					&\multicolumn{1}{p{1.5cm}}{\centering [Feirer \emph{et al.} 2017]\cite{feirer2017seamless}}
					&\multicolumn{1}{p{1.5cm}}{\centering IEEE 802.11k-based handover}
					&\multicolumn{1}{p{1.2cm}}{\centering Industrial WLAN}
					&\multicolumn{1}{p{2.5cm}}{\centering Message overhead}
					&\multicolumn{1}{p{1.2cm}}{\centering \xmark}
					&\multicolumn{1}{p{1cm}}{\centering Low }
					&\multicolumn{1}{p{1.2cm}}{\centering -}
					&\multicolumn{1}{p{1.2cm}}{\centering -}
					&\multicolumn{1}{p{1.2cm}}{\centering \cmark } \\[6pt]\\
					
					\multicolumn{1}{p{1.5cm}}{\centering  Authentication protocol}
					&\multicolumn{1}{p{1.5cm}}{\centering [Saxena \emph{et al.} 2016]\cite{saxena2016authentication}}
					&\multicolumn{1}{p{1.5cm}}{\centering Symmetric key cryptosystem}
					&\multicolumn{1}{p{1.2cm}}{\centering LTE Networks}
					&\multicolumn{1}{p{2.5cm}}{\centering Storage overhead, Computation overhead, Bandwidth consumption}
					&\multicolumn{1}{p{1.2cm}}{\centering \cmark}
					&\multicolumn{1}{p{1cm}}{\centering -}
					&\multicolumn{1}{p{1.2cm}}{\centering Low }
					&\multicolumn{1}{p{1.2cm}}{\centering -}
					&\multicolumn{1}{p{1.2cm}}{\centering \cmark } \\[6pt]\\
					
					\multicolumn{1}{p{1.5cm}}{\centering  Mutual authentication handoff}
					&\multicolumn{1}{p{1.5cm}}{\centering [Ndibanje  \emph{et al.} 2017]\cite{ndibanje2017secure}}
					&\multicolumn{1}{p{1.5cm}}{\centering RSS and PKC-based protocol}
					&\multicolumn{1}{p{1.2cm}}{\centering IoT-Sensor networks}
					&\multicolumn{1}{p{2.5cm}}{\centering RSS}
					&\multicolumn{1}{p{1.2cm}}{\centering \cmark}
					&\multicolumn{1}{p{1cm}}{\centering Low }
					&\multicolumn{1}{p{1.2cm}}{\centering -}
					&\multicolumn{1}{p{1.2cm}}{\centering \cmark}
					&\multicolumn{1}{p{1.2cm}}{\centering \cmark } \\[6pt]\\
					
					\hline
				\end{longtable}
			\end{center}
			\twocolumn
		\end{landscape}
		
		\item Uniform: Such types of handovers authentication are more prominent in LTE and LTE-A networks as these can be used for all types of networks~\cite{cao2012uniform,haddad2016secure}. This is one of the most suitable handovers procedures for smart M-IoT networks. Such mechanisms are low-complex, computationally-inexpensive and highly secure solutions for mobile security.
		\item Media-Independent: Such types of handovers rely on the security governed by IEEE 802.21a-2012 for supporting security along with media independence while shifting services from one entity to another in an inter-handover mode~\cite{chiang2017forward,ameur2017enhanced,sharma2017IEEE,sharma2017IEEE2}. Amalgamation of MIH solutions with F-PMIPv6 techniques is gaining popularity because of their low complexity and high security~\cite{guan2017extension}.
	\end{itemize}
	
	\subsection{Summary and Insights}
	In this section, we surveyed solutions for secure handover of smart M-IoT devices. The devices can perform intra- or inter-handover depending on the layout of the network. Proactive authentication plays a key role in securing service layoffs between the devices and can ensure long-sessions without disrupting the services of a user under movement. Distributed security protocols play a considerable role in managing nodes under high mobility scenarios by preventing unnecessary passes to the core for re-authentication of devices.
	
	Handoff latency, discovery time, bandwidth support, mutual authentication, and overheads are some of the key metrics to be considered for selecting an efficient handover scheme for M-IoT, as shown in Tables~\ref{Table5} and~\ref{Table6}. There are plenty of solutions which have diversified the security aspects of handovers and provide a wide range of services for handling billions of IoT devices. Despite this, the majority of them fails on the aspect of performance and does not account for the tradeoff between the security and Quality of Experience (QoE). Thus, new approaches are required that can take into account these requirements of security as well as the performance before their final deployment and testing while causing minimum overheads during handoffs.
	
	\section{Research Challenges, Open Issues and Future Directions}
	Security, privacy, and trust are supported through specific requirements of a system, which are the open challenges to be resolved in M-IoT. Most of the challenges and issues can be acquired from the studies presented in~\cite{roman2018mobile,7902207,7903611,8119706,feng2017survey,zhou2017security,mollah2017security,li2016internet,malina2016perspective,palattella2016internet,lin2016iot,sicari2015security,al2015internet,arias2015privacy,yan2014survey,jing2014security,keoh2014securing,roman2013features,bhattasali2013study,chang2012survey,bonetto2012secure,koien2011reflections,medaglia2010overview}. From these studies, it is noticeable that the major open issues to be resolved for M-IoT are:
	\begin{itemize}
		\item Satisfaction of the security requirements: It is of utmost importance that any approach which aims to facilitate security, privacy and trust in M-IoT must satisfy certain security requirements that are listed below:
		\begin{itemize}
			\item \emph{Mutual Authentication}: Security agreement between each entity in M-IoT is of utmost importance. Each device must be able to identify the correctness of every other device involved in transmission. The trust relationship between the devices can help to attain the requirements of mutual authentication.
			\item \emph{Secure Key Exchange}: Security keys are the pillar for preventing attacks in a network. It is a must that keys are exchanged secretly over a secure channel and must not reveal at any instance of operations.
			\item \emph{Session Key Management}: This is a requirement which helps to secure the communication between the M-IoT devices. It is necessary for an approach to use a secure key which is different from other keys while communicating with a particular device in a network. Session keys must be renewed consistently for preventing any attacks because of lack of key freshness.
			\item \emph{Perfect Forward Secrecy}: In a communication setup, capturing of long-term keys should not be able to generate past session keys. This helps to secure previous contents and also protect future compromises and password sharing.
			\item \emph{Defense against a Replay Attack}: Repetition of valid data can reveal the security policies as well as lead to overconsumption of resources in the protection of the system. Such kinds of attacks are caused by interceptions and must be avoided as the traffic in M-IoT is very sensitive and crucial.
			\item \emph{Access Control and Authorization}: It is required that the new solutions are able to provide control on the accessibility limits of each device and also provide policies for authorization and management of content along with session formations.
			\item \emph{Defense against a Resource Exhaustion Attack}: This type of attack should be prevented as resource exhaustion attacks can exploit the network and the M-IoT devices through excessive key operations. Such an attack may lead to the shutdown of the entire network.
		\end{itemize}
		\item Performance tradeoff: Apart from the security requirement, it is required that a solution should not compromise the performance of the system and must be capable of handling the performance tradeoffs due to computational burden of security mechanisms. The approaches must be able to handle the implementation overheads during continuous operations.
		\item Platform compatibility: Due to a difference in the types of devices and their configurations, it is difficult to support platform compatibility in M-IoT. However, there is still a strong requirement of such solutions which can be operated irrespective of the types of technologies being operated over M-IoT devices. Platform compatibility can be obtained by defining security mechanisms which rely on operations that have lesser variations when shifted across devices.
		\item Resource utilization: Efficient utilization of resources like memory and power and prevention of their overconsumption can save the operations up to a longer duration. Resource utilization can be attained by using novel network architectures as well as independent layers for each operation in M-IoT. Such a facility can be obtained through SDN-NFV technologies. As discussed earlier, facilities like osmotic computing, fog computing, catalytic computing and edge-crowd modeling can be used for handling resource utilization while providing security and privacy solutions for M-IoT.
		\item Insider threat management: Prevention of theft, fraud, and damage through non-compromising models is required as this can help to manage the false-occurrences caused by the criminal aspects of M-IoT users. Models like blockchain, distributed mobility management, and crowdsourcing can be used for management of insider threats in a system.
	\end{itemize}
	
	Future aspects of M-IoT are quite vast as it has to deal with a lot of dependencies of the underlaid architecture. Network designing and placement of components play a key role in providing security in M-IoT; whereas privacy has a lot to do with an individual as well as the service providers. Trust is built on the backbone of security and privacy and its management is as crucial as other services. Till date, two of the major aspects to achieve in trust management is its visualization and formal way of expressing for a large set of users. Even in the lights of different solutions, there are no standard mechanisms which can help to visualize trust as a property of a device. Thus, future approaches must consider formally defining trust and building some standard rules which should operate together with the security and privacy considerations for enhancing the practicality of M-IoT services to users. In lieu of various properties of existing solutions as discussed throughout this article, following key points can be used for directing further research on different aspects of smart M-IoT.
	\subsection{Security related future research directions}
	\begin{itemize}
		\item Network Monitoring in M-IoT: M-IoT security relies on the true operations of the entities involved in providing services to the users. Any faulty equipment can result in sets of failures which may compromise the operations in M-IoT leading to the devastation of infrastructure as well as data. Futuristic approaches must ensure efficient deployment of solutions like IDS, network monitors, and ethical packet sniffers for enhancing the security requirements. Network monitoring should emphasize the resource-based evaluation of the involved devices so as to prevent service halts and offer ultra-reliable QoE to its users. New tools can be developed which can analyze the traffic passes between the devices. In addition, security of network monitors is to be considered for preventing any eavesdropping on the ethically gathered data. Monitoring tools and procedures should possess encapsulation as a key property and prevent and disclosure of type and make of equipment even if the attacker possesses maximum data~\cite{lee2018monitoring}~\cite{garcia2018wireless}.
		\item Vulnerability Assessment: For secure operations, it is of utmost importance that the entire network is consistently monitored for potential vulnerabilities that may lead to different types of threats. Such a task can be attained by defining security policies for each entity in the network and building profilers which can help to assess devices in case of weird behavior or functioning~\cite{wang2018vulnerability}. Vulnerability assessment can help to determine the influence of attack on a particular set of entities~\cite{samtani2018identifying}. The vulnerability assessments should be conducted at both the user-side and network-side. User-side evaluations should be abstracted and must not consume excessive operations and must be low on overheads; network-side evaluations should be conducted with zero-maintenance time and any service halts. Anomaly detection, community classification and attacker marking are the main targets of vulnerability assessment~\cite{8322278}. All of these are open issues and their applicability are subject to application and operational scenarios.
		\item Policies for Zero-day Attack: Zero-day attacks in software modules of M-IoT are the key threats to its security. It is difficult to identify such possibilities unless made public by the attacker. Most of these are identified during the development stage, but some of these are marked during the regular testing operations. It becomes the liability of service providers and software-distributors to provide security patches as soon as vulnerabilities are identified. Further, providing customer knowledge and making mandatory to download and install security updates should be considered for effective countermeasures against such attacks~\cite{sharma2018framework,abeshu2018deep,lobato2018adaptive}.
		\item Hacking and Accessibility: Despite always being a hot topic, hacking and accessibility are yet open future challenges in smart M-IoT. It is required that new solutions are developed for code obfuscation and new policies are made for controlling the accessibility to M-IoT components and its services~\cite{weinberg2015internet}~\cite{you2010malware}. Pre-authentication mechanisms and multi-registration phases can help to attain these requirements. However, performance and overheads are the major issues attached to such provisioning, and any approach controlling the accessibility must not cause performance overheads and should not disturb the regular operations of the network.
	\end{itemize}
	
	\subsection{Privacy related future research directions}
	\begin{itemize}
		\item Prevention of Device Profiling: Data gathering is one of the key requirements of modern day organizations to provide a personalized experience to its users. However, the process of data gathering and information analysis may cause different types of threats by deliberately breaching the privacy of users. Collection of data and using it for evaluating user behavior and controlling the preferences may allow a threat to confidentiality and integrity of an individual; further, hold on information by an eavesdropper leads to vulnerable conditions, which violates the network policies~\cite{lee2017profiot}~\cite{shaashua2018physical}. Thus, to overcome such issues, it is required that futuristic solutions should not allow unauthorized device profiling and information gathering procedures must be controlled by the service providers. In addition, no selling of data should be done as this violates the personal space of an individual. Use of device profiling for advertisements for generating revenues is fine, but it should not affect the preferences of an individual.
		\item Control over Data Gathering: M-IoT devices are sensitive to information and data across their network is delicate to threats. Classification of data and generating knowledge by data-processing disclose different types of vulnerabilities, which are the tools of hackers for exploiting the network and its users. Approaches are required that prohibits uncontrolled data gathering and limits the service providing apps from collecting excessive information other than the required ones. Data gathering procedures should be controlled by app hosting platforms and as per the individual is concerned, they must be provided with knowledge of using authenticated sources to prevent any enforced data gatherings~\cite{jayaraman2017privacy}~\cite{doukas2012enabling}.
		\item Personalized Settings: Every application, be it open source or proprietary, must provide preferential settings to its users, where they can manage and control the amount of information to be shared across the M-IoT platforms. It is necessary that every user should be able to monitor the amount of information and extend up to which his/her information is used and for what purposes. Personalized settings should be supported by access management, accountability and authorization controls.
		\item Managing Information Flow: For sufficiently high privacy settings, every entity in M-IoT must be provided with facilities for managing information flow. These information flows should be manageable remotely, thus, different techniques and solutions can be developed for such requirements which pave a way for controlling the information flow even being present on-site. Development of toolkits and apps for information flow are other future research challenges in smart M-IoT. Further, these can be used with AI techniques to perform a priori probabilistic checks on the occurrence of attacks for a particular set of settings.
	\end{itemize}
	\subsection{Trust related future research directions}
	\begin{itemize}
		\item Dedicated Node Management: Trust is a compliance degree between the entities to ensure accurate operational behavior in the network. M-IoT is dedicated to operating networks which will heavily depend on the crowd sources for the majority of their operations. Such a dependency raises a crucial requirement of node management and control over the service-relationships between the devices. Research must be conducted in this direction while ensuring how the devices will interact on basis of what policies they can accurately judge each others' correctness~\cite{al2017trust}~\cite{abderrahim2017tmcoi}. It is required that certain solutions must be developed that can provide dedicated node management at a fine granular level while leveraging the properties of existing solutions for trust management. Different type of protocols can be designed that takes situational awareness as one of the key properties for ensuring trust-aware communications in M-IoT. In addition, contextual behavior monitoring and aspect-based classification can help to ensure trust-compliance between the entities of M-IoT.
		\item Trust Visualization and Markings:  There are a huge set of applications and approaches which emphasize on computing trust in different types of network as per the requirements of the applications. But the majority of these fail to provide any conceptualization on the visualization process which helps to easy identification of service-law violators. It is required that research must be conducted in this direction while finding a benchmark which can be used as a backbone for trust-visualization and markings~\cite{sharma2017computational}. In addition, facilities must be provided to check trust roles and authorization activities across the network.
		\item Anomaly Detection and Recovery: Anomaly detection is the other key aspect of trust maintenance solutions. Futuristic research must focus on providing enhanced, on-demand and real-time facilities for detecting anomalies. This must accompany the solutions which can help to recover the users which are marked as anomalies by allowing them to re-justify their associations with the networks' terms and conditions and their flow control~\cite{8322278}. It is required that trust evaluations must lift themselves from the traditional reputation-based systems as such facilities can easily fall prey to Sybil attacks and may mislead the trust-maintenance process.
		\item Distributed Evaluations and Trust offloading: Apart from trust-management, the approaches are required which can operate in a distributed manner and yet provide competitive results as that of centralized solutions. This will help to prevent any single point of failure~\cite{khan2017trust,khan2017trusta,sharma2017managing,sharma2017computational}. Such solutions can be fixated on different offloading techniques which can be operated in parallel to data evaluations and does not interfere with the regular network operations. Development of distributed IDS and crowd-sourced IDS can be crucial solutions for attaining distributed evaluations as well as trust offloading.
	\end{itemize}
	\subsection{Necessity of Amalgamation}
	Security, privacy, and trust in M-IoT go hand in hand. A breach of policies of one may lead to attack through other. Security policies must be strong enough to prevent any unauthorized access to the personal information of an individual in M-IoT and privacy policies must ensure that the data is always shared with the trusted party. Such an activity is also operational in reverse and holds true for any sort of network formations in smart M-IoT. The necessities for amalgamating these solutions can be accounted for following points:
	\begin{itemize}
		\item Prevention against Cyber Spies: Combining all the aspects of security, privacy and trust for smart M-IoT ensure protection against cyber bullies, spies, and service breachers~\cite{oravec2017emerging,chowdhury2017cyber,kim2017national}. These three requirements ensure that the network is operating in closed perimeter even its operations are distributed across the huge cyber network. Here, close perimeter refers to the path lengths and routes which can be tracked down easily and conterminously without many overheads.
		\item Risk Assessment and Mitigation: A network should be assessed for potential risks in its operations. It is necessary that the risk evaluations are conducted on the basis of combined rules for security, privacy, and trust. Risk evaluations are usually probabilistic, however, with complete details of all possible rules, these can be used for generating a particular output that yields visible results for risk assessments~\cite{chochliouros2017enabling,jiang2012attack,zheng2013iot}.
		\item Reliable Communications: Modern network services, especially the ones operating for smart M-IoT, require reliable connections for their continuous operations. Such a requirement can be ensured only if the network components and their services satisfy the requirements associated with security, privacy, and trust. In fact, the upcoming applications in smart M-IoT not only demand reliable communications, instead their focus is on ultra-reliable communications~\cite{liao2018eavesdropping}~\cite{soldani20185g} with lower dependencies and controlled cohesion and coupling amongst their software solutions.
	\end{itemize}
	Thus, it becomes inevitably important to develop solutions, which hold true justifications for security, privacy, and trust at the same instance and at the same level.
	
	\section{Conclusions}
	\textcolor{black}{Security solutions must be able to fortify authentication, confidentiality, integrity, freshness, access control and authorizations for M-IoT devices and its platforms, whereas privacy must support information protection for every device and its users. Both of these requisites can be obtained by building trust relationships across the networks. However, there exists a mixture of approaches that consider one of these requisites but ignore the other requirements. Previous studies have lighted such issues and withal compared the majority of them on the substructure of different parameters. However, prior studies have shown a constrained role in evaluating security, privacy and trust especially for keenly intellective and connected M-IoT networks. This paper considered the shortcomings of existing literature and provided an in-depth evaluation of different approaches which fixates on the crucial aspects of security, privacy, and trust.}
	
	\textcolor{black}{This article covered the concept and ideology of smart M-IoT networks and its devices followed by their applications, advances, challenges, characteristics, technologies, and standards. Then the literature evaluations were presented for approaches which emphasized secure frameworks, data-privacy, secure protocols, physical layer security, and handover protections for smart M-IoT. Next, different ways for analyzing the security, privacy, and trust in M-IoT were discussed followed by roadmap and open issues along with highlights of some pertinent materials which can be followed for improving understandings in this direction of research.  This study has highlighted the requirements of new solutions, which can collectively resolve the issues related to security, privacy, and trust in smart M-IoT without compromising the performance and complexity of operations.}
	\bibliographystyle{ieeetr}
	\bibliography{related}

\begin{thebibliography}{100}

\bibitem{farris2018federated}
I.~Farris, A.~Orsino, L.~Militano, A.~Iera, and G.~Araniti, ``Federated iot
  services leveraging 5g technologies at the edge,'' {\em Ad Hoc Networks},
  vol.~68, pp.~58--69, 2018.

\bibitem{liu2018neighbor}
W.~Liu, K.~Nakauchi, and Y.~Shoji, ``A neighbor-based probabilistic broadcast
  protocol for data dissemination in mobile iot networks,'' {\em IEEE Access},
  2018.

\bibitem{ghasempour2016optimizing}
A.~Ghasempour and T.~K. Moon, ``Optimizing the number of collectors in
  machine-to-machine advanced metering infrastructure architecture for internet
  of things-based smart grid,'' in {\em Green Technologies Conference
  (GreenTech)}, pp.~51--55, IEEE, 2016.

\bibitem{misra2019detour}
S.~Misra and N.~Saha, ``Detour: Dynamic task offloading in software-defined fog
  for iot applications,'' {\em IEEE Journal on Selected Areas in
  Communications}, vol.~37, no.~5, pp.~1159--1166, 2019.

\bibitem{afzal2019enabling}
B.~Afzal, M.~Umair, G.~A. Shah, and E.~Ahmed, ``Enabling iot platforms for
  social iot applications: vision, feature mapping, and challenges,'' {\em
  Future Generation Computer Systems}, vol.~92, pp.~718--731, 2019.

\bibitem{celik2019program}
Z.~B. Celik, E.~Fernandes, E.~Pauley, G.~Tan, and P.~McDaniel, ``Program
  analysis of commodity iot applications for security and privacy: Challenges
  and opportunities,'' {\em ACM Computing Surveys (CSUR)}, vol.~52, no.~4,
  p.~74, 2019.

\bibitem{cheng2017traffic}
S.-M. Cheng, P.-Y. Chen, C.-C. Lin, and H.-C. Hsiao, ``Traffic-aware patching
  for cyber security in mobile iot,'' {\em IEEE Communications Magazine},
  vol.~55, no.~7, pp.~29--35, 2017.

\bibitem{goudos2017survey}
S.~K. Goudos, P.~I. Dallas, S.~Chatziefthymiou, and S.~Kyriazakos, ``A survey
  of iot key enabling and future technologies: 5g, mobile iot, sematic web and
  applications,'' {\em Wireless Personal Communications}, vol.~97, no.~2,
  pp.~1645--1675, 2017.

\bibitem{ghasempour2016optimum}
A.~Ghasempour, ``Optimum number of aggregators based on power consumption,
  cost, and network lifetime in advanced metering infrastructure architecture
  for smart grid internet of things,'' in {\em 13th IEEE Annual Consumer
  Communications \& Networking Conference (CCNC)}, pp.~295--296, IEEE, 2016.

\bibitem{3G3}
``Low power, wide area networks (lpwan).''
  https://www.link-labs.com/blog/past-present-future-lpwan [Last Accessed -
  September 2018].

\bibitem{sharma2018lorawan}
V.~Sharma, I.~You, G.~Pau, M.~Collotta, J.~D. Lim, and J.~N. Kim,
  ``Lorawan-based energy-efficient surveillance by drones for intelligent
  transportation systems,'' {\em Energies}, vol.~11, no.~3, p.~573, 2018.

\bibitem{adelantado2017understanding}
F.~Adelantado, X.~Vilajosana, P.~Tuset-Peiro, B.~Martinez, J.~Melia-Segui, and
  T.~Watteyne, ``Understanding the limits of lorawan,'' {\em IEEE
  Communications Magazine}, vol.~55, no.~9, pp.~34--40, 2017.

\bibitem{neumann2016indoor}
P.~Neumann, J.~Montavont, and T.~No{\"e}l, ``Indoor deployment of low-power
  wide area networks (lpwan): A lorawan case study,'' in {\em 12th
  International Conference on Wireless and Mobile Computing, Networking and
  Communications (WiMob)}, pp.~1--8, IEEE, 2016.

\bibitem{jermyn2015scalability}
J.~Jermyn, R.~P. Jover, I.~Murynets, M.~Istomin, and S.~Stolfo, ``Scalability
  of machine to machine systems and the internet of things on lte mobile
  networks,'' in {\em 16th International Symposium on a World of Wireless,
  Mobile and Multimedia Networks (WoWMoM)}, pp.~1--9, IEEE, 2015.

\bibitem{jover2015connection}
R.~P. Jover and I.~Murynets, ``Connection-less communication of iot devices
  over lte mobile networks,'' in {\em 12th Annual International Conference on
  Sensing, Communication, and Networking (SECON)}, pp.~247--255, IEEE, 2015.

\bibitem{chakrabarty2015black}
S.~Chakrabarty, D.~W. Engels, and S.~Thathapudi, ``Black sdn for the internet
  of things,'' in {\em 12th International Conference on Mobile Ad Hoc and
  Sensor Systems (MASS)}, pp.~190--198, IEEE, 2015.

\bibitem{ojo2016sdn}
M.~Ojo, D.~Adami, and S.~Giordano, ``A sdn-iot architecture with nfv
  implementation,'' in {\em Globecom Workshops (GC Wkshps)}, pp.~1--6, IEEE,
  2016.

\bibitem{sharma2017efficient}
V.~Sharma, F.~Song, I.~You, and H.-C. Chao, ``Efficient management and fast
  handovers in software defined wireless networks using uavs,'' {\em IEEE
  Network}, vol.~31, no.~6, pp.~78--85, 2017.

\bibitem{bi2019software}
Y.~Bi, G.~Han, S.~Xu, X.~Wang, C.~Lin, Z.~Yu, and P.~Sun, ``Software defined
  space-terrestrial integrated networks: Architecture, challenges, and
  solutions,'' {\em IEEE Network}, vol.~33, no.~1, pp.~22--28, 2019.

\bibitem{muthanna2019secure}
A.~Muthanna, A.~A~Ateya, A.~Khakimov, I.~Gudkova, A.~Abuarqoub, K.~Samouylov,
  and A.~Koucheryavy, ``Secure and reliable iot networks using fog computing
  with software-defined networking and blockchain,'' {\em Journal of Sensor and
  Actuator Networks}, vol.~8, no.~1, p.~15, 2019.

\bibitem{aksu2018advertising}
H.~Aksu, L.~Babun, M.~Conti, G.~Tolomei, and A.~S. Uluagac, ``Advertising in
  the iot era: Vision and challenges,'' {\em arXiv preprint arXiv:1802.04102},
  2018.

\bibitem{ammar2018internet}
M.~Ammar, G.~Russello, and B.~Crispo, ``Internet of things: A survey on the
  security of iot frameworks,'' {\em Journal of Information Security and
  Applications}, vol.~38, pp.~8--27, 2018.

\bibitem{sawng2017technology}
Y.-W. Sawng, H.-W. Kim, S.-J. Lee, and J.-W. Choi, ``Technology forecasting of
  iot healthcare with big data analysis,'' {\em ICCC Society of Korea},
  pp.~89--90, 2017.

\bibitem{ghasempour2015optimized}
A.~Ghasempour, ``Optimized scalable decentralized hybrid advanced metering
  infrastructure for smart grid,'' in {\em International Conference on Smart
  Grid Communications (SmartGridComm)}, pp.~223--228, IEEE, 2015.

\bibitem{ghasempour2016boptimum}
A.~Ghasempour, ``Optimum packet service and arrival rates in advanced metering
  infrastructure architecture of smart grid,'' in {\em Green Technologies
  Conference (GreenTech)}, pp.~1--5, IEEE, 2016.

\bibitem{ghasempour2016boptimized}
A.~Ghasempour, ``Optimized advanced metering infrastructure architecture of
  smart grid based on total cost, energy, and delay,'' in {\em Power \& Energy
  Society Innovative Smart Grid Technologies Conference (ISGT)}, pp.~1--6,
  IEEE, 2016.

\bibitem{chasempour2016boptimizing}
A.~Chasempour, ``Optimizing the advanced metering infrastructure architecture
  in smart grid,'' All Graduate Theses and Dissertations, 5023, 2016.

\bibitem{sharma2017energy}
V.~Sharma, F.~Song, I.~You, and M.~Atiquzzaman, ``Energy efficient device
  discovery for reliable communication in 5g-based iot and bsns using unmanned
  aerial vehicles,'' {\em Journal of Network and Computer Applications},
  vol.~97, pp.~79--95, 2017.

\bibitem{medaglia2010overview}
C.~M. Medaglia and A.~Serbanati, ``An overview of privacy and security issues
  in the internet of things,'' in {\em The Internet of Things}, pp.~389--395,
  Springer, 2010.

\bibitem{koien2011reflections}
G.~M. K{\o}ien, ``Reflections on trust in devices: an informal survey of human
  trust in an internet-of-things context,'' {\em Wireless Personal
  Communications}, vol.~61, no.~3, pp.~495--510, 2011.

\bibitem{bonetto2012secure}
R.~Bonetto, N.~Bui, V.~Lakkundi, A.~Olivereau, A.~Serbanati, and M.~Rossi,
  ``Secure communication for smart iot objects: Protocol stacks, use cases and
  practical examples,'' in {\em International Symposium on a World of Wireless,
  Mobile and Multimedia Networks (WoWMoM)}, pp.~1--7, IEEE, 2012.

\bibitem{chang2012survey}
K.-D. Chang and J.-L. Chen, ``A survey of trust management in wsns, internet of
  things and future internet,'' {\em KSII Transactions on Internet \&
  Information Systems}, vol.~6, no.~1, 2012.

\bibitem{bhattasali2013study}
T.~Bhattasali, R.~Chaki, and N.~Chaki, ``Study of security issues in pervasive
  environment of next generation internet of things,'' in {\em Computer
  Information Systems and Industrial Management}, pp.~206--217, Springer, 2013.

\bibitem{roman2013features}
R.~Roman, J.~Zhou, and J.~Lopez, ``On the features and challenges of security
  and privacy in distributed internet of things,'' {\em Computer Networks},
  vol.~57, no.~10, pp.~2266--2279, 2013.

\bibitem{yan2014survey}
Z.~Yan, P.~Zhang, and A.~V. Vasilakos, ``A survey on trust management for
  internet of things,'' {\em Journal of network and computer applications},
  vol.~42, pp.~120--134, 2014.

\bibitem{jing2014security}
Q.~Jing, A.~V. Vasilakos, J.~Wan, J.~Lu, and D.~Qiu, ``Security of the internet
  of things: perspectives and challenges,'' {\em Wireless Networks}, vol.~20,
  no.~8, pp.~2481--2501, 2014.

\bibitem{sicari2015security}
S.~Sicari, A.~Rizzardi, L.~A. Grieco, and A.~Coen-Porisini, ``Security, privacy
  and trust in internet of things: The road ahead,'' {\em Computer networks},
  vol.~76, pp.~146--164, 2015.

\bibitem{arias2015privacy}
O.~Arias, J.~Wurm, K.~Hoang, and Y.~Jin, ``Privacy and security in internet of
  things and wearable devices,'' {\em IEEE Transactions on Multi-Scale
  Computing Systems}, vol.~1, no.~2, pp.~99--109, 2015.

\bibitem{malina2016perspective}
L.~Malina, J.~Hajny, R.~Fujdiak, and J.~Hosek, ``On perspective of security and
  privacy-preserving solutions in the internet of things,'' {\em Computer
  Networks}, vol.~102, pp.~83--95, 2016.

\bibitem{li2016internet}
S.~Li, T.~Tryfonas, and H.~Li, ``The internet of things: a security point of
  view,'' {\em Internet Research}, vol.~26, no.~2, pp.~337--359, 2016.

\bibitem{zhou2017security}
J.~Zhou, Z.~Cao, X.~Dong, and A.~V. Vasilakos, ``Security and privacy for
  cloud-based iot: Challenges,'' {\em IEEE Communications Magazine}, vol.~55,
  no.~1, pp.~26--33, 2017.

\bibitem{7902207}
Y.~Yang, L.~Wu, G.~Yin, L.~Li, and H.~Zhao, ``A survey on security and privacy
  issues in internet-of-things,'' {\em IEEE Internet of Things Journal},
  vol.~4, pp.~1250--1258, Oct 2017.

\bibitem{7903611}
L.~Chen, S.~Thombre, K.~Järvinen, E.~S. Lohan, A.~Alén-Savikko,
  H.~Leppäkoski, M.~Z.~H. Bhuiyan, S.~Bu-Pasha, G.~N. Ferrara, S.~Honkala,
  J.~Lindqvist, L.~Ruotsalainen, P.~Korpisaari, and H.~Kuusniemi, ``Robustness,
  security and privacy in location-based services for future iot: A survey,''
  {\em IEEE Access}, vol.~5, pp.~8956--8977, 2017.

\bibitem{feng2017survey}
W.~Feng, Z.~Yan, H.~Zhang, K.~Zeng, Y.~Xiao, and Y.~T. Hou, ``A survey on
  security, privacy and trust in mobile crowdsourcing,'' {\em IEEE Internet of
  Things Journal}, 2017.

\bibitem{8119706}
K.~Yang, D.~Blaauw, and D.~Sylvester, ``Hardware designs for security in
  ultra-low-power iot systems: An overview and survey,'' {\em IEEE Micro},
  vol.~37, pp.~72--89, November 2017.

\bibitem{sen2018trifecta}
S.~Sen, J.~Koo, and S.~Bagchi, ``Trifecta: Security, energy efficiency, and
  communication capacity comparison for wireless iot devices,'' {\em IEEE
  Internet Computing}, vol.~22, no.~1, pp.~74--81, 2018.

\bibitem{tang2018jamming}
X.~Tang, P.~Ren, and Z.~Han, ``Jamming mitigation via hierarchical security
  game for iot communications,'' {\em IEEE Access}, 2018.

\bibitem{parne2018segb}
B.~L. Parne, S.~Gupta, and N.~S. Chaudhari, ``Segb: Security enhanced group
  based aka protocol for m2m communication in an iot enabled lte/lte-a
  network,'' {\em IEEE Access}, vol.~6, pp.~3668--3684, 2018.

\bibitem{dao2017achievable}
N.-N. Dao, Y.~Kim, S.~Jeong, M.~Park, and S.~Cho, ``Achievable multi-security
  levels for lightweight iot-enabled devices in infrastructureless peer-aware
  communications,'' {\em IEEE Access}, vol.~5, pp.~26743--26753, 2017.

\bibitem{mangia2018low}
M.~Mangia, F.~Pareschi, R.~Rovatti, and G.~Setti, ``Low-cost security of iot
  sensor nodes with rakeness-based compressed sensing: Statistical and
  known-plaintext attacks,'' {\em IEEE Transactions on Information Forensics
  and Security}, vol.~13, no.~2, pp.~327--340, 2018.

\bibitem{wang2018enabling}
J.~Wang, Z.~Hong, Y.~Zhang, and Y.~Jin, ``Enabling security-enhanced
  attestation with intel sgx for remote terminal and iot,'' {\em IEEE
  Transactions on Computer-Aided Design of Integrated Circuits and Systems},
  vol.~37, no.~1, pp.~88--96, 2018.

\bibitem{sedjelmaci2017accurate}
H.~Sedjelmaci, S.~M. Senouci, and T.~Taleb, ``An accurate security game for
  low-resource iot devices,'' {\em IEEE Transactions on Vehicular Technology},
  vol.~66, no.~10, pp.~9381--9393, 2017.

\bibitem{giuliano2017security}
R.~Giuliano, F.~Mazzenga, A.~Neri, and A.~M. Vegni, ``Security access protocols
  in iot capillary networks,'' {\em IEEE Internet of Things Journal}, vol.~4,
  no.~3, pp.~645--657, 2017.

\bibitem{yu2017lightweight}
W.~Yu and S.~K{\"o}se, ``A lightweight masked aes implementation for securing
  iot against cpa attacks,'' {\em IEEE Transactions on Circuits and Systems I:
  Regular Papers}, vol.~64, no.~11, pp.~2934--2944, 2017.

\bibitem{ulz2017secured}
T.~Ulz, T.~Pieber, A.~H{\"o}ller, S.~Haas, and C.~Steger, ``Secured and
  easy-to-use nfc-based device configuration for the internet of things,'' {\em
  IEEE Journal of Radio Frequency Identification}, vol.~1, no.~1, pp.~75--84,
  2017.

\bibitem{xu2017network}
G.~Xu, Y.~Cao, Y.~Ren, X.~Li, and Z.~Feng, ``Network security situation
  awareness based on semantic ontology and user-defined rules for internet of
  things,'' {\em IEEE Access}, vol.~5, pp.~21046--21056, 2017.

\bibitem{luo2018privacyprotector}
E.~Luo, M.~Z.~A. Bhuiyan, G.~Wang, M.~A. Rahman, J.~Wu, and M.~Atiquzzaman,
  ``Privacyprotector: Privacy-protected patient data collection in iot-based
  healthcare systems,'' {\em IEEE Communications Magazine}, vol.~56, no.~2,
  pp.~163--168, 2018.

\bibitem{ghasempour2016finding}
A.~Ghasempour and J.~H. Gunther, ``Finding the optimal number of aggregators in
  machine-to-machine advanced metering infrastructure architecture of smart
  grid based on cost, delay, and energy consumption,'' in {\em 13th Annual
  Consumer Communications \& Networking Conference (CCNC)}, pp.~960--963, IEEE,
  2016.

\bibitem{sharma2017consensus}
V.~Sharma, K.~Lee, S.~Kwon, J.~Kim, H.~Park, K.~Yim, and S.-Y. Lee, ``A
  consensus framework for reliability and mitigation of zero-day attacks in
  iot,'' {\em Security and Communication Networks}, vol.~2017, 2017.

\bibitem{kammuller2017insider}
F.~Kamm{\"u}ller, M.~Kerber, and C.~W. Probst, ``Insider threats and auctions:
  Formalization, mechanized proof, and code generation,'' {\em Journal of
  Wireless Mobile Networks, Ubiquitous Computing, and Dependable Applications},
  vol.~8, no.~1, pp.~44--78, 2017.

\bibitem{sharma2017socializing}
V.~Sharma, I.~You, and G.~Kul, ``Socializing drones for inter-service
  operability in ultra-dense wireless networks using blockchain,'' in {\em
  Proceedings of the International Workshop on Managing Insider Security
  Threats}, pp.~81--84, ACM, 2017.

\bibitem{li2017application}
G.~Li, H.~Zhou, G.~Li, and B.~Feng, ``Application-aware and dynamic security
  function chaining for mobile networks,'' {\em Journal of Internet Services
  and Information Security (JISIS)}, vol.~7, no.~4, pp.~21--34, 2017.

\bibitem{sharma2017isma}
V.~Sharma, I.~You, and R.~Kumar, ``Isma: Intelligent sensing model for
  anomalies detection in cross platform osns with a case study on iot,'' {\em
  IEEE Access}, vol.~5, pp.~3284--3301, 2017.

\bibitem{3G1}
``Iot standards and protocols.''
  https://www.postscapes.com/internet-of-things-protocols/ [Last Accessed -
  September 2018].

\bibitem{3G2}
``Gsma iot security guidelines.''
  https://www.gsma.com/iot/iot-security/iot-security-guidelines/ [Last Accessed
  - September 2018].

\bibitem{mekki2018comparative}
K.~Mekki, E.~Bajic, F.~Chaxel, and F.~Meyer, ``A comparative study of lpwan
  technologies for large-scale iot deployment,'' {\em ICT Express}, 2018.

\bibitem{petajajarvi2016evaluation}
J.~Pet{\"a}j{\"a}j{\"a}rvi, K.~Mikhaylov, M.~H{\"a}m{\"a}l{\"a}inen, and
  J.~Iinatti, ``Evaluation of lora lpwan technology for remote health and
  wellbeing monitoring,'' in {\em 10th International Symposium on Medical
  Information and Communication Technology (ISMICT)}, pp.~1--5, IEEE, 2016.

\bibitem{shelby20116lowpan}
Z.~Shelby and C.~Bormann, {\em 6LoWPAN: The wireless embedded Internet},
  vol.~43.
\newblock John Wiley \& Sons, 2011.

\bibitem{trasvina2016network}
C.~A. Trasvi{\~n}a-Moreno, R.~Blasco, R.~Casas, and {\'A}.~Asensio, ``A network
  performance analysis of lora modulation for lpwan sensor devices,'' in {\em
  Ubiquitous Computing and Ambient Intelligence}, pp.~174--181, Springer, 2016.

\bibitem{ratasuk2016nb}
R.~Ratasuk, B.~Vejlgaard, N.~Mangalvedhe, and A.~Ghosh, ``Nb-iot system for m2m
  communication,'' in {\em Wireless Communications and Networking Conference
  (WCNC)}, pp.~1--5, IEEE, 2016.

\bibitem{IEEEturl}
Standards, ``http://standards.ieee.org/innovate/iot/stds.html,'' {\em [last
  accessed March 10, 2018]}.

\bibitem{ioturl}
IoT, ``https://www.postscapes.com/internet-of-things-protocols/,'' {\em [last
  accessed March 10, 2018]}.

\bibitem{palattella2013standardized}
M.~R. Palattella, N.~Accettura, X.~Vilajosana, T.~Watteyne, L.~A. Grieco,
  G.~Boggia, and M.~Dohler, ``Standardized protocol stack for the internet of
  (important) things,'' {\em IEEE communications surveys \& tutorials},
  vol.~15, no.~3, pp.~1389--1406, 2013.

\bibitem{SG1}
``Ibm-watson iot.'' https://developer.ibm.com/iotplatform/ [Last Accessed -
  September 2018].

\bibitem{SG2}
``Microsoft azure iot suite.''
  https://www.microsoft.com/en-us/internet-of-things [Last Accessed - September
  2018].

\bibitem{SG3}
``Openiot.'' http://www.openiot.eu/ [Last Accessed - September 2018].

\bibitem{SG4}
``Ocf.'' https://openconnectivity.org/ [Last Accessed - September 2018].

\bibitem{amit2013pinpointing}
Y.~Amit, R.~Hay, R.~Saltzman, and A.~Sharabani, ``Pinpointing security
  vulnerabilities in computer software applications,'' Aug.~13 2013.
\newblock US Patent 8,510,842.

\bibitem{alhazmi2007measuring}
O.~H. Alhazmi, Y.~K. Malaiya, and I.~Ray, ``Measuring, analyzing and predicting
  security vulnerabilities in software systems,'' {\em Computers \& Security},
  vol.~26, no.~3, pp.~219--228, 2007.

\bibitem{benton2013openflow}
K.~Benton, L.~J. Camp, and C.~Small, ``Openflow vulnerability assessment,'' in
  {\em Proceedings of the second ACM SIGCOMM workshop on Hot topics in software
  defined networking}, pp.~151--152, ACM, 2013.

\bibitem{WG1}
``Open web application security project (owasp).''
  https://www.owasp.org/index.php/Main\_Page [Last Accessed - October 2018].

\bibitem{peguero2018empirical}
K.~Peguero, N.~Zhang, and X.~Cheng, ``An empirical study of the framework
  impact on the security of javascript web applications,'' in {\em Companion of
  the The Web Conference 2018 on The Web Conference 2018}, pp.~753--758,
  International World Wide Web Conferences Steering Committee, 2018.

\bibitem{fang2018deepxss}
Y.~Fang, Y.~Li, L.~Liu, and C.~Huang, ``Deepxss: Cross site scripting detection
  based on deep learning,'' in {\em Proceedings of the 2018 International
  Conference on Computing and Artificial Intelligence}, pp.~47--51, ACM, 2018.

\bibitem{sagar2018studying}
D.~Sagar, S.~Kukreja, J.~Brahma, S.~Tyagi, and P.~Jain, ``Studying open source
  vulnerability scanners for vulnerabilities in web applications,'' {\em IIOAB
  JOURNAL}, vol.~9, no.~2, pp.~43--49, 2018.

\bibitem{lancioni2018method}
G.~Lancioni, S.~Hunt, and M.~D. Wood, ``Method and system to accelerate iot
  patch propagation and reduce security vulnerabilities exposure time,'' Oct.~4
  2018.
\newblock US Patent App. 15/476,219.

\bibitem{samtani2018identifying}
S.~Samtani, S.~Yu, H.~Zhu, M.~Patton, J.~Matherly, and H.~Chen, ``Identifying
  supervisory control and data acquisition (scada) devices and their
  vulnerabilities on the internet of things (iot): A text mining approach,''
  {\em IEEE Intelligent Systems}, 2018.

\bibitem{sharma2018behavior}
V.~Sharma, G.~Choudhary, Y.~Ko, and I.~You, ``Behavior and vulnerability
  assessment of drones-enabled industrial internet of things (iiot),'' {\em
  IEEE Access}, vol.~6, pp.~43368--43383, 2018.

\bibitem{kim2017method}
K.~Kim, J.~Lee, and W.~Jung, ``Method of building a security vulnerability
  information collection and management system for analyzing the security
  vulnerabilities of iot devices,'' in {\em Advanced Multimedia and Ubiquitous
  Engineering}, pp.~205--210, Springer, 2017.

\bibitem{frustaci2018evaluating}
M.~Frustaci, P.~Pace, G.~Aloi, and G.~Fortino, ``Evaluating critical security
  issues of the iot world: present and future challenges,'' {\em IEEE Internet
  of Things Journal}, vol.~5, no.~4, pp.~2483--2495, 2018.

\bibitem{kim2017national}
K.~Kim, I.~Kim, and J.~Lim, ``National cyber security enhancement scheme for
  intelligent surveillance capacity with public iot environment,'' {\em The
  Journal of Supercomputing}, vol.~73, no.~3, pp.~1140--1151, 2017.

\bibitem{stellios2018survey}
I.~Stellios, P.~Kotzanikolaou, M.~Psarakis, C.~Alcaraz, and J.~Lopez, ``A
  survey of iot-enabled cyberattacks: Assessing attack paths to critical
  infrastructures and services,'' {\em IEEE Communications Surveys \&
  Tutorials}, 2018.

\bibitem{xie2017vulnerability}
W.~Xie, Y.~Jiang, Y.~Tang, N.~Ding, and Y.~Gao, ``Vulnerability detection in
  iot firmware: A survey,'' in {\em Parallel and Distributed Systems (ICPADS),
  2017 IEEE 23rd International Conference on}, pp.~769--772, IEEE, 2017.

\bibitem{huang2017insight}
Z.~Huang, S.~Liu, X.~Mao, K.~Chen, and J.~Li, ``Insight of the protection for
  data security under selective opening attacks,'' {\em Information Sciences},
  vol.~412, pp.~223--241, 2017.

\bibitem{li2016secure}
J.~Li, J.~Li, D.~Xie, and Z.~Cai, ``Secure auditing and deduplicating data in
  cloud,'' {\em IEEE Transactions on Computers}, vol.~65, no.~8,
  pp.~2386--2396, 2016.

\bibitem{li2017privacy}
P.~Li, J.~Li, Z.~Huang, C.-Z. Gao, W.-B. Chen, and K.~Chen,
  ``Privacy-preserving outsourced classification in cloud computing,'' {\em
  Cluster Computing}, pp.~1--10, 2017.

\bibitem{dsouza2014policy}
C.~Dsouza, G.-J. Ahn, and M.~Taguinod, ``Policy-driven security management for
  fog computing: Preliminary framework and a case study,'' in {\em 15th
  International Conference on Information Reuse and Integration (IRI)},
  pp.~16--23, IEEE, 2014.

\bibitem{cirani2015iot}
S.~Cirani, M.~Picone, P.~Gonizzi, L.~Veltri, and G.~Ferrari, ``Iot-oas: An
  oauth-based authorization service architecture for secure services in iot
  scenarios,'' {\em IEEE sensors journal}, vol.~15, no.~2, pp.~1224--1234,
  2015.

\bibitem{sahoo2015secured}
K.~S. Sahoo, B.~Sahoo, and A.~Panda, ``A secured sdn framework for iot,'' in
  {\em International Conference on Man and Machine Interfacing (MAMI)},
  pp.~1--4, IEEE, 2015.

\bibitem{pacheco2016iota}
J.~Pacheco, S.~Satam, S.~Hariri, C.~Grijalva, and H.~Berkenbrock, ``Iot
  security development framework for building trustworthy smart car services,''
  in {\em IEEE Conference on Intelligence and Security Informatics (ISI)},
  pp.~237--242, IEEE, 2016.

\bibitem{pereira2014authentication}
P.~P. Pereira, J.~Eliasson, and J.~Delsing, ``An authentication and access
  control framework for coap-based internet of things,'' in {\em 40th Annual
  Conference of the Industrial Electronics Society}, pp.~5293--5299, IEEE,
  2014.

\bibitem{gonzalez2016sdn}
C.~Gonzalez, S.~M. Charfadine, O.~Flauzac, and F.~Nolot, ``Sdn-based security
  framework for the iot in distributed grid,'' in {\em International
  Multidisciplinary Conference on Computer and Energy Science (SpliTech)},
  pp.~1--5, IEEE, 2016.

\bibitem{seitz2013authorization}
L.~Seitz, G.~Selander, and C.~Gehrmann, ``Authorization framework for the
  internet-of-things,'' in {\em 14th International Symposium and Workshops on a
  World of Wireless, Mobile and Multimedia Networks (WoWMoM)}, pp.~1--6, IEEE,
  2013.

\bibitem{hernandez2015safir}
J.~L. Hern{\'a}ndez-Ramos, M.~V. Moreno, J.~B. Bernab{\'e}, D.~G. Carrillo, and
  A.~F. Skarmeta, ``Safir: Secure access framework for iot-enabled services on
  smart buildings,'' {\em Journal of Computer and System Sciences}, vol.~81,
  no.~8, pp.~1452--1463, 2015.

\bibitem{guchhait2018hybrid}
A.~Guchhait, ``A hybrid v2v system for collision-free high-speed internet
  access in intelligent transportation system,'' {\em Transactions on Emerging
  Telecommunications Technologies}, vol.~29, no.~3, p.~e3282, 2018.

\bibitem{dagdee2009credential}
N.~Dagdee and R.~Vijaywargiya, ``Credential based hybrid access control
  methodology for shared electronic health records,'' in {\em International
  Conference on Information Management and Engineering}, pp.~624--628, IEEE,
  2009.

\bibitem{abie2012risk}
H.~Abie and I.~Balasingham, ``Risk-based adaptive security for smart iot in
  ehealth,'' in {\em Proceedings of the 7th International Conference on Body
  Area Networks}, pp.~269--275, ICST (Institute for Computer Sciences,
  Social-Informatics and Telecommunications Engineering), 2012.

\bibitem{ge2015framework}
M.~Ge and D.~S. Kim, ``A framework for modeling and assessing security of the
  internet of things,'' in {\em 21st International Conference on Parallel and
  Distributed Systems (ICPADS)}, pp.~776--781, IEEE, 2015.

\bibitem{ray2014scalable}
B.~R. Ray, J.~Abawajy, and M.~Chowdhury, ``Scalable rfid security framework and
  protocol supporting internet of things,'' {\em Computer Networks}, vol.~67,
  pp.~89--103, 2014.

\bibitem{kebande2016generic}
V.~R. Kebande and I.~Ray, ``A generic digital forensic investigation framework
  for internet of things (iot),'' in {\em 4th International Conference on
  Future Internet of Things and Cloud (FiCloud)}, pp.~356--362, IEEE, 2016.

\bibitem{wang2014vecure}
Q.~Wang and S.~Sawhney, ``Vecure: A practical security framework to protect the
  can bus of vehicles,'' in {\em International Conference on the Internet of
  Things (IOT)}, pp.~13--18, IEEE, 2014.

\bibitem{huang2016seciot}
X.~Huang, P.~Craig, H.~Lin, and Z.~Yan, ``Seciot: a security framework for the
  internet of things,'' {\em Security and communication networks}, vol.~9,
  no.~16, pp.~3083--3094, 2016.

\bibitem{mclachlan2015adaptive}
J.~G. Mclachlan, A.~J. Farrugia, and N.~T. Sullivan, ``Adaptive secondary
  authentication criteria based on account data,'' May~26 2015.
\newblock US Patent 9,043,887.

\bibitem{tao2018multi}
M.~Tao, J.~Zuo, Z.~Liu, A.~Castiglione, and F.~Palmieri, ``Multi-layer cloud
  architectural model and ontology-based security service framework for
  iot-based smart homes,'' {\em Future Generation Computer Systems}, vol.~78,
  pp.~1040--1051, 2018.

\bibitem{rahman2016securing}
A.~F.~A. Rahman, M.~Daud, and M.~Z. Mohamad, ``Securing sensor to cloud
  ecosystem using internet of things (iot) security framework,'' in {\em
  Proceedings of the International Conference on Internet of things and Cloud
  Computing}, p.~79, ACM, 2016.

\bibitem{ahmad2017unsupervised}
S.~Ahmad, A.~Lavin, S.~Purdy, and Z.~Agha, ``Unsupervised real-time anomaly
  detection for streaming data,'' {\em Neurocomputing}, vol.~262, pp.~134--147,
  2017.

\bibitem{8322278}
v.~sharma, R.~KUMAR, W.~H. Cheng, M.~Atiquzzaman, K.~Srinivasan, and A.~Zomaya,
  ``Nhad: Neuro-fuzzy based horizontal anomaly detection in online social
  networks,'' {\em IEEE Transactions on Knowledge and Data Engineering},
  pp.~1--1, 2018.

\bibitem{toledano2018real}
M.~Toledano, I.~Cohen, Y.~Ben-Simhon, and I.~Tadeski, ``Real-time anomaly
  detection system for time series at scale,'' in {\em KDD-Workshop on Anomaly
  Detection in Finance}, pp.~56--65, 2018.

\bibitem{ahmed2014network}
M.~Ahmed and A.~N. Mahmood, ``Network traffic pattern analysis using improved
  information theoretic co-clustering based collective anomaly detection,'' in
  {\em International Conference on Security and Privacy in Communication
  Systems}, pp.~204--219, Springer, 2014.

\bibitem{monniaux1999decision}
D.~Monniaux, ``Decision procedures for the analysis of cryptographic protocols
  by logics of belief,'' in {\em Proceedings of the 12th Computer Security
  Foundations Workshop}, pp.~44--54, IEEE, 1999.

\bibitem{cohen2005logical}
M.~Cohen and M.~Dam, ``Logical omniscience in the semantics of ban logic,'' in
  {\em Proceedings of the Foundations of Computer Security}, pp.~121--132,
  2005.

\bibitem{matsuo2010evaluate}
S.~Matsuo, K.~Miyazaki, A.~Otsuka, and D.~Basin, ``How to evaluate the security
  of real-life cryptographic protocols?,'' in {\em International Conference on
  Financial Cryptography and Data Security}, pp.~182--194, Springer, 2010.

\bibitem{bleumer2011random}
G.~Bleumer, ``Random oracle model,'' in {\em Encyclopedia of Cryptography and
  Security}, pp.~1027--1028, Springer, 2011.

\bibitem{mao2003modern}
W.~Mao, {\em Modern cryptography: theory and practice}.
\newblock Prentice Hall Professional Technical Reference, 2003.

\bibitem{paulson1999inductive}
L.~C. Paulson, ``Inductive analysis of the internet protocol tls,'' {\em ACM
  Transactions on Information and System Security (TISSEC)}, vol.~2, no.~3,
  pp.~332--351, 1999.

\bibitem{armando2005avispa}
A.~Armando, D.~Basin, Y.~Boichut, Y.~Chevalier, L.~Compagna, J.~Cu{\'e}llar,
  P.~H. Drielsma, P.-C. H{\'e}am, O.~Kouchnarenko, J.~Mantovani, {\em et~al.},
  ``The avispa tool for the automated validation of internet security protocols
  and applications,'' in {\em International conference on computer aided
  verification}, pp.~281--285, Springer, 2005.

\bibitem{kaufmann2013computer}
M.~Kaufmann, P.~Manolios, and J.~S. Moore, {\em Computer-aided reasoning: ACL2
  case studies}, vol.~4.
\newblock Springer Science \& Business Media, 2013.

\bibitem{kusters2009using}
R.~K{\"u}sters and T.~Truderung, ``Using proverif to analyze protocols with
  diffie-hellman exponentiation,'' in {\em 22nd Computer Security Foundations
  Symposium, CSF'09}, pp.~157--171, IEEE, 2009.

\bibitem{cremers2008scyther}
C.~J. Cremers, ``The scyther tool: Verification, falsification, and analysis of
  security protocols,'' in {\em International Conference on Computer Aided
  Verification}, pp.~414--418, Springer, 2008.

\bibitem{urien2013llcps}
P.~Urien, ``Llcps: A new security framework based on tls for nfc p2p
  applications in the internet of things,'' in {\em Consumer Communications and
  Networking Conference (CCNC)}, pp.~845--846, IEEE, 2013.

\bibitem{park2017security}
K.~C. Park and D.-H. Shin, ``Security assessment framework for iot service,''
  {\em Telecommunication Systems}, vol.~64, no.~1, pp.~193--209, 2017.

\bibitem{ma2018learning}
C.~Ma, S.~Kulshrestha, W.~Shi, Y.~Okada, and R.~Bose, ``E-learning material
  development framework supporting vr/ar based on linked data for iot security
  education,'' in {\em International Conference on Emerging Internetworking,
  Data \& Web Technologies}, pp.~479--491, Springer, 2018.

\bibitem{siboni2016advanced}
S.~Siboni, A.~Shabtai, N.~O. Tippenhauer, J.~Lee, and Y.~Elovici, ``Advanced
  security testbed framework for wearable iot devices,'' {\em ACM Transactions
  on Internet Technology (TOIT)}, vol.~16, no.~4, p.~26, 2016.

\bibitem{ning2012cyber}
H.~Ning and H.~Liu, ``Cyber-physical-social based security architecture for
  future internet of things,'' {\em Advances in Internet of Things}, vol.~2,
  no.~01, p.~1, 2012.

\bibitem{bernabe2014privacy}
J.~B. Bernabe, J.~L. Hern{\'a}ndez, M.~V. Moreno, and A.~F.~S. Gomez,
  ``Privacy-preserving security framework for a social-aware internet of
  things,'' in {\em International conference on ubiquitous computing and
  ambient intelligence}, pp.~408--415, Springer, 2014.

\bibitem{lake2014internet}
D.~Lake, R.~M.~R. Milito, M.~Morrow, and R.~Vargheese, ``Internet of things:
  Architectural framework for ehealth security,'' {\em Journal of ICT
  Standardization}, vol.~1, no.~3, pp.~301--328, 2014.

\bibitem{wang2013system}
F.~Wang, B.~Ge, L.~Zhang, Y.~Chen, Y.~Xin, and X.~Li, ``A system framework of
  security management in enterprise systems,'' {\em Systems Research and
  Behavioral Science}, vol.~30, no.~3, pp.~287--299, 2013.

\bibitem{olivier2015new}
F.~Olivier, G.~Carlos, and N.~Florent, ``New security architecture for iot
  network,'' {\em Procedia Computer Science}, vol.~52, pp.~1028--1033, 2015.

\bibitem{shafagh2014security}
H.~Shafagh and A.~Hithnawi, ``Security comes first, a public-key cryptography
  framework for the internet of things,'' in {\em IEEE International Conference
  on Distributed Computing in Sensor Systems (DCOSS)}, pp.~135--136, IEEE,
  2014.

\bibitem{kasinathan2013ids}
P.~Kasinathan, G.~Costamagna, H.~Khaleel, C.~Pastrone, and M.~A. Spirito, ``An
  ids framework for internet of things empowered by 6lowpan,'' in {\em
  Proceedings of the ACM SIGSAC conference on Computer \& communications
  security}, pp.~1337--1340, ACM, 2013.

\bibitem{flauzac2015sdn}
O.~Flauzac, C.~Gonzalez, A.~Hachani, and F.~Nolot, ``Sdn based architecture for
  iot and improvement of the security,'' in {\em 29th International Conference
  on Advanced Information Networking and Applications Workshops (WAINA)},
  pp.~688--693, IEEE, 2015.

\bibitem{sehgal2015smart}
V.~K. Sehgal, A.~Patrick, A.~Soni, and L.~Rajput, ``Smart human security
  framework using internet of things, cloud and fog computing,'' in {\em
  Intelligent distributed computing}, pp.~251--263, Springer, 2015.

\bibitem{gharibi2016internet}
M.~Gharibi, R.~Boutaba, and S.~L. Waslander, ``Internet of drones,'' {\em IEEE
  Access}, vol.~4, pp.~1148--1162, 2016.

\bibitem{sharma2017managing}
V.~Sharma, K.~Srinivasan, D.~N.~K. Jayakody, O.~Rana, and R.~Kumar, ``Managing
  service-heterogeneity using osmotic computing,'' {\em arXiv preprint
  arXiv:1704.04213}, 2017.

\bibitem{sharma2017resource}
V.~Sharma, I.~You, and R.~Kumar, ``Resource-based mobility management for video
  users in 5g using catalytic computing,'' {\em Computer Communications}, 2017.

\bibitem{li2018secure}
J.~Li, Y.~Zhang, X.~Chen, and Y.~Xiang, ``Secure attribute-based data sharing
  for resource-limited users in cloud computing,'' {\em Computers \& Security},
  vol.~72, pp.~1--12, 2018.

\bibitem{shin2017secure}
D.~Shin, V.~Sharma, J.~Kim, S.~Kwon, and I.~You, ``Secure and efficient
  protocol for route optimization in pmipv6-based smart home iot networks,''
  {\em IEEE Access}, vol.~5, pp.~11100--11117, 2017.

\bibitem{sharma2017cooperative}
V.~Sharma, I.~You, D.~N.~K. Jayakody, and M.~Atiquzzaman, ``Cooperative trust
  relaying and privacy preservation via edge-crowdsourcing in social internet
  of things,'' {\em Future Generation Computer Systems}, 2017.

\bibitem{deepa2017hhsrp}
C.~Deepa and B.~Latha, ``Hhsrp: a cluster based hybrid hierarchical secure
  routing protocol for wireless sensor networks,'' {\em Cluster Computing},
  pp.~1--17, 2017.

\bibitem{ullah2014hybrid}
S.~Ullah, M.~Imran, and M.~Alnuem, ``A hybrid and secure priority-guaranteed
  mac protocol for wireless body area network,'' {\em International Journal of
  Distributed Sensor Networks}, vol.~10, no.~2, p.~481761, 2014.

\bibitem{yang2016secure}
Y.~Yang and S.~Roy, ``Secure mac protocol for periodic smart metering data
  communication with compressive sensing,'' in {\em Globecom Workshops (GC
  Wkshps)}, pp.~1--6, IEEE, 2016.

\bibitem{li2017intelligent}
N.~Li, J.-F. Martinez-Ortega, and V.~H. Diaz, ``Intelligent cross-layer and
  reliable opportunistic routing algorithm for internet of things,'' {\em arXiv
  preprint arXiv:1710.00105}, 2017.

\bibitem{vinayagam2017adopting}
J.~K. Vinayagam, C.~Balaswamy, and K.~Soundararajan, ``Adopting cross layer
  approach for detecting and segregating malicious nodes in manet,'' in {\em
  International Conference on Signal Processing and Communication (ICSPC)},
  pp.~457--461, IEEE, 2017.

\bibitem{adibi2017novel}
S.~Adibi, ``A novel energy-efficient cross-application-layer platform with
  qos-security support,'' {\em International Journal of Communication Systems},
  vol.~30, no.~2, 2017.

\bibitem{chze2014secure}
P.~L.~R. Chze and K.~S. Leong, ``A secure multi-hop routing for iot
  communication,'' in {\em World Forum on Internet of Things (WF-IoT)},
  pp.~428--432, IEEE, 2014.

\bibitem{raza2012lightweight}
S.~Raza, T.~Voigt, and V.~Jutvik, ``Lightweight ikev2: a key management
  solution for both the compressed ipsec and the ieee 802.15. 4 security,'' in
  {\em Proceedings of the IETF workshop on smart object security}, vol.~23,
  Citeseer, 2012.

\bibitem{hummen2013tailoring}
R.~Hummen, H.~Wirtz, J.~H. Ziegeldorf, J.~Hiller, and K.~Wehrle, ``Tailoring
  end-to-end ip security protocols to the internet of things,'' in {\em 21st
  IEEE International Conference on Network Protocols (ICNP)}, pp.~1--10, IEEE,
  2013.

\bibitem{you2017spfp}
I.~You and J.-H. Lee, ``Spfp: Ticket-based secure handover for fast proxy
  mobile ipv6 in 5g networks,'' {\em Computer Networks}, vol.~129,
  pp.~363--372, 2017.

\bibitem{sharma2018secure}
V.~Sharma, I.~You, F.-Y. Leu, and M.~Atiquzzaman, ``Secure and efficient
  protocol for fast handover in 5g mobile xhaul networks,'' {\em Journal of
  Network and Computer Applications}, vol.~102, pp.~38--57, 2018.

\bibitem{guan2017extension}
J.~Guan, V.~Sharma, I.~You, and M.~Atiquzzaman, ``Extension of mih to support
  fpmipv6 for optimized heterogeneous handover,'' {\em arXiv preprint
  arXiv:1705.09835}, 2017.

\bibitem{xu2014ticket}
L.~Xu, Y.~He, X.~Chen, and X.~Huang, ``Ticket-based handoff authentication for
  wireless mesh networks,'' {\em Computer Networks}, vol.~73, pp.~185--194,
  2014.

\bibitem{yadav2017secure}
P.~Yadav and M.~Hussain, ``A secure aodv routing protocol with node
  authentication,'' in {\em International conference of Electronics,
  Communication and Aerospace Technology (ICECA)}, vol.~1, pp.~489--493, IEEE,
  2017.

\bibitem{kang2017practical}
K.~D. Kang {\em et~al.}, {\em A Practical and Lightweight Source Authentication
  Protocol Using One-Way Hash Chain in CAN}.
\newblock PhD thesis, DGIST, 2017.

\bibitem{kothmayr2013dtls}
T.~Kothmayr, C.~Schmitt, W.~Hu, M.~Br{\"u}nig, and G.~Carle, ``Dtls based
  security and two-way authentication for the internet of things,'' {\em Ad Hoc
  Networks}, vol.~11, no.~8, pp.~2710--2723, 2013.

\bibitem{porambage2014two}
P.~Porambage, C.~Schmitt, P.~Kumar, A.~Gurtov, and M.~Ylianttila, ``Two-phase
  authentication protocol for wireless sensor networks in distributed iot
  applications,'' in {\em Wireless Communications and Networking Conference
  (WCNC)}, pp.~2728--2733, IEEE, 2014.

\bibitem{amin2018light}
R.~Amin, N.~Kumar, G.~Biswas, R.~Iqbal, and V.~Chang, ``A light weight
  authentication protocol for iot-enabled devices in distributed cloud
  computing environment,'' {\em Future Generation Computer Systems}, vol.~78,
  pp.~1005--1019, 2018.

\bibitem{gope2017lightweight}
P.~Gope, R.~Amin, S.~H. Islam, N.~Kumar, and V.~K. Bhalla, ``Lightweight and
  privacy-preserving rfid authentication scheme for distributed iot
  infrastructure with secure localization services for smart city
  environment,'' {\em Future Generation Computer Systems}, 2017.

\bibitem{kalra2015secure}
S.~Kalra and S.~K. Sood, ``Secure authentication scheme for iot and cloud
  servers,'' {\em Pervasive and Mobile Computing}, vol.~24, pp.~210--223, 2015.

\bibitem{mishra2017efficient}
D.~Mishra, P.~Vijayakumar, V.~Sureshkumar, R.~Amin, S.~H. Islam, and P.~Gope,
  ``Efficient authentication protocol for secure multimedia communications in
  iot-enabled wireless sensor networks,'' {\em Multimedia Tools and
  Applications}, pp.~1--31, 2017.

\bibitem{alkuhlani2017lightweight}
A.~M.~I. Alkuhlani and S.~Thorat, ``Lightweight anonymity-preserving
  authentication and key agreement protocol for the internet of things
  environment,'' in {\em International Conference on Intelligent Information
  Technologies}, pp.~108--125, Springer, 2017.

\bibitem{sharma2017secure}
G.~Sharma and S.~Kalra, ``A secure remote user authentication scheme for smart
  cities e-governance applications,'' {\em Journal of Reliable Intelligent
  Environments}, vol.~3, no.~3, pp.~177--188, 2017.

\bibitem{rahman2017anonpri}
F.~Rahman, M.~E. Hoque, and S.~I. Ahamed, ``Anonpri: A secure anonymous private
  authentication protocol for rfid systems,'' {\em Information Sciences},
  vol.~379, pp.~195--210, 2017.

\bibitem{ermics2017key}
O.~Ermi{\c{s}}, {\c{S}}.~Bahtiyar, E.~Anar{\i}m, and M.~U.
  {\c{C}}a{\u{g}}layan, ``A key agreement protocol with partial backward
  confidentiality,'' {\em Computer Networks}, vol.~129, pp.~159--177, 2017.

\bibitem{dhillon2017lightweight}
P.~K. Dhillon and S.~Kalra, ``A lightweight biometrics based remote user
  authentication scheme for iot services,'' {\em Journal of Information
  Security and Applications}, vol.~34, pp.~255--270, 2017.

\bibitem{raza2013lithe}
S.~Raza, H.~Shafagh, K.~Hewage, R.~Hummen, and T.~Voigt, ``Lithe: Lightweight
  secure coap for the internet of things,'' {\em IEEE Sensors Journal},
  vol.~13, no.~10, pp.~3711--3720, 2013.

\bibitem{lee2014lightweight}
J.-Y. Lee, W.-C. Lin, and Y.-H. Huang, ``A lightweight authentication protocol
  for internet of things,'' in {\em International Symposium on Next-Generation
  Electronics (ISNE)}, pp.~1--2, IEEE, 2014.

\bibitem{garcia2016lightweight}
D.~Garcia-Carrillo and R.~Marin-Lopez, ``Lightweight coap-based bootstrapping
  service for the internet of things,'' {\em Sensors}, vol.~16, no.~3, p.~358,
  2016.

\bibitem{heer2011security}
T.~Heer, O.~Garcia-Morchon, R.~Hummen, S.~L. Keoh, S.~S. Kumar, and K.~Wehrle,
  ``Security challenges in the ip-based internet of things,'' {\em Wireless
  Personal Communications}, vol.~61, no.~3, pp.~527--542, 2011.

\bibitem{granjal2015security}
J.~Granjal, E.~Monteiro, and J.~S. Silva, ``Security for the internet of
  things: a survey of existing protocols and open research issues,'' {\em IEEE
  Communications Surveys \& Tutorials}, vol.~17, no.~3, pp.~1294--1312, 2015.

\bibitem{guo2017generic}
Q.~Guo, Y.~Cui, X.~Zou, and Q.~Huang, ``Generic construction of
  privacy-preserving optimistic fair exchange protocols,'' {\em Journal of
  Internet Services and Information Security (JISIS)}, vol.~7, no.~2,
  pp.~44--56, 2017.

\bibitem{accettura2014optimal}
N.~Accettura and G.~Piro, ``Optimal and secure protocols in the ietf 6tisch
  communication stack,'' in {\em 23rd international symposium on Industrial
  electronics (ISIE)}, pp.~1469--1474, IEEE, 2014.

\bibitem{syafruddin2010performance}
H.~Syafruddin and A.~S.~J. Putra, ``Performance analysis of using a reliable
  transport layer protocol for transmitting eap message over radius in
  inter-domain wlan roaming,'' in {\em International Conference on Information
  and Communication Technology for the Muslim World (ICT4M)}, pp.~G1--G5, IEEE,
  2010.

\bibitem{nakhjiri2005aaa}
M.~Nakhjiri and M.~Nakhjiri, {\em AAA and network security for mobile access:
  radius, diameter, EAP, PKI and IP mobility}.
\newblock John Wiley \& Sons, 2005.

\bibitem{irazabal2019blockchain}
J.~Irazabal, R.~O. Laprida, D.~A. Masini, and D.~B. Ponceleon, ``Blockchain
  enabled crowdsourcing,'' Apr.~25 2019.
\newblock US Patent App. 15/789,635.

\bibitem{8734799}
V.~{Sharma}, I.~{You}, D.~N.~K. {Jayakody}, D.~G. {Reina}, and K.~R. {Choo},
  ``Neural-blockchain based ultra-reliable caching for edge-enabled uav
  networks,'' {\em IEEE Transactions on Industrial Informatics}, pp.~1--1,
  2019.

\bibitem{8743548}
T.~{Hardjono}, A.~{Lipton}, and A.~{Pentland}, ``Toward an interoperability
  architecture for blockchain autonomous systems,'' {\em IEEE Transactions on
  Engineering Management}, pp.~1--12, 2019.

\bibitem{sharma2018block}
V.~Sharma, I.~You, F.~Palmieri, D.~N.~K. Jayakody, and J.~Li, ``Secure and
  energy efficient handover in fog networks using blockchain-based dmm,'' {\em
  IEEE Communications Magazine}, pp.~2--11, 10.1109/MCOM.2018.1700863, 2018.

\bibitem{alcaide2013anonymous}
A.~Alcaide, E.~Palomar, J.~Montero-Castillo, and A.~Ribagorda, ``Anonymous
  authentication for privacy-preserving iot target-driven applications,'' {\em
  computers \& security}, vol.~37, pp.~111--123, 2013.

\bibitem{evans2012efficient}
D.~Evans and D.~M. Eyers, ``Efficient data tagging for managing privacy in the
  internet of things,'' in {\em IEEE International Conference on Green
  Computing and Communications (GreenCom)}, pp.~244--248, IEEE, 2012.

\bibitem{ukil2012negotiation}
A.~Ukil, S.~Bandyopadhyay, J.~Joseph, V.~Banahatti, and S.~Lodha,
  ``Negotiation-based privacy preservation scheme in internet of things
  platform,'' in {\em Proceedings of the First International Conference on
  Security of Internet of Things}, pp.~75--84, ACM, 2012.

\bibitem{perez2017towards}
S.~P{\'e}rez, D.~Rotondi, D.~Pedone, L.~Straniero, M.~J. N{\'u}{\~n}ez, and
  F.~Gigante, ``Towards the cp-abe application for privacy-preserving secure
  data sharing in iot contexts,'' in {\em International Conference on
  Innovative Mobile and Internet Services in Ubiquitous Computing},
  pp.~917--926, Springer, 2017.

\bibitem{jayaraman2017privacy}
P.~P. Jayaraman, X.~Yang, A.~Yavari, D.~Georgakopoulos, and X.~Yi, ``Privacy
  preserving internet of things: From privacy techniques to a blueprint
  architecture and efficient implementation,'' {\em Future Generation Computer
  Systems}, vol.~76, pp.~540--549, 2017.

\bibitem{belguith2018phoabe}
S.~Belguith, N.~Kaaniche, M.~Laurent, A.~Jemai, and R.~Attia, ``Phoabe:
  Securely outsourcing multi-authority attribute based encryption with policy
  hidden for cloud assisted iot,'' {\em Computer Networks}, vol.~133,
  pp.~141--156, 2018.

\bibitem{bamasag2015lightweight}
O.~Bamasag, ``A lightweight privacy and integrity preserving data communication
  in smart grid,'' {\em European Journal of Computer Science and Information
  Technology}, vol.~3, no.~4, pp.~21--30, 2015.

\bibitem{hu2011identity}
C.~Hu, J.~Zhang, and Q.~Wen, ``An identity-based personal location system with
  protected privacy in iot,'' in {\em 4th IEEE International Conference on
  Broadband Network and Multimedia Technology (IC-BNMT)}, pp.~192--195, IEEE,
  2011.

\bibitem{doukas2012enabling}
C.~Doukas, I.~Maglogiannis, V.~Koufi, F.~Malamateniou, and G.~Vassilacopoulos,
  ``Enabling data protection through pki encryption in iot m-health devices,''
  in {\em 12th International Conference on Bioinformatics \& Bioengineering
  (BIBE)}, pp.~25--29, IEEE, 2012.

\bibitem{wang2014performance}
X.~Wang, J.~Zhang, E.~M. Schooler, and M.~Ion, ``Performance evaluation of
  attribute-based encryption: Toward data privacy in the iot,'' in {\em IEEE
  International Conference on Communications (ICC)}, pp.~725--730, IEEE, 2014.

\bibitem{li2014p3}
D.~Li, Z.~Aung, J.~Williams, and A.~Sanchez, ``P3: Privacy preservation
  protocol for automatic appliance control application in smart grid,'' {\em
  IEEE Internet of Things Journal}, vol.~1, no.~5, pp.~414--429, 2014.

\bibitem{bao2016lightweight}
H.~Bao and L.~Chen, ``A lightweight privacy-preserving scheme with data
  integrity for smart grid communications,'' {\em Concurrency and Computation:
  Practice and Experience}, vol.~28, no.~4, pp.~1094--1110, 2016.

\bibitem{bao2015new}
H.~Bao and R.~Lu, ``A new differentially private data aggregation with fault
  tolerance for smart grid communications,'' {\em IEEE Internet of Things
  Journal}, vol.~2, no.~3, pp.~248--258, 2015.

\bibitem{gong2015medical}
T.~Gong, H.~Huang, P.~Li, K.~Zhang, and H.~Jiang, ``A medical healthcare system
  for privacy protection based on iot,'' in {\em Seventh International
  Symposium on Parallel Architectures, Algorithms and Programming (PAAP)},
  pp.~217--222, IEEE, 2015.

\bibitem{lu2017lightweight}
R.~Lu, K.~Heung, A.~H. Lashkari, and A.~A. Ghorbani, ``A lightweight
  privacy-preserving data aggregation scheme for fog computing-enhanced iot,''
  {\em IEEE Access}, vol.~5, pp.~3302--3312, 2017.

\bibitem{mascetti2009longitude}
S.~Mascetti, C.~Bettini, and D.~Freni, ``Longitude: Centralized
  privacy-preserving computation of users’ proximity,'' in {\em Workshop on
  Secure Data Management}, pp.~142--157, Springer, 2009.

\bibitem{addo2014reference}
I.~D. Addo, S.~I. Ahamed, S.~S. Yau, and A.~Buduru, ``A reference architecture
  for improving security and privacy in internet of things applications,'' in
  {\em IEEE International Conference on Mobile Services (MS)}, pp.~108--115,
  IEEE, 2014.

\bibitem{SG5}
``Spore.'' http://www.lsv.fr/Software/spore/table.html [Last Accessed -
  September 2018].

\bibitem{vasilomanolakis2015security}
E.~Vasilomanolakis, J.~Daubert, M.~Luthra, V.~Gazis, A.~Wiesmaier, and
  P.~Kikiras, ``On the security and privacy of internet of things architectures
  and systems,'' in {\em International Workshop on Secure Internet of Things
  (SIoT)}, pp.~49--57, IEEE, 2015.

\bibitem{alqassem2014taxonomy}
I.~Alqassem and D.~Svetinovic, ``A taxonomy of security and privacy
  requirements for the internet of things (iot),'' in {\em IEEE International
  Conference on Industrial Engineering and Engineering Management (IEEM)},
  pp.~1244--1248, IEEE, 2014.

\bibitem{patil2014big}
H.~K. Patil and R.~Seshadri, ``Big data security and privacy issues in
  healthcare,'' in {\em IEEE International Congress on Big Data (BigData
  Congress)}, pp.~762--765, IEEE, 2014.

\bibitem{bertino2016data}
E.~Bertino, ``Data security and privacy in the iot.,'' in {\em EDBT},
  vol.~2016, pp.~1--3, 2016.

\bibitem{perera2015big}
C.~Perera, R.~Ranjan, L.~Wang, S.~U. Khan, and A.~Y. Zomaya, ``Big data privacy
  in the internet of things era,'' {\em IT Professional}, vol.~17, no.~3,
  pp.~32--39, 2015.

\bibitem{hwang2015iot}
Y.~H. Hwang, ``Iot security \& privacy: threats and challenges,'' in {\em
  Proceedings of the 1st ACM Workshop on IoT Privacy, Trust, and Security},
  pp.~1--1, ACM, 2015.

\bibitem{zhang2014iot}
Z.-K. Zhang, M.~C.~Y. Cho, C.-W. Wang, C.-W. Hsu, C.-K. Chen, and S.~Shieh,
  ``Iot security: ongoing challenges and research opportunities,'' in {\em 7th
  International Conference on Service-Oriented Computing and Applications
  (SOCA)}, pp.~230--234, IEEE, 2014.

\bibitem{jiang2017efficient}
X.~Jiang, X.~Ge, J.~Yu, F.~Kong, X.~Cheng, and R.~Hao, ``An efficient symmetric
  searchable encryption scheme for cloud storage,'' {\em Journal of Internet
  Services and Information Security (JISIS)}, vol.~7, no.~2, pp.~1--18, 2017.

\bibitem{sadeghi2015security}
A.-R. Sadeghi, C.~Wachsmann, and M.~Waidner, ``Security and privacy challenges
  in industrial internet of things,'' in {\em Proceedings of the 52nd annual
  design automation conference}, p.~54, ACM, 2015.

\bibitem{skarmeta2014decentralized}
A.~F. Skarmeta, J.~L. Hernandez-Ramos, and M.~V. Moreno, ``A decentralized
  approach for security and privacy challenges in the internet of things,'' in
  {\em World Forum on Internet of Things (WF-IoT)}, pp.~67--72, IEEE, 2014.

\bibitem{lin2015insecurity}
X.-J. Lin, L.~Sun, and H.~Qu, ``Insecurity of an anonymous authentication for
  privacy-preserving iot target-driven applications,'' {\em computers \&
  security}, vol.~48, pp.~142--149, 2015.

\bibitem{schurgot2015experiments}
M.~R. Schurgot, D.~A. Shinberg, and L.~G. Greenwald, ``Experiments with
  security and privacy in iot networks,'' in {\em 16th International Symposium
  on a World of Wireless, Mobile and Multimedia Networks (WoWMoM)}, pp.~1--6,
  IEEE, 2015.

\bibitem{campan2009data}
A.~Campan and T.~M. Truta, ``Data and structural k-anonymity in social
  networks,'' in {\em Privacy, Security, and Trust in KDD}, pp.~33--54,
  Springer, 2009.

\bibitem{cheng2010k}
J.~Cheng, A.~W.-c. Fu, and J.~Liu, ``K-isomorphism: privacy preserving network
  publication against structural attacks,'' in {\em Proceedings of the 2010 ACM
  SIGMOD International Conference on Management of data}, pp.~459--470, ACM,
  2010.

\bibitem{zou2009k}
L.~Zou, L.~Chen, and M.~T. {\"O}zsu, ``K-automorphism: A general framework for
  privacy preserving network publication,'' {\em Proceedings of the VLDB
  Endowment}, vol.~2, no.~1, pp.~946--957, 2009.

\bibitem{abomhara2014security}
M.~Abomhara and G.~M. K{\o}ien, ``Security and privacy in the internet of
  things: Current status and open issues,'' in {\em International Conference on
  Privacy and Security in Mobile Systems (PRISMS)}, pp.~1--8, IEEE, 2014.

\bibitem{weber2010internet}
R.~H. Weber, ``Internet of things--new security and privacy challenges,'' {\em
  Computer law \& security review}, vol.~26, no.~1, pp.~23--30, 2010.

\bibitem{kozlov2012security}
D.~Kozlov, J.~Veijalainen, and Y.~Ali, ``Security and privacy threats in iot
  architectures,'' in {\em Proceedings of the 7th International Conference on
  Body Area Networks}, pp.~256--262, ICST (Institute for Computer Sciences,
  Social-Informatics and Telecommunications Engineering), 2012.

\bibitem{ukil2015privacy}
A.~Ukil, S.~Bandyopadhyay, and A.~Pal, ``Privacy for iot: Involuntary privacy
  enablement for smart energy systems,'' in {\em 2015 IEEE International
  Conference on Communications (ICC)}, pp.~536--541, IEEE, 2015.

\bibitem{ukil2014iot}
A.~Ukil, S.~Bandyopadhyay, and A.~Pal, ``Iot-privacy: To be private or not to
  be private,'' in {\em IEEE Conference on Computer Communications Workshops
  (INFOCOM WKSHPS)}, pp.~123--124, IEEE, 2014.

\bibitem{dorri2017blockchain}
A.~Dorri, S.~S. Kanhere, R.~Jurdak, and P.~Gauravaram, ``Blockchain for iot
  security and privacy: The case study of a smart home,'' in {\em IEEE
  International Conference on Pervasive Computing and Communications Workshops
  (PerCom Workshops)}, pp.~618--623, IEEE, 2017.

\bibitem{abhishta2017comparing}
A.~Abhishta, R.~Joosten, and L.~J. Nieuwenhuis, ``Comparing alternatives to
  measure the impact of ddos attack announcements on target stock prices,''
  {\em Journal of wireless mobile networks, ubiquitous computing, and
  dependable applications}, vol.~8, no.~4, pp.~1--18, 2017.

\bibitem{kotenko2017parallel}
I.~Kotenko, I.~Saenko, and A.~Kushnerevich, ``Parallel big data processing
  system for security monitoring in internet of things networks,'' {\em Journal
  of Wireless Mobile Networks, Ubiquitous Computing, and Dependable
  Applications}, vol.~8, no.~4, pp.~60--74, 2017.

\bibitem{zhou2014security}
W.~Zhou and S.~Piramuthu, ``Security/privacy of wearable fitness tracking iot
  devices,'' in {\em 9th Iberian Conference on Information Systems and
  Technologies (CISTI)}, pp.~1--5, IEEE, 2014.

\bibitem{pohls2014rerum}
H.~C. P{\"o}hls, V.~Angelakis, S.~Suppan, K.~Fischer, G.~Oikonomou, E.~Z.
  Tragos, R.~D. Rodriguez, and T.~Mouroutis, ``Rerum: Building a reliable iot
  upon privacy-and security-enabled smart objects,'' in {\em Wireless
  Communications and Networking Conference Workshops (WCNCW)}, pp.~122--127,
  IEEE, 2014.

\bibitem{sivaraman2015network}
V.~Sivaraman, H.~H. Gharakheili, A.~Vishwanath, R.~Boreli, and O.~Mehani,
  ``Network-level security and privacy control for smart-home iot devices,'' in
  {\em 11th International Conference on Wireless and Mobile Computing,
  Networking and Communications (WiMob)}, pp.~163--167, IEEE, 2015.

\bibitem{wrona2016data}
K.~Wrona, A.~de~Castro, and B.~Vasilache, ``Data-centric security in military
  applications of commercial iot technology,'' in {\em 3rd World Forum on
  Internet of Things (WF-IoT)}, pp.~239--244, IEEE, 2016.

\bibitem{alshehri2017centralized}
M.~D. Alshehri and F.~K. Hussain, ``A centralized trust management mechanism
  for the internet of things (ctm-iot),'' in {\em International Conference on
  Broadband and Wireless Computing, Communication and Applications},
  pp.~533--543, Springer, 2017.

\bibitem{chen2016trust}
R.~Chen, J.~Guo, and F.~Bao, ``Trust management for soa-based iot and its
  application to service composition,'' {\em IEEE Transactions on Services
  Computing}, vol.~9, no.~3, pp.~482--495, 2016.

\bibitem{wu2017multi}
X.~Wu and F.~Li, ``A multi-domain trust management model for supporting rfid
  applications of iot,'' {\em PloS one}, vol.~12, no.~7, p.~e0181124, 2017.

\bibitem{guo2017mobile}
J.~Guo, R.~Chen, and J.~J. Tsai, ``A mobile cloud hierarchical trust management
  protocol for iot systems,'' in {\em 5th IEEE International Conference on
  Mobile Cloud Computing, Services, and Engineering (MobileCloud)},
  pp.~125--130, IEEE, 2017.

\bibitem{peshwe2017algorithm}
N.~Peshwe and D.~Das, ``Algorithm for trust based policy hidden communication
  in the internet of things,'' in {\em 42nd Conference on Local Computer
  Networks Workshops (LCN Workshops)}, pp.~148--153, IEEE, 2017.

\bibitem{ziegler2017anastacia}
S.~Ziegler, A.~Skarmeta, J.~Bernal, E.~E. Kim, and S.~Bianchi, ``Anastacia:
  Advanced networked agents for security and trust assessment in cps iot
  architectures,'' in {\em Global Internet of Things Summit (GIoTS)}, pp.~1--6,
  IEEE, 2017.

\bibitem{chahal2017fuzzy}
R.~K. Chahal and S.~Singh, ``Fuzzy rule-based expert system for determining
  trustworthiness of cloud service providers,'' {\em International Journal of
  Fuzzy Systems}, vol.~19, no.~2, pp.~338--354, 2017.

\bibitem{mahalle2013fuzzy}
P.~N. Mahalle, P.~A. Thakre, N.~R. Prasad, and R.~Prasad, ``A fuzzy approach to
  trust based access control in internet of things,'' in {\em 3rd International
  Conference on Wireless Communications, Vehicular Technology, Information
  Theory and Aerospace \& Electronic Systems (VITAE)}, pp.~1--5, IEEE, 2013.

\bibitem{son2017adaptive}
H.~Son, N.~Kang, B.~Gwak, and D.~Lee, ``An adaptive iot trust estimation scheme
  combining interaction history and stereotypical reputation,'' in {\em 14th
  IEEE Annual Consumer Communications \& Networking Conference (CCNC)},
  pp.~349--352, IEEE, 2017.

\bibitem{sharma2017computational}
V.~Sharma, I.~You, R.~Kumar, and P.~Kim, ``Computational offloading for
  efficient trust management in pervasive online social networks using osmotic
  computing,'' {\em IEEE Access}, vol.~5, pp.~5084--5103, 2017.

\bibitem{chen2011trm}
D.~Chen, G.~Chang, D.~Sun, J.~Li, J.~Jia, and X.~Wang, ``Trm-iot: A trust
  management model based on fuzzy reputation for internet of things,'' {\em
  Computer Science and Information Systems}, vol.~8, no.~4, pp.~1207--1228,
  2011.

\bibitem{khan2017trusta}
Z.~A. Khan, J.~Ullrich, A.~G. Voyiatzis, and P.~Herrmann, ``A trust-based
  resilient routing mechanism for the internet of things,'' in {\em Proceedings
  of the 12th International Conference on Availability, Reliability and
  Security}, p.~27, ACM, 2017.

\bibitem{kravari2017ordain}
K.~Kravari and N.~Bassiliades, ``Ordain: An ontology for trust management in
  the internet of things,'' in {\em OTM Confederated International Conferences
  On the Move to Meaningful Internet Systems}, pp.~216--223, Springer, 2017.

\bibitem{santos2013opcode}
I.~Santos, F.~Brezo, X.~Ugarte-Pedrero, and P.~G. Bringas, ``Opcode sequences
  as representation of executables for data-mining-based unknown malware
  detection,'' {\em Information Sciences}, vol.~231, pp.~64--82, 2013.

\bibitem{zhou2008application}
R.~Zhou, J.~Pan, X.~Tan, and H.~Xi, ``Application of clips expert system to
  malware detection system,'' in {\em International Conference on Computational
  Intelligence and Security}, vol.~1, pp.~309--314, IEEE, 2008.

\bibitem{niki2009drive}
A.~Niki, ``Drive-by download attacks: Effects and detection methods,'' in {\em
  3rd IT Security Conference for the Next Generation}, 2009.

\bibitem{gai2017multidimensional}
F.~Gai, J.~Zhang, P.~Zhu, and X.~Jiang, ``Multidimensional trust-based anomaly
  detection system in internet of things,'' in {\em International Conference on
  Wireless Algorithms, Systems, and Applications}, pp.~302--313, Springer,
  2017.

\bibitem{wang2018trust}
J.~Wang, R.~Chen, J.~J. Tsai, and D.-C. Wang, ``Trust-based mechanism design
  for cooperative spectrum sensing in cognitive radio networks,'' {\em Computer
  Communications}, vol.~116, pp.~90--100, 2018.

\bibitem{ozcelik2017hybrid}
M.~M. Ozcelik, E.~Irmak, and S.~Ozdemir, ``A hybrid trust based intrusion
  detection system for wireless sensor networks,'' in {\em International
  Symposium on Networks, Computers and Communications (ISNCC)}, pp.~1--6, IEEE,
  2017.

\bibitem{hinarejos2018risklaine}
M.~F. Hinarejos, F.~Almen{\'a}rez, P.~Arias-Cabarcos, J.-L. Ferrer-Gomila, and
  A.~Mar{\'\i}n, ``Risklaine: A probabilistic approach for assessing risk in
  certificate-based security,'' {\em IEEE Transactions on Information Forensics
  and Security}, 2018.

\bibitem{obaidi2017persona}
A.~A. Obaidi and E.~W. Yocam, ``Persona and device based certificate
  management,'' July~6 2017.
\newblock US Patent App. 14/985,273.

\bibitem{namal2015autonomic}
S.~Namal, H.~Gamaarachchi, G.~MyoungLee, and T.-W. Um, ``Autonomic trust
  management in cloud-based and highly dynamic iot applications,'' in {\em ITU
  Kaleidoscope: Trust in the Information Society (K-2015)}, pp.~1--8, IEEE,
  2015.

\bibitem{nitti2012subjective}
M.~Nitti, R.~Girau, L.~Atzori, A.~Iera, and G.~Morabito, ``A subjective model
  for trustworthiness evaluation in the social internet of things,'' in {\em
  23rd International Symposium on Personal Indoor and Mobile Radio
  Communications (PIMRC)}, pp.~18--23, IEEE, 2012.

\bibitem{mhetre2016trust}
N.~A. Mhetre, A.~V. Deshpande, and P.~N. Mahalle, ``Trust management model
  based on fuzzy approach for ubiquitous computing,'' {\em International
  Journal of Ambient Computing and Intelligence (IJACI)}, vol.~7, no.~2,
  pp.~33--46, 2016.

\bibitem{nitti2014trustworthiness}
M.~Nitti, R.~Girau, and L.~Atzori, ``Trustworthiness management in the social
  internet of things,'' {\em IEEE Transactions on knowledge and data
  engineering}, vol.~26, no.~5, pp.~1253--1266, 2014.

\bibitem{lize2014trust}
G.~Lize, W.~Jingpei, and S.~Bin, ``Trust management mechanism for internet of
  things,'' {\em China Communications}, vol.~11, no.~2, pp.~148--156, 2014.

\bibitem{duan2014energy}
J.~Duan, D.~Gao, D.~Yang, C.~H. Foh, and H.-H. Chen, ``An energy-aware trust
  derivation scheme with game theoretic approach in wireless sensor networks
  for iot applications,'' {\em IEEE Internet of Things Journal}, vol.~1, no.~1,
  pp.~58--69, 2014.

\bibitem{chen2014trust}
R.~Chen, J.~Guo, and F.~Bao, ``Trust management for service composition in
  soa-based iot systems,'' in {\em Wireless Communications and Networking
  Conference (WCNC)}, pp.~3444--3449, IEEE, 2014.

\bibitem{wang2013distributed}
J.~P. Wang, S.~Bin, Y.~Yu, and X.~X. Niu, ``Distributed trust management
  mechanism for the internet of things,'' in {\em Applied Mechanics and
  Materials}, vol.~347, pp.~2463--2467, Trans Tech Publ, 2013.

\bibitem{saied2013trust}
Y.~B. Saied, A.~Olivereau, D.~Zeghlache, and M.~Laurent, ``Trust management
  system design for the internet of things: A context-aware and multi-service
  approach,'' {\em Computers \& Security}, vol.~39, pp.~351--365, 2013.

\bibitem{yusof2017timely}
S.~A.~M. Yusof, N.~Zakaria, and N.~A.~R. Muton, ``Timely trust: The use of iot
  and cultural effects on swift trust formation within global virtual teams,''
  in {\em 8th International Conference on Information Technology (ICIT)},
  pp.~297--303, IEEE, 2017.

\bibitem{al2017trust}
H.~Al-Hamadi and R.~Chen, ``Trust-based decision making for health iot
  systems,'' {\em IEEE Internet of Things Journal}, vol.~4, no.~5,
  pp.~1408--1419, 2017.

\bibitem{abderrahim2017tmcoi}
O.~B. Abderrahim, M.~H. Elhdhili, and L.~Saidane, ``Tmcoi-siot: A trust
  management system based on communities of interest for the social internet of
  things,'' in {\em 13th International Wireless Communications and Mobile
  Computing Conference (IWCMC)}, pp.~747--752, IEEE, 2017.

\bibitem{khan2017trust}
Z.~A. Khan and P.~Herrmann, ``A trust based distributed intrusion detection
  mechanism for internet of things,'' in {\em IEEE 31st International
  Conference on Advanced Information Networking and Applications (AINA)},
  pp.~1169--1176, IEEE, 2017.

\bibitem{bao2012trust}
F.~Bao and R.~Chen, ``Trust management for the internet of things and its
  application to service composition,'' in {\em International Symposium on a
  World of Wireless, Mobile and Multimedia Networks (WoWMoM)}, pp.~1--6, IEEE,
  2012.

\bibitem{bao2012dynamic}
F.~Bao and I.-R. Chen, ``Dynamic trust management for internet of things
  applications,'' in {\em Proceedings of the 2012 international workshop on
  Self-aware internet of things}, pp.~1--6, ACM, 2012.

\bibitem{liu2010securing}
R.~Liu and W.~Trappe, {\em Securing wireless communications at the physical
  layer}, vol.~7.
\newblock Springer, 2010.

\bibitem{zaman2017polarization}
I.~U. Zaman, A.~B. Lopez, M.~A. Al~Faruque, and O.~Boyraz, ``Polarization mode
  dispersion-based physical layer key generation for optical fiber link
  security,'' in {\em Optical Sensors}, pp.~JTu4A--20, Optical Society of
  America, 2017.

\bibitem{xu2016security}
Q.~Xu, P.~Ren, H.~Song, and Q.~Du, ``Security enhancement for iot
  communications exposed to eavesdroppers with uncertain locations,'' {\em IEEE
  Access}, vol.~4, pp.~2840--2853, 2016.

\bibitem{chen2016securing}
B.~Chen, C.~Zhu, L.~Shu, M.~Su, J.~Wei, V.~C. Leung, and J.~J. Rodrigues,
  ``Securing uplink transmission for lightweight single-antenna ues in the
  presence of a massive mimo eavesdropper,'' {\em IEEE Access}, vol.~4,
  pp.~5374--5384, 2016.

\bibitem{zhang2016secure}
G.~Zhang and H.~Sun, ``Secure distributed detection under energy constraint in
  iot-oriented sensor networks,'' {\em Sensors}, vol.~16, no.~12, p.~2152,
  2016.

\bibitem{hu2017secure}
H.~Hu, Z.~Gao, X.~Liao, and V.~Leung, ``Secure communications in ciot networks
  with a wireless energy harvesting untrusted relay,'' {\em Sensors}, vol.~17,
  no.~9, p.~2023, 2017.

\bibitem{islam2017secured}
S.~N. Islam, M.~A. Mahmud, and A.~Oo, ``Secured communication among iot devices
  in the presence of cellular interference,'' in {\em VTC2017-Spring: Light up
  vehicular innovation: Proceedings of the 85th Vehicular Technology
  Conference}, pp.~1--6, Institute of Electrical and Electronics Engineers,
  2017.

\bibitem{choi2017physical}
J.~Choi, ``Physical layer security for channel-aware random access with
  opportunistic jamming,'' {\em IEEE Transactions on Information Forensics and
  Security}, vol.~12, no.~11, pp.~2699--2711, 2017.

\bibitem{hu2017cooperative}
L.~Hu, H.~Wen, B.~Wu, F.~Pan, R.-F. Liao, H.~Song, J.~Tang, and X.~Wang,
  ``Cooperative jamming for physical layer security enhancement in internet of
  things,'' {\em IEEE Internet of Things Journal}, 2017.

\bibitem{li2016worst}
Z.~Li, T.~Jing, L.~Ma, Y.~Huo, and J.~Qian, ``Worst-case cooperative jamming
  for secure communications in ciot networks,'' {\em Sensors}, vol.~16, no.~3,
  p.~339, 2016.

\bibitem{wei2016polarization}
D.~Wei, L.~Liang, M.~Zhang, R.~Qiao, and W.~Huang, ``A polarization state
  modulation based physical layer security scheme for wireless
  communications,'' in {\em Military Communications Conference, MILCOM},
  pp.~1195--1201, IEEE, 2016.

\bibitem{gao2016physical}
Z.~Gao, H.~Hu, D.~Cheng, J.~Xu, and X.~Sun, ``Physical layer security based on
  artificial noise and spatial modulation,'' in {\em 8th International
  Conference on Wireless Communications \& Signal Processing (WCSP)}, pp.~1--5,
  IEEE, 2016.

\bibitem{li2015compressed}
Y.~Li, T.~Jiang, and J.~Huang, ``Compressed sensing method for secret key
  generation based on mimo channel estimation,'' in {\em The Proceedings of the
  Third International Conference on Communications, Signal Processing, and
  Systems}, pp.~419--428, Springer, 2015.

\bibitem{limmanee2010secure}
A.~Limmanee and W.~Henkel, ``Secure physical-layer key generation protocol and
  key encoding in wireless communications,'' in {\em GLOBECOM Workshops (GC
  Wkshps)}, pp.~94--98, IEEE, 2010.

\bibitem{trappe2015challenges}
W.~Trappe, ``The challenges facing physical layer security,'' {\em IEEE
  communications magazine}, vol.~53, no.~6, pp.~16--20, 2015.

\bibitem{mukherjee2015physical}
A.~Mukherjee, ``Physical-layer security in the internet of things: Sensing and
  communication confidentiality under resource constraints,'' {\em Proceedings
  of the IEEE}, vol.~103, no.~10, pp.~1747--1761, 2015.

\bibitem{pecorella2016role}
T.~Pecorella, L.~Brilli, and L.~Mucchi, ``The role of physical layer security
  in iot: A novel perspective,'' {\em Information}, vol.~7, no.~3, p.~49, 2016.

\bibitem{zhang2017securing}
J.~Zhang, T.~Q. Duong, R.~Woods, and A.~Marshall, ``Securing wireless
  communications of the internet of things from the physical layer, an
  overview,'' {\em Entropy}, vol.~19, no.~8, p.~420, 2017.

\bibitem{kitana2016impact}
A.~Kitana, I.~Traore, and I.~Woungang, ``Impact study of a mobile botnet over
  lte networks.,'' {\em J. Internet Serv. Inf. Secur.}, vol.~6, no.~2,
  pp.~1--22, 2016.

\bibitem{zeng2015physical}
K.~Zeng, ``Physical layer key generation in wireless networks: challenges and
  opportunities,'' {\em IEEE Communications Magazine}, vol.~53, no.~6,
  pp.~33--39, 2015.

\bibitem{altolini2013low}
D.~Altolini, V.~Lakkundi, N.~Bui, C.~Tapparello, and M.~Rossi, ``Low power link
  layer security for iot: Implementation and performance analysis,'' in {\em
  9th International Wireless Communications and Mobile Computing Conference
  (IWCMC)}, pp.~919--925, IEEE, 2013.

\bibitem{lee2014securing}
C.~Lee, L.~Zappaterra, K.~Choi, and H.-A. Choi, ``Securing smart home:
  Technologies, security challenges, and security requirements,'' in {\em IEEE
  Conference on Communications and Network Security (CNS)}, pp.~67--72, IEEE,
  2014.

\bibitem{brilli2016physical}
L.~Brilli, T.~Pecorella, and L.~Mucchi, ``Physical layer security for iot
  devices configuration and key management-a proof of concept,'' in {\em AEIT
  International Annual Conference (AEIT)}, pp.~1--6, IEEE, 2016.

\bibitem{lee2012host}
J.-H. Lee, J.-M. Bonnin, and X.~Lagrange, ``Host-based distributed mobility
  management support protocol for ipv6 mobile networks,'' in {\em 8th
  International Conference on Wireless and Mobile Computing, Networking and
  Communications (WiMob)}, pp.~61--68, IEEE, 2012.

\bibitem{lee2013distributed}
J.-H. Lee, J.-M. Bonnin, P.~Seite, and H.~A. Chan, ``Distributed ip mobility
  management from the perspective of the ietf: motivations, requirements,
  approaches, comparison, and challenges,'' {\em IEEE Wireless Communications},
  vol.~20, no.~5, pp.~159--168, 2013.

\bibitem{ghahfarokhi2012context}
B.~S. Ghahfarokhi and N.~MOVAHHEDINIA, ``Context gathering and management for
  centralized context-aware handover in heterogeneous mobile networks,'' {\em
  Turkish Journal of Electrical Engineering \& Computer Sciences}, vol.~20,
  no.~6, pp.~914--933, 2012.

\bibitem{chai2017security}
H.-S. Chai, J.~Jeong, and C.-H. Cho, ``Security analysis of fast inter-lma
  domain handover scheme in proxy mobile ipv6 networks,'' {\em Pervasive and
  Mobile Computing}, vol.~39, pp.~100--116, 2017.

\bibitem{chai2015enhanced}
H.-S. Chai, J.-Y. Choi, and J.~Jeong, ``An enhanced secure mobility management
  scheme for building iot applications,'' {\em Procedia Computer Science},
  vol.~56, pp.~586--591, 2015.

\bibitem{ndibanje2017secure}
B.~Ndibanje, K.~Kim, Y.~Kang, H.~Kim, T.~Kim, and H.~Lee, ``A secure and
  efficient mutual authentication hand-off protocol for sensor device support
  in internet of things,'' {\em Sensors and Materials}, vol.~29, no.~7,
  pp.~953--960, 2017.

\bibitem{saxena2016authentication}
N.~Saxena, S.~Grijalva, and N.~S. Chaudhari, ``Authentication protocol for an
  iot-enabled lte network,'' {\em ACM Transactions on Internet Technology
  (TOIT)}, vol.~16, no.~4, p.~25, 2016.

\bibitem{cao2012uniform}
J.~Cao, M.~Ma, and H.~Li, ``An uniform handover authentication between e-utran
  and non-3gpp access networks,'' {\em IEEE Transactions on Wireless
  Communications}, vol.~11, no.~10, pp.~3644--3650, 2012.

\bibitem{haddad2016secure}
Z.~Haddad, M.~Mahmoud, I.~A. Saroit, and S.~Taha, ``Secure and efficient
  uniform handover scheme for lte-a networks,'' in {\em Wireless Communications
  and Networking Conference (WCNC), 2016 IEEE}, pp.~1--6, IEEE, 2016.

\bibitem{chiang2017forward}
M.-S. Chiang, C.-M. Huang, P.~B. Chau, S.~Xu, H.~Zhou, and D.~Ren, ``A forward
  fast media independent handover control scheme for proxy mobile ipv6
  (ffmih-pmipv6) over heterogeneous wireless mobile network,'' {\em
  Telecommunication Systems}, vol.~65, no.~4, pp.~699--715, 2017.

\bibitem{ameur2017enhanced}
H.~Ameur, M.~Esseghir, L.~Khoukhi, and L.~Merghem-Boulahia, ``Enhanced mih
  (media independent handover) for collaborative green wireless
  communications,'' {\em International Journal of Communication Systems},
  vol.~30, no.~7, 2017.

\bibitem{raza2014secure}
S.~Raza, S.~Duquennoy, J.~H{\"o}glund, U.~Roedig, and T.~Voigt, ``Secure
  communication for the internet of things—a comparison of link-layer
  security and ipsec for 6lowpan,'' {\em Security and Communication Networks},
  vol.~7, no.~12, pp.~2654--2668, 2014.

\bibitem{swetina2014toward}
J.~Swetina, G.~Lu, P.~Jacobs, F.~Ennesser, and J.~Song, ``Toward a standardized
  common m2m service layer platform: Introduction to onem2m,'' {\em IEEE
  Wireless Communications}, vol.~21, no.~3, pp.~20--26, 2014.

\bibitem{rajaram2017opportunistic}
A.~Rajaram, D.~N.~K. Jayakody, K.~Srinivasan, B.~Chen, and V.~Sharma,
  ``Opportunistic-harvesting: Rf wireless power transfer scheme for multiple
  access relays system,'' {\em IEEE Access}, vol.~5, pp.~16084--16099, 2017.

\bibitem{centenaro2016long}
M.~Centenaro, L.~Vangelista, A.~Zanella, and M.~Zorzi, ``Long-range
  communications in unlicensed bands: The rising stars in the iot and smart
  city scenarios,'' {\em IEEE Wireless Communications}, vol.~23, no.~5,
  pp.~60--67, 2016.

\bibitem{xu2017security}
Q.~Xu, P.~Ren, H.~Song, and Q.~Du, ``Security-aware waveforms for enhancing
  wireless communications privacy in cyber-physical systems via multipath
  receptions,'' {\em IEEE Internet of Things Journal}, vol.~4, no.~6,
  pp.~1924--1933, 2017.

\bibitem{khan2017enabling}
M.~Khan, S.~Din, M.~Gohar, A.~Ahmad, S.~Cuomo, F.~Piccialli, and G.~Jeon,
  ``Enabling multimedia aware vertical handover management in internet of
  things based heterogeneous wireless networks,'' {\em Multimedia Tools and
  Applications}, vol.~76, no.~24, pp.~25919--25941, 2017.

\bibitem{ju2015efficient}
H.~Ju and Y.~Yoo, ``Efficient packet transmission utilizing vertical handover
  in iot environment,'' {\em Journal of KIISE}, vol.~42, no.~6, pp.~807--816,
  2015.

\bibitem{luzuriaga2015handling}
J.~E. Luzuriaga, J.~C. Cano, C.~Calafate, P.~Manzoni, M.~Perez, and P.~Boronat,
  ``Handling mobility in iot applications using the mqtt protocol,'' in {\em
  Internet Technologies and Applications (ITA)}, pp.~245--250, IEEE, 2015.

\bibitem{valera2010architecture}
A.~J.~J. Valera, M.~A. Zamora, and A.~F. Skarmeta, ``An architecture based on
  internet of things to support mobility and security in medical
  environments,'' in {\em 7th IEEE Consumer Communications and Networking
  Conference (CCNC)}, pp.~1--5, IEEE, 2010.

\bibitem{gaur2017iot}
A.~S. Gaur, J.~Budakoti, C.-H. Lung, and A.~Redmond, ``Iot-equipped uav
  communications with seamless vertical handover,'' in {\em IEEE Conference on
  Dependable and Secure Computing}, pp.~459--465, IEEE, 2017.

\bibitem{baek2017spatially}
K.-D. Baek and I.-Y. Ko, ``Spatially cohesive service discovery and dynamic
  service handover for distributed iot environments,'' in {\em International
  Conference on Web Engineering}, pp.~60--78, Springer, 2017.

\bibitem{li2017sat}
T.~Li, H.~Zhou, H.~Luo, I.~You, and Q.~Xu, ``Sat-flow: multi-strategy flow
  table management for software defined satellite networks,'' {\em IEEE
  Access}, vol.~5, pp.~14952--14965, 2017.

\bibitem{lee2013comparative}
J.-H. Lee, J.-M. Bonnin, I.~You, and T.-M. Chung, ``Comparative handover
  performance analysis of ipv6 mobility management protocols,'' {\em IEEE
  Transactions on Industrial Electronics}, vol.~60, no.~3, pp.~1077--1088,
  2013.

\bibitem{sharma2017saca}
V.~Sharma, J.~D. Lim, J.~N. Kim, and I.~You, ``Saca: Self-aware communication
  architecture for iot using mobile fog servers,'' {\em Mobile Information
  Systems}, vol.~2017, 2017.

\bibitem{khan2014sensor}
R.~A. Khan and A.~Mir, ``Sensor fast proxy mobile ipv6 (sfpmipv6)-a framework
  for mobility supported ip-wsn for improving qos and building iot,'' in {\em
  International Conference on Communications and Signal Processing (ICCSP)},
  pp.~1593--1598, IEEE, 2014.

\bibitem{chen2017robustness}
L.~Chen, S.~Thombre, K.~Jarvinen, E.~S. Lohan, A.~Alen-Savikko, H.~Leppakoski,
  M.~Z.~H. Bhuiyan, S.~Bu-Pasha, G.~N. Ferrara, S.~Honkala, {\em et~al.},
  ``Robustness, security and privacy in location-based services for future iot:
  A survey,'' {\em IEEE Access}, vol.~5, pp.~8956--8977, 2017.

\bibitem{ni2018location}
L.~Ni, Y.~Yuan, X.~Wang, M.~Zhang, and J.~Zhang, ``A location privacy
  preserving scheme based on repartitioning anonymous region in mobile social
  network,'' {\em Procedia Computer Science}, vol.~129, pp.~368--371, 2018.

\bibitem{han2018caslp}
G.~Han, H.~Wang, J.~Jiang, W.~Zhang, and S.~Chan, ``Caslp: A confused arc-based
  source location privacy protection scheme in wsns for iot,'' {\em IEEE
  Communications Magazine}, vol.~56, no.~9, pp.~42--47, 2018.

\bibitem{zakhary2018location}
S.~Zakhary and A.~Benslimane, ``On location-privacy in opportunistic mobile
  networks, a survey,'' {\em Journal of Network and Computer Applications},
  vol.~103, pp.~157--170, 2018.

\bibitem{liao2017framework}
D.~Liao, G.~Sun, H.~Li, H.~Yu, and V.~Chang, ``The framework and algorithm for
  preserving user trajectory while using location-based services in iot-cloud
  systems,'' {\em Cluster Computing}, vol.~20, no.~3, pp.~2283--2297, 2017.

\bibitem{mirzamohammadi2017ditio}
S.~Mirzamohammadi, J.~A. Chen, A.~A. Sani, S.~Mehrotra, and G.~Tsudik, ``Ditio:
  Trustworthy auditing of sensor activities in mobile \& iot devices,'' in {\em
  Proceedings of the 15th ACM Conference on Embedded Network Sensor Systems},
  p.~28, ACM, 2017.

\bibitem{mao2018privacy}
T.~Mao, C.~Cao, X.~Peng, and W.~Han, ``A privacy preserving data aggregation
  scheme to investigate apps installment in massive mobile devices,'' {\em
  Procedia Computer Science}, vol.~129, pp.~331--340, 2018.

\bibitem{wang2018truthful}
Y.~Wang, Z.~Cai, X.~Tong, Y.~Gao, and G.~Yin, ``Truthful incentive mechanism
  with location privacy-preserving for mobile crowdsourcing systems,'' {\em
  Computer Networks}, vol.~135, pp.~32--43, 2018.

\bibitem{ullah2018esot}
I.~Ullah, M.~A. Shah, A.~Wahid, A.~Mehmood, and H.~Song, ``Esot: a new privacy
  model for preserving location privacy in internet of things,'' {\em
  Telecommunication Systems}, vol.~67, no.~4, pp.~553--575, 2018.

\bibitem{sun2017efficient}
G.~Sun, V.~Chang, M.~Ramachandran, Z.~Sun, G.~Li, H.~Yu, and D.~Liao,
  ``Efficient location privacy algorithm for internet of things (iot) services
  and applications,'' {\em Journal of Network and Computer Applications},
  vol.~89, pp.~3--13, 2017.

\bibitem{lopez2007network}
R.~M. Lopez, A.~Dutta, Y.~Ohba, H.~Schulzrinne, and A.~F.~G. Skarmeta,
  ``Network-layer assisted mechanism to optimize authentication delay during
  handoff in 802.11 networks,'' in {\em Fourth Annual International Conference
  on Mobile and Ubiquitous Systems: Networking \& Services}, pp.~1--8, IEEE,
  2007.

\bibitem{he2008bash}
Y.~He and D.~Perkins, ``Bash: A backhaul-aided seamless handoff scheme for
  wireless mesh networks,'' in {\em International Symposium on a World of
  Wireless, Mobile and Multimedia Networks}, pp.~1--8, IEEE, 2008.

\bibitem{fu2012efficient}
A.~Fu, Y.~Zhang, Z.~Zhu, Q.~Jing, and J.~Feng, ``An efficient handover
  authentication scheme with privacy preservation for ieee 802.16 m network,''
  {\em Computers \& Security}, vol.~31, no.~6, pp.~741--749, 2012.

\bibitem{zhang2014generic}
Y.~Zhang, X.~Chen, J.~Li, and H.~Li, ``Generic construction for secure and
  efficient handoff authentication schemes in eap-based wireless networks,''
  {\em Computer Networks}, vol.~75, pp.~192--211, 2014.

\bibitem{chien2008fast}
H.-Y. Chien, T.-H. Hsu, and Y.-L. Tang, ``Fast pre-authentication with
  minimized overhead and high security for wlan handoff,'' {\em WSEAS
  Transactions on Computers}, no.~2, pp.~46--51, 2008.

\bibitem{choi2010handover}
J.~Choi and S.~Jung, ``A handover authentication using credentials based on
  chameleon hashing,'' {\em IEEE communications letters}, vol.~14, no.~1,
  pp.~54--56, 2010.

\bibitem{al2011fast}
A.~A. Al~Shidhani and V.~C. Leung, ``Fast and secure reauthentications for 3gpp
  subscribers during wimax-wlan handovers,'' {\em IEEE transactions on
  dependable and secure computing}, vol.~8, no.~5, pp.~699--713, 2011.

\bibitem{kalong2010dynamic}
M.~Kalong, S.~Ngamsuriyaroj, and V.~Visoottiviseth, ``Dynamic key management
  for secure continuous handoff in wireless lan,'' in {\em 6th Workshop on
  Secure Network Protocols (NPSec)}, pp.~7--12, IEEE, 2010.

\bibitem{saxena2011novel}
N.~Saxena and A.~Roy, ``Novel framework for proactive handover with seamless
  multimedia over wlans,'' {\em IET communications}, vol.~5, no.~9,
  pp.~1204--1212, 2011.

\bibitem{jing2011privacy}
Q.~Jing, Y.~Zhang, A.~Fu, and X.~Liu, ``A privacy preserving handover
  authentication scheme for eap-based wireless networks,'' in {\em Global
  Telecommunications Conference (GLOBECOM 2011)}, pp.~1--6, IEEE, 2011.

\bibitem{nguyen2012enhanced}
T.~N. Nguyen and M.~Ma, ``Enhanced eap-based pre-authentication for fast and
  secure inter-asn handovers in mobile wimax networks,'' {\em IEEE transactions
  on wireless communications}, vol.~11, no.~6, pp.~2173--2181, 2012.

\bibitem{lai2014cpal}
C.~Lai, H.~Li, X.~Liang, R.~Lu, K.~Zhang, and X.~Shen, ``Cpal: A conditional
  privacy-preserving authentication with access linkability for roaming
  service,'' {\em IEEE Internet of Things Journal}, vol.~1, no.~1, pp.~46--57,
  2014.

\bibitem{chien2009secure}
H.-Y. Chien and T.-H. Hsu, ``Secure fast wlan handoff using time-bound
  delegated authentication,'' {\em International Journal of Communication
  Systems}, vol.~22, no.~5, pp.~565--584, 2009.

\bibitem{ma2013proxy}
C.~Ma, K.~Xue, and P.~Hong, ``A proxy signature based re-authentication scheme
  for secure fast handoff in wireless mesh networks.,'' {\em IJ Network
  Security}, vol.~15, no.~2, pp.~122--132, 2013.

\bibitem{he2013handauth}
D.~He, J.~Bu, S.~Chan, and C.~Chen, ``Handauth: Efficient handover
  authentication with conditional privacy for wireless networks,'' {\em IEEE
  Transactions on Computers}, vol.~62, no.~3, pp.~616--622, 2013.

\bibitem{wang2017efficient}
C.~Wang, M.~Ma, and L.~Zhang, ``An efficient eap-based pre-authentication for
  inter-wran handover in tv white space,'' {\em IEEE Access}, vol.~5,
  pp.~9785--9796, 2017.

\bibitem{cao2015ugha}
J.~Cao, H.~Li, M.~Ma, and F.~Li, ``Ugha: Uniform group-based handover
  authentication for mtc within e-utran in lte-a networks,'' in {\em
  International Conference on Communications (ICC)}, pp.~7246--7251, IEEE,
  2015.

\bibitem{kong2017achieve}
Q.~Kong, R.~Lu, S.~Chen, and H.~Zhu, ``Achieve secure handover session key
  management via mobile relay in lte-advanced networks,'' {\em IEEE Internet of
  Things Journal}, vol.~4, no.~1, pp.~29--39, 2017.

\bibitem{feirer2017seamless}
S.~Feirer and T.~Sauter, ``Seamless handover in industrial wlan using ieee
  802.11 k,'' in {\em 26th International Symposium on Industrial Electronics
  (ISIE)}, pp.~1234--1239, IEEE, 2017.

\bibitem{sharma2017IEEE}
V.~Sharma, J.~Kim, S.~Kwon, I.~You, and F.-Y. Leu, ``An overview of
  802.21a-2012 and its incorporation into iot-fog networks using osmotic
  framework,'' in {\em 3rd EAI International Conference on IoT as a Service},
  pp.~1--6, EAI, 2017.

\bibitem{sharma2017IEEE2}
V.~Sharma, J.~Kim, S.~Kwon, I.~You, and H.-C. Chen, ``Fuzzy-based protocol for
  secure remote diagnosis of iot devices in 5g networks,'' in {\em 3rd EAI
  International Conference on IoT as a Service}, pp.~1--6, EAI, 2017.

\bibitem{roman2018mobile}
R.~Roman, J.~Lopez, and M.~Mambo, ``Mobile edge computing, fog et al.: A survey
  and analysis of security threats and challenges,'' {\em Future Generation
  Computer Systems}, vol.~78, pp.~680--698, 2018.

\bibitem{mollah2017security}
M.~B. Mollah, M.~A.~K. Azad, and A.~Vasilakos, ``Security and privacy
  challenges in mobile cloud computing: Survey and way ahead,'' {\em Journal of
  Network and Computer Applications}, vol.~84, pp.~38--54, 2017.

\bibitem{palattella2016internet}
M.~R. Palattella, M.~Dohler, A.~Grieco, G.~Rizzo, J.~Torsner, T.~Engel, and
  L.~Ladid, ``Internet of things in the 5g era: Enablers, architecture, and
  business models,'' {\em IEEE Journal on Selected Areas in Communications},
  vol.~34, no.~3, pp.~510--527, 2016.

\bibitem{lin2016iot}
H.~Lin and N.~W. Bergmann, ``Iot privacy and security challenges for smart home
  environments,'' {\em Information}, vol.~7, no.~3, p.~44, 2016.

\bibitem{al2015internet}
A.~Al-Fuqaha, M.~Guizani, M.~Mohammadi, M.~Aledhari, and M.~Ayyash, ``Internet
  of things: A survey on enabling technologies, protocols, and applications,''
  {\em IEEE Communications Surveys \& Tutorials}, vol.~17, no.~4,
  pp.~2347--2376, 2015.

\bibitem{keoh2014securing}
S.~L. Keoh, S.~S. Kumar, and H.~Tschofenig, ``Securing the internet of things:
  A standardization perspective,'' {\em IEEE Internet of Things Journal},
  vol.~1, no.~3, pp.~265--275, 2014.

\bibitem{lee2018monitoring}
H.-C. Lee and K.-H. Ke, ``Monitoring of large-area iot sensors using a lora
  wireless mesh network system: Design and evaluation,'' {\em IEEE Transactions
  on Instrumentation and Measurement}, 2018.

\bibitem{garcia2018wireless}
G.~T. Garcia, V.~M. Sanchez, C.~N.~L. Marin, J.~I. Cortez, C.~A.~R. Acevedo,
  G.~S. Gonzalez, J.~L.~H. Ameca, and M.~d. C.~M. Garcia, ``Wireless sensor
  network for monitoring physical variables applied to green technology (iot
  green technology),'' {\em European Journal of Electrical Engineering and
  Computer Science}, vol.~2, no.~2, 2018.

\bibitem{wang2018vulnerability}
H.~Wang, Z.~Chen, J.~Zhao, X.~Di, and D.~Liu, ``A vulnerability assessment
  method in industrial internet of things based on attack graph and maximum
  flow,'' {\em IEEE ACCESS}, vol.~6, pp.~8599--8609, 2018.

\bibitem{sharma2018framework}
V.~Sharma, J.~Kim, S.~Kwon, I.~You, K.~Lee, and K.~Yim, ``A framework for
  mitigating zero-day attacks in iot,'' {\em arXiv preprint arXiv:1804.05549},
  2018.

\bibitem{abeshu2018deep}
A.~Abeshu and N.~Chilamkurti, ``Deep learning: The frontier for distributed
  attack detection in fog-to-things computing,'' {\em IEEE Communications
  Magazine}, vol.~56, no.~2, pp.~169--175, 2018.

\bibitem{lobato2018adaptive}
A.~G.~P. Lobato, M.~A. Lopez, I.~SANZ, A.~A. Cardenas, O.~C.~M. Duarte, and
  G.~Pujolle, ``An adaptive real-time architecture for zero-day threat
  detection,'' in {\em International Conference on Communications-ICC},
  pp.~1--6, IEEE, 2018.

\bibitem{weinberg2015internet}
B.~D. Weinberg, G.~R. Milne, Y.~G. Andonova, and F.~M. Hajjat, ``Internet of
  things: Convenience vs. privacy and secrecy,'' {\em Business Horizons},
  vol.~58, no.~6, pp.~615--624, 2015.

\bibitem{you2010malware}
I.~You and K.~Yim, ``Malware obfuscation techniques: A brief survey,'' in {\em
  International Conference on Broadband, Wireless Computing, Communication and
  Applications (BWCCA)}, pp.~297--300, IEEE, 2010.

\bibitem{lee2017profiot}
S.-Y. Lee, S.-r. Wi, E.~Seo, J.-K. Jung, and T.-M. Chung, ``Profiot: Abnormal
  behavior profiling (abp) of iot devices based on a machine learning
  approach,'' in {\em Telecommunication Networks and Applications Conference
  (ITNAC), 2017 27th International}, pp.~1--6, IEEE, 2017.

\bibitem{shaashua2018physical}
T.~M. Shaashua and O.~Shaashua, ``Physical environment profiling through
  internet of things integration platform,'' Jan.~16 2018.
\newblock US Patent 9,871,865.

\bibitem{oravec2017emerging}
J.~A. Oravec, ``Emerging “cyber hygiene” practices for the internet of
  things (iot): Professional issues in consulting clients and educating users
  on iot privacy and security,'' in {\em IEEE International Professional
  Communication Conference (ProComm)}, pp.~1--5, IEEE, 2017.

\bibitem{chowdhury2017cyber}
A.~Chowdhury, ``Cyber attacks in mechatronics systems based on internet of
  things,'' in {\em IEEE International Conference on Mechatronics (ICM)},
  pp.~476--481, IEEE, 2017.

\bibitem{chochliouros2017enabling}
I.~Chochliouros, S.~Ziegler, L.~Bolognini, N.~Alonistioti, M.~Stamatelatos,
  P.~Kontopoulos, G.~Mourikas, V.~Vlachos, N.~Gligoric, and M.~Holst,
  ``Enabling crowd-sourcing-based privacy risk assessment in eu: the privacy
  flag project,'' in {\em Proceedings of the 21st Pan-Hellenic Conference on
  Informatics}, p.~31, ACM, 2017.

\bibitem{jiang2012attack}
R.~Jiang, J.~Luo, and X.~Wang, ``An attack tree based risk assessment for
  location privacy in wireless sensor networks,'' in {\em 8th International
  Conference on Wireless Communications, Networking and Mobile Computing
  (WiCOM)}, pp.~1--4, IEEE, 2012.

\bibitem{zheng2013iot}
R.~Zheng, M.~Zhang, Q.~Wu, C.~Yang, W.~Wei, D.~Zhang, and Z.~Ma, ``An iot
  security risk autonomic assessment algorithm,'' {\em Indonesian Journal of
  Electrical Engineering and Computer Science}, vol.~11, no.~2, pp.~819--826,
  2013.

\bibitem{liao2018eavesdropping}
C.-H. Liao, H.-H. Shuai, and L.-C. Wang, ``Eavesdropping prevention for
  heterogeneous internet of things systems,'' in {\em 15th Annual Consumer
  Communications \& Networking Conference (CCNC)}, pp.~1--2, IEEE, 2018.

\bibitem{soldani20185g}
D.~Soldani, Y.~J. Guo, B.~Barani, P.~Mogensen, I.~Chih-Lin, and S.~K. Das, ``5g
  for ultra-reliable low-latency communications,'' {\em IEEE Network}, vol.~32,
  no.~2, pp.~6--7, 2018.

\end{thebibliography}

\end{document}